UNIVERSIDAD ANA G. MÉNDEZ

ESCUELA DE NEGOCIOS Y EMPRESARISMO

# IMPACTO DE LA ACTIVIDAD ECONÓMICA DIGITAL EN EL CRECIMIENTO ECONÓMICO REGIONAL: UN ESTUDIO DE CASO DEL NORTE DE MINAS GERAIS ENTRE 2009 AL 2018

por

Cesar Rommel Salas Guerra

DISERTACIÓN DOCTORAL

Presentada como Requisito para Obtención del Grado
de Doctor en Administración de Empresas
Con especialidad en Sistemas de Información

UNIVERSIDAD ANA G. MÉNDEZ

CERTIFICACIÓN DE APROBACIÓN DE DISERTACIÓN

La disertación de Cesar Rommel Salas Guerra fue revisada y aprobada por los miembros del Comité de Disertación. El formulario de Cumplimiento de Requisitos Académicos Doctorales con las firmas de los miembros del comité se encuentra depositado en el Registrador y en el Centro de Estudios Graduados e Investigación de la Universidad Ana G. Méndez.

MIEMBROS DEL COMITÉ DE DISERTACIÓN

Miembros del Comité de Disertación

Dr. Juan Carlos Sosa – Director

Dra. Maribel Ortiz Soto – Lectora

Dr. Ángel Ojeda – Lector

Dr. Luis Esteves – Lector

# RESUMEN

En la actualidad la medición económica de las oficinas nacionales de estadística no ha definido o capturado los beneficios de la actividad economía digital, debido a la baja calidad o inexistencia de metodologías. En la actualidad existe un debate relevante sobre la capacidad que tiene la actividad economía digital para generar productividad, crecimiento económico y bienestar a través de la innovación y conocimiento. Por tal razón, esta investigación identificó y estudió el conocimiento especializado, asentamiento humano y actividad económica digital como los factores que influencian en el crecimiento económico regional. Como resultado se midió el impacto que generan los nuevos modelos de operación de negocios basados en tecnología de la información. Esta investigación usó un modelo empírico de medición que permitió identificar ciertos fenómenos como los polos regionales de desarrollo económico regional (PRDE) que envuelven a las regiones económicamente florecientes. Además, demostró que los municipios con altos grados de crecimiento económico fueron impactados por la actividad económica digital y conocimiento especializado, hallazgo que es cónsono con las teorías de crecimiento económico que apuntan a la evolución tecnológica como el principal factor del crecimiento económico moderno. Por consiguiente, este estudio contribuyó con resultados beneficiosos a los gobiernos y empresas municipales para el desarrollo de estrategias enmarcadas en solventar los problemas de cooperación industrial de regiones económicamente florecientes con sus vecinos, afrontando el problema de aglomeración de recursos y capital reflejados en el asentamiento humano los cuales fomentan el desequilibrio en el crecimiento económico y desarrollo social.

# AGRADECIMIENTOS

El reto más grande luego de terminar los requerimientos académicos fue el desarrollo de la Tesis Doctoral, una etapa llena de mucho sacrificio y dedición. El apoyo innegable del comité fue de trascendental importancia por lo tanto quiero agradecer el apoyo recibido por: El Dr. Juan Carlos Sosa quien, con su sabiduría y disposición supo dirigir acertadamente el trabajo que contempla el diseño, la propuesta y disertación doctoral. Al Dr. Luiz Esteves quien, con su paciencia y buena voluntad, elementos característicos de la cultura brasileña, contribuyo en los componentes econométricos de este trabajo de investigación. Al Dr. Ángel Ojeda, quien con su tiempo y dedicación dio un impulso significativo al estudio en el área de Sistemas de información. A la Dra. Maribel Ortiz por su preocupación y cuidado en los detalles relevantes a la buena ejecución de este estudio. Por último, quiero agradecer la hospitalidad del gobierno y el pueblo de Brasil, ya que en las estancias Doctorales en Minas Gerais y Sao Paulo fue grato sentir el apoyo desinteresado y generoso tanto de las autoridades como de su ciudadanía. Queda un agradecimiento profundo y perpetuo en mi corazón para el noble pueblo brasileño.

TABLA DE CONTENIDO

LISTA DE TABLAS

LISTA DE FIGURAS

LISTA DE ABREVIACIONES

| | |
|---|---|
| AED | Actividad económica digital |
| AHU | Asentamiento humano |
| CE | Conocimiento especializado |
| CER | Crecimiento económico regional |
| CBO | Clasificación brasilera de ocupaciones |
| CNAE | Clasificación Nacional de Actividades Económicas |
| DCNT | Docentes con maestría y doctorado |
| FUNDEB | Aportación económica gubernamental en el sistema educativo |
| IA | Inteligencia artificial |
| IUPP | Instituciones universitarias públicas y privadas |
| IBGE | Instituto Brasileño de Geografía y Estadística |
| IRB | Comité de ética de la Universidad Ana G. Méndez |
| IoT | Internet de las cosas |
| LISA | Asociación local espacial |
| NOAA | Centro de Información Ambiental de los Estados Unidos |
| O2O | Factor de proximidad geográfica comercio "Online2Offline” |
| PYTHON | Lenguaje de programación interpretado de tipado dinámico |
| PLS | Análisis de mínimos cuadrados parciales |
| PIB | Producto interno bruto municipal |
| POB | Población municipal |
| PRDE | Polos regionales de desarrollo económico |
| TIC | Tecnología de información y comunicaciones |
| VAB | Valor añadido bruto |

LISTA DE MAPAS

LISTA DE APÉNDICES

LISTA DE ECUACIONES

LISTA DE GRAFICOS ESTADISTICOS

# CAPITULO I

# INTRODUCCIÓN

## 1.1 Antecedentes

Desde sus inicios las teorías económicas establecieron que el nivel de producción aumenta por la transferencia del conocimiento y tecnologías, así como la inversión en el capital humano (Cvetanović et al., 2015). La automatización de los procesos productivos ha sido fundamental en el crecimiento económico desde la revolución industrial (Agrawal et al., 2019), tomando auge con el desarrollo de la convergencia tecno-social.

Este progreso se debe al establecimiento de una nueva forma de interacción humano – información – máquina (Salas, 2017), la cual parte de nuevas características adaptativas del ser humano (Engen et al., 2017). La nueva forma de interacción del ser humano, así como sus características adaptativas han generado nuevas oportunidades en las empresas dedicadas a las tecnologías y comunicaciones, generando el desarrollo de las economías regionales (Boisier, 2016; Boschma, 2017).

Las tecnologías de la información y comunicaciones (TIC), se han integrado profundamente en los últimos años a las infraestructuras urbanas existentes (Kumar & Dahiya, 2016). Por consiguiente, Li & Nuccialleri (2016) afirman que esta integración se compone de tres efectos principales: 1) Mejora la eficiencia y la eficacia de los servicios públicos. 2) Permite la creación de nuevos servicios innovadores. 3) Fomenta la actualización de los modelos de operación existentes al desbloquear el valor económico de recursos generando el desarrollo de nuevos modelos de negocios a nivel regional (Meijer & Bolívar, 2016).

## 1.2 La arquitectura de información y agrupamiento espacial regional

Los nuevos modelos de negocios que usan la tecnología dentro de los procesos de producción están basados en las arquitecturas de la información (Paea & Baird, 2018). Los mismos, representan las estructuras subyacentes que dan forma al significado y al contenido en el manejo de datos mediante el uso de algoritmos de inteligencia artificial (IA) y modelos de análisis predictivos (Espino, 2017). Estos son usados actualmente dentro de la inteligencia de negocios (Hruschka, Campello & Freitas, 2010).

Los nuevos procesos avanzados de gestión de datos han permitido a las empresas, gobierno y organizaciones distinguir entre la concentración geográfica y la estructura regional (Forga & Valiente, 2014), vislumbrándose la relevancia de la toma de decisiones basadas en el análisis de datos (Rahman et al., 2018). Por consiguiente, la literatura establece que las prácticas de clasificación de ciudades mediante el análisis espacial de datos pueden ser útiles para la descentralización urbana y la reducción de los problemas enmarcados en la aglomeración (Lu, Zhang, Sun, & Li, 2018). Esto aliviaría la presión del aglomeramiento urbano sobre las ciudades más grandes promoviendo de tal forma el desarrollo de ciudades pequeñas (Manuel et al., 2016). No obstante, es importante resaltar que este desarrollo parte de una característica relevante de la distribución espacial en la actividad económica, la cual es la tendencia hacia la concentración (Manuel et al., 2016).

Estas tendencias en la actividad económica responden a varios factores como el tecnológico, ambiental, social, educativo, económico y espacial (Méndez, 2016; Stern, Wares, & Epner, 2017). Estos factores influyen en el desarrollo de ciertas actividades económicas concentradas en determinados espacios geográficos como por ejemplo las universidades, centros gubernamentales y empresas multinacionales que tienden hacer lugares de concentración.

### 1.3 Innovación tecnológica urbana y la economía de la información

De acuerdo con Kumar & Dahiya (2016) los efectos de la aglomeración pueden ser subsanados mediante nuevas estrategias que vinculen a las ciudades con el desarrollo de procesos de innovación tecnológica. Dado que, como lo afirma Agrawal et al. (2019) la revolución de las (TIC) ha producido efectos de difusión del conocimiento, que a su vez ha inducido una importante reasignación sectorial.

Esta transferencia de conocimiento minimiza la dependencia a externalidades tecnológicas (Ma & Zhao, 2019), ya que los procesos de industrialización y urbanización, en las ciudades vinculadas al desarrollo de innovación tecnológica, poseen ventaja competitiva en comparación con las ciudades que no lo están. Este hecho permite que las ciudades se transformen a futuro en unidades regionales con desarrollo económico (Nagy et al., 2018a).

En los últimos años las ciudades han integrado progresivamente servicios de tecnologías de la información y comunicaciones, así como otras infraestructuras y servicios urbanos. Esto ha permitido a los gobiernos locales, empresas e instituciones académicas puedan explorar nuevos enfoques para el diseño y la prestación de servicios (Li & Nuccialleri, 2016). En las ciudades al incorporar la innovación tecnológica e inversión de capital en sus infraestructuras pernoctan como motores de crecimiento y desarrollo económico a nivel regional (Rahman et al., 2018).

Las ciudades que incorporan los nuevos modelos de negocios mediante el uso de plataformas, así como el de modalidades en línea y físico, enfrentan el factor de proximidad geográfica fortalecido con el denominado comercio "Online2Offline (O2O)" (Tsai & Wang, 2015). Como ejemplos se puede observar lo que sucede con Airbnb o Uber quienes buscan encontrar consumidores en línea y llevarlos hacia una atención física.

De la misma forma las tiendas conocidas como “Marketplace” son otro ejemplo importante, donde los consumidores pueden comprar productos en línea y al mismo tiempo recibir los productos o servicios en la tienda física (Roth, 2018). Este es el caso de Amazon en los Estados Unidos y Magazine Luiza en la meso región norte de Minas Gerais en Brasil. Consecuentemente, como lo menciona Li, Nirei, & Yamana (2019); la revolución digital ha afectado la innovación y el crecimiento, así como los grados de competencia del mercado de productos en relación con el desarrollo de estos nuevos modelos de negocio.

Finalmente, al ser definida la digitalización por Neffati & Gouidar (2019) como la adopción masiva de servicios digitales conectados por parte de consumidores, empresas y gobiernos, y al ser considerada como un medio principal para mejorar la productividad (Nagy et al., 2018a). La revisión de literatura presenta a la actividad económica digital como esa capacidad necesaria para generar productividad, crecimiento económico y bienestar, mediante la construcción de economías regionales basadas en innovación y conocimiento (Brynjolfsson, Collis, Diewert, Eggers, & Fox, 2018; Méndez, 2016).

## 1.4 Planteamiento del problema de investigación

En las actuales agendas de investigación existe una relevante discusión sobre la capacidad e implicaciones que tiene la actividad economía digital en el crecimiento económico regional (Brynjolfsson et al., 2018). La necesidad e importancia de medir al ser humano con conocimiento especializado es fundamental, para obtener éxito en la actividad económica digital a través de la innovación (Mcadam & Mcadam, 2016; Vega, Miranda, Mera, & Mayo, 2018). El nuevo conocimiento generado por la innovación y la tecnología, requieren proveedores especializados y mano de obra calificada (Grillitsch & Trippl, 2018).

Otro problema identificado en la revisión de literatura son las escasas investigaciones relacionadas al impacto de la actividad económica digital en el crecimiento económico regional debido a la falta de estructuras de medición consistentes (Mcadam & Mcadam, 2016; Feliu & Rodríguez, 2017; Vega et al., 2018). Además de la necesidad de realizar estudios sobre los fenómenos socioeconómicos asociados al asentamiento humano regional y los problemas de aglomeración de ciudades prosperas las cuales dificultan el crecimiento económico regional de sus vecinos (Li & Nuccialleri, 2016; Lu & Coops, 2018; Zhao et al., 2019).

Por consiguiente, estas metodologías de medición permitirían encontrar respuesta a los problemas que envuelven a la capacidad e implicaciones que tuvo la actividad economía digital en el crecimiento económico regional (Brynjolfsson et al., 2018), y a los problemas asociados a la “aglomeración” (Forga & Valiente, 2014) de ciudades prosperas (Cordeiro, 2016), las cuales dificultan el crecimiento y desarrollo regional de sus vecinos (Ma & Zhao, 2019).

## 1.5 Objetivo de investigación

Cónsono a lo antes expuesto, esta investigación tiene como objetivo:

a) Crear un modelo empírico apoyado en la función Cobb-Douglas que permita determinar la existencia de cambios en la economía de los municipios que forman la mesoregión norte del estado de Minas Gerais en Brasil entre el 2009 al 2018.

b) Medir las implicaciones que tiene el conocimiento especializado en la actividad económica digital de la mesoregión norte del estado de Minas Gerais en Brasil entre el 2009 al 2018.

c) Medir las implicaciones que tiene la actividad económica digital en el asentamiento humano de la mesoregión norte del estado de Minas Gerais en Brasil entre el 2009 al 2018.

d) Determinar el efecto de la actividad económica digital en el crecimiento económico de la mesoregión norte del estado de Minas Gerais en Brasil entre el 2009 al 2017.

## 1.6 Preguntas de investigación

1. ¿Cuáles son los factores que están relacionados al crecimiento económico de la meso región norte de Minas Gerais en Brasil?

2. ¿El conocimiento especializado, permitirá incrementar la actividad económica digital en la meso región norte de Minas Gerais en Brasil?

3. ¿La actividad económica digital promueve el asentamiento humano en la mesoregión norte de Minas Gerais en Brasil?

4. ¿La actividad económica digital tendrá un impacto en el crecimiento económico de la meso región norte de Minas Gerais en Brasil?

### 1.7 Justificación del estudio

La realización de esta investigación servirá de base para entender los factores que se relacionan con la actividad económica digital, mediante un nuevo modelo de medición del crecimiento económico regional. Además, esta investigación permitirá obtener hallazgos de las implicaciones que tiene la actividad economía digital en el crecimiento económico regional y en ciertos fenómenos socioeconómicos que son abordados dentro del estudio como el asentamiento humano regional. De igual forma, esta investigación proveerá información importante sobre la relación que tuvo el ser humano con conocimiento especializado en la generación de innovación y productividad dentro de la actividad económica digital.

### 1.8 Clasificación de las variables de estudio

Las variables de este estudio se componen de constructos e indicadores, que se describirán a continuación según el orden que fueron considerados en el modelo conceptual. Primer constructo: Conocimiento especializado (CE), incluye los siguientes indicadores: cantidad de instituciones universitarias públicas y privadas (IUPP) que comprenden a instituciones que ofrecen educación técnica profesional, instituciones de educación tecnológica, instituciones educativas de educación subgraduada, instituciones educativas de educación graduada e instituciones educativas de educación extendida por municipio (2009-2018). Cantidad de docentes (DCNT) con doctorado y maestría, instructores de estudios profesionales, profesores de informática, profesores de ciencias, profesores de ciencias económicas y administrativas, profesores de ciencias biológicas, profesores de ingeniería, profesores de matemática y estadística, profesores de ciencias humanas de instituciones universitarias públicas y privadas por municipio (2009-2018). Cantidad de aportación económica gubernamental en el sistema educativo municipal (2009-2018).

El segundo constructo es Actividad económica digital (AED), el cual cuenta con los siguientes indicadores: cantidad de empresas (TIC) por municipio (2009-2018) y la clasificación brasileña de ocupaciones (CBO) compuesta por: cantidad de analista de redes de comunicación de datos, director de servicios de informática, gerente de producción de tecnología de información, gerente de proyectos de tecnología de información, gerente de soporte técnico de tecnología de información, administrador de base de datos, electricista de mantenimiento de líneas eléctricas y telefónicas de comunicación de datos, instalador-reparador de redes telefónicas y comunicación de datos, técnico de comunicación de datos, administrador en seguridad de información, director de servicios de informática, analista de información  (investigador de redes de información), gerente de proyectos de tecnología de información, gerente de producción en tecnología de información, gerente de soporte técnico de tecnología de información, programador de sistemas de información, técnico en mantenimiento de equipos de informática, técnico de apoyo al usuario de informática, tecnólogo en gestión de tecnología de información, ingeniero en programación de computadoras, ingeniero en equipos de computación, ingeniero de sistemas operativos en computación, instalador de equipos electrónicos (computadores y equipos auxiliares), operador de computador (inclusive micro computador), programador de internet, todo por municipio desde (2009-2018).

El tercer constructo es asentamiento humano (AHU), el cual incluye los siguientes indicadores: Población absoluta municipal (POB) (2009-2018). Por último, tenemos a la variable dependiente de este estudio, compuesta por el constructo crecimiento económico regional (CER), el cual se compone del indicador producto interno bruto municipal (PIB) (2009-2018).

## 1.9 Modelo Sugerido en la Investigación

El modelo sugerido en esta investigación busca medir las relaciones de dependencia entre las variables cocimiento especializado, actividad económica digital y asentamiento humano y crecimiento económico regional considerada en este estudio como una variable endógena (Ávila & Moreno, 2018a). Estas relaciones multivariantes están basadas en una extensa revisión de literatura y un marco teórico explicado en el capítulo posterior.

Ilustración 1

*Modelo sugerido de investigación*

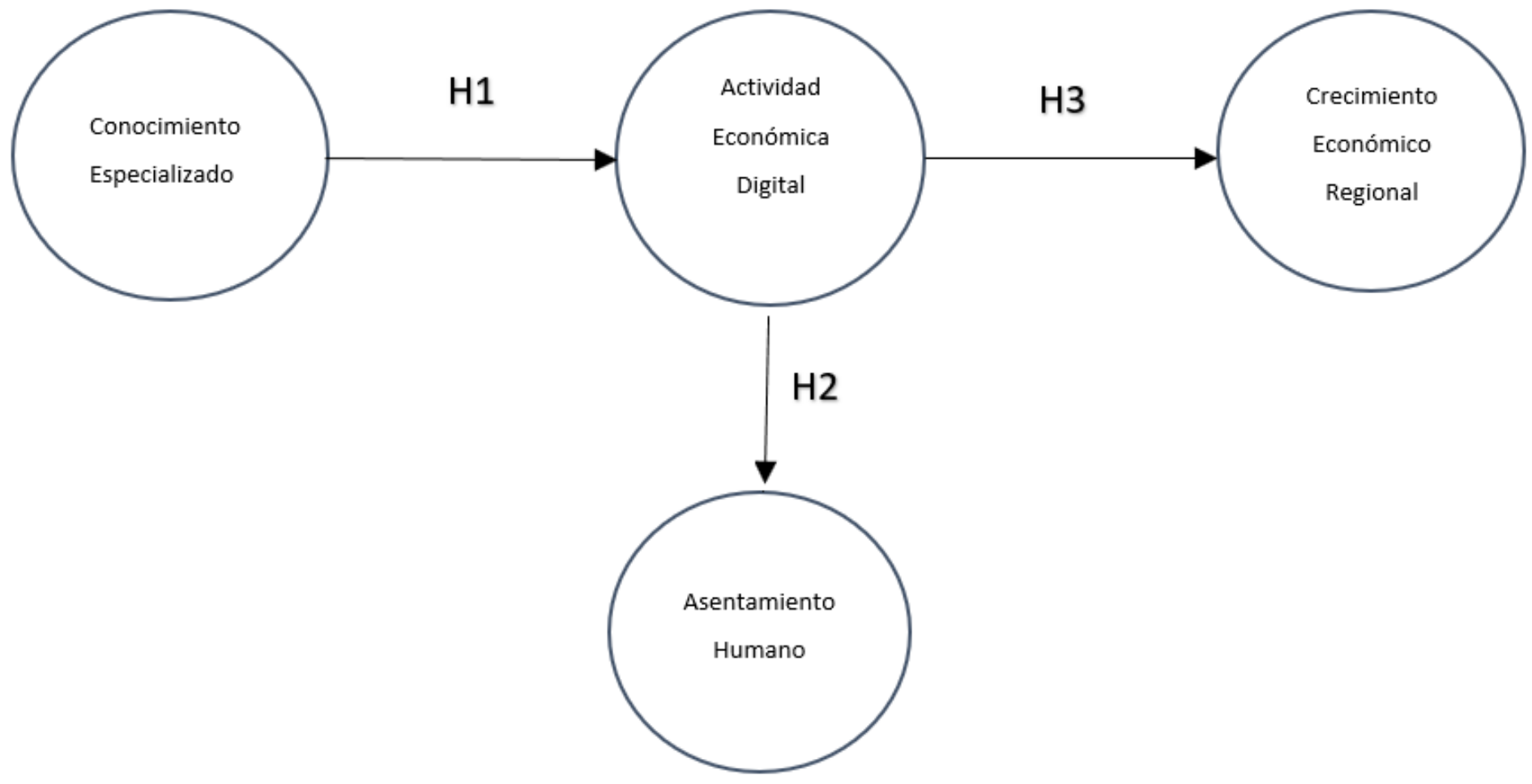

## 1.10 Matriz teórica conceptual de variables

Para demostrar la capacidad explicativa que tuvo la revisión de literatura, y con el propósito de comprobar empíricamente la nueva teoría del crecimiento, se ha elaborado una matriz que incorpora la descripción de cada componente del modelo estructural de esta investigación.

Tabla 1

*Matriz de constructos y definición de variables*

| **Código** | **Variable** | **Base Teórica** | **Definición Variables** | **Referencia** |
|---|---|---|---|---|
| **CE** | **Conocimiento Especializado** | *Nueva Teoría del Crecimiento* | **El conocimiento especializado:** Está compuesto por instituciones de nivel superior con gran importancia en el desarrollo económico regional e individuos especializados y relacionados con el conocimiento científico tecnológico con el fin de crear y capturar valor para una empresa. | (Giones & Brem, 2018; Bialetti, 2012; Ogundari & Awokuse, 2018; Cortright, 2001) |
| **AHU** | **Asentamiento Humano** | *Nueva Teoría del Crecimiento* | **El asentamiento humano**: Permite observar al crecimiento de zonas pobladas como lugares de innovación social, mediante el impulso de los servicios de las tecnologías de la información que contribuyen a la obtención de ventaja competitiva, transformando las ciudades en unidades regionales con desarrollo económico. | (Ma & Zhao, 2019; Chowdhury, Bhaduri, & McKee, 2018; Rahman et al., 2018; Meijer & Bolívar, 2016; Li, & Nuccialleri, 2016) |
| **AED** | **Actividad Económica Digital** | *Nueva Teoría del Crecimiento* | **La actividad económica digital**: Es la capacidad para generar, productividad, crecimiento económico y bienestar, mediante la construcción de economías regionales basadas en innovación y conocimiento. | (Brynjolfsson et al., 2018; Mendez, 2016; Cortright, 2001; Gomez, 2017; Manuel et al., 2016; dos Santos, 2017; Century et al., 2007) |
| **CER** | **Crecimiento Económico Regional** | *Teoría del Crecimiento* | **El crecimiento económico regional**: Se manifiesta como crecimiento de la renta per cápita y el incremento de nuevas empresas como polos de atracción para inversionistas. | (Gomez, 2017; dos Santos, 2017; Manuel et al., 2016; Hernández, 2016 Ramfla, 2015; Sierra, 2015; Century et al., 2007) |

# CAPITULO II

# REVISION DE LITERATURA

## 2.1 Introducción a la economía digital

La dificultad de definir a la economía digital es la naturaleza cambiante de la tecnología y su adopción por parte de empresas y consumidores como lo mencionan Barefoot, Curtis, Jolliff, Nicholson & Omohundro (2018). La computación en la nube, el Internet de las cosas (IoT), la analítica de los grandes datos, el aprendizaje de máquina y la inteligencia artificial son un componente dinámico de la actividad económica digital (Núñez, 2018).

De acuerdo con Brynjolfsson, Collis, Diewert, Eggers, & Fox (2018) los bienes y servicios nuevos y "gratuitos", así como su proliferación digital son una característica de la naturaleza de la economía digital. Esto ha permitido el surgimiento de nuevos modelos de negocios que incorporan la gestión de información mediante el uso de plataformas digitales (Li et al., 2019).

Entre las características que distinguen a la economía digital está el uso de la información (Gontareva et al., 2018) y la manera como las organizaciones se comunican y realizan sus transacciones (Sutherland & Jarrahi, 2018). No obstante, los individuos relacionados con el conocimiento especializado "científico-tecnológico" aportan también de valor sustancial para el desarrollo de los nuevos modelos de negocio inmersos en la actividad económica digital (Giones & Brem, 2018).

Muchos de estos nuevos modelos de negocio comercializan bienes y servicios digitales a costos mínimos, por lo tanto, investigaciones recientes examinan como la tecnología promueve a la actividad económica digital (Goldfarb, 2017; Brynjolfsson et al., 2018; Li et al., 2019; Brynjolfsson, Collis, Diewert, Eggers, & Fox, 2019). Muchos de estos estudios afirman lo expuesto por la nueva teoría del crecimiento, conocida también como la teoría del crecimiento "endógeno", la cual busca internalizar a la tecnología en un modelo de cómo funcionan actualmente los mercados (Cortright, 2001).

## 2.2 La cuarta revolución industrial

La industria es conocida como el proceso económico de producción de bienes materiales altamente automatizados (Lasi et al., 2014). No obstante, la cuarta revolución industrial ha transformado la producción gracias al uso de la tecnología en los procesos productivos, la misma incluye, pero no se limita a la inteligencia artificial y el internet de las cosas (Vasin et al., 2018).

La Industria 4.0, como se conoce a la cuarta revolución industrial, ha permitido el desarrollo de un entorno donde todos los elementos coexisten unidos de una manera continua y ubicua (Zambon et al., 2019). Aunque las computadoras y automatización existieron desde décadas anteriores este proceso de interacción busca establecer capacidades de coordinación para la mejora de la gestión productiva mediante el uso correcto de los datos (Nagy et al., 2018b).

El flujo de datos gestionados local y remotamente entre los diferentes elementos de este nuevo ecosistema de creación de valor llamado “Industria 4.0” (Stock & Seliger, 2016), permite a las empresas conectar sus maquinarias en el espectro físico-cibernético, para de esta manera desatar acciones mediante el intercambio de información de forma autónoma (Du et al., 2018).

De acuerdo con Manesh, Pellegrini & Marzi (2019) la presión de la competencia hacia las organizaciones motiva a que las empresas confíen más en el conocimiento y su aprovechamiento buscando mantener una ventaja competitiva a largo plazo generando de esta forma una prioridad para muchos centros de investigación universitarios (Du et al., 2018) que apoyan a los profesionales en el desarrollo de soluciones apropiadas.

## 2.3 La nueva teoría del crecimiento

Desde sus inicios esta teoría económica incorpora dos puntos importantes de acuerdo con Cortright (2001); en primero lugar, describe al progreso tecnológico como un producto de la actividad económica; en segundo lugar, esta teoría sostiene que, a diferencia de los objetos físicos, el conocimiento y la tecnología aportan al crecimiento económico.

Aunque la escuela neoclásica manifiesta que el crecimiento en el valor de la producción a nivel regional es el resultado de un aumento en la cantidad y calidad de trabajo (Cvetanović et al., 2015); para la nueva teoría de crecimiento la naturaleza del cambio tecnológico marca la diferencia (Cvetanović et al., 2015). Esto debido a que el nivel tecnológico de producción aumenta a través de las actividades de transferencia de conocimiento e innovación mediante la aportación de sistemas educativos innovadores (Giones & Brem, 2018).

El crecimiento endógeno es el crecimiento económico a largo plazo, y a un ritmo determinado por las fuerzas internas del sistema económico (Howitt, 2001), principalmente las que están basadas en propiciar estímulos para crear conocimiento tecnológico.  La base de la teoría del crecimiento endógeno se refleja en la ecuación $y =$ $AK$ mencionada por Pack (1994). Donde, $A$ debe entenderse como una expresión que representa a los factores tecnológicos, mientras que $K$ incluye al capital humano como el conocimiento.

El modelo más usado en los análisis teóricos y empíricos de crecimiento y productividad a lo largo de muchos años asido la función de Cobb-Douglas según Adams (2005). Debido a que la estimación de los parámetros de esta función es fundamental para gran parte de los estudios actuales, especialmente en ciertos determinantes de entrada como: el conocimiento, las habilidades del trabajo, la distinción entre capital (TIC), capital no (TIC) y el factor del progreso tecnológico (Hanclova et al., 2015).

Para Akcigit (2013) las regiones alejadas de la frontera de tecnología disfrutan de una "ventaja de atraso", esta desventaja implica que, a la larga, una región con una baja tasa de innovación quedará rezagada (Paunov, 2017). Esto debido a que las innovaciones en una región, sector o país a menudo se basan en el conocimiento creado por innovaciones en otra región, sector o país.

## 2.4 Conocimiento especializado

La revisión de literatura actual establece que el conocimiento especializado reúne a individuos que están estrechamente relacionados con el conocimiento científico tecnológico los cuales aportan en el desarrollo de la actividad económica digital (Giones & Brem, 2018). Estableciendo de una manera notable el papel que tienen las instituciones académicas en el desarrollo económico regional (Feliu & Rodríguez, 2017).

Como base para el desarrollo regional, Vega-Gomez et al. (2018) mencionan que el conocimiento especializado busca a nivel regional alterar las estructuras económicas mediante la a) innovación, b) detección de oportunidades y c) creación de empresas; produciendo de esta forma proveedores especializados, demanda sofisticada, mano de obra calificada, actividades educativas de investigación y servicios de apoyo creando así ventaja competitiva (Grillitsch & Trippl, 2018).

Por consiguiente, el impulso que ejercen las universidades en el desarrollo económico mediante la transferencia tecnológica universitaria (Mcadam & Mcadam, 2016) involucra una alta gama de intercambio de conocimiento universitario-comunitario. Estas nuevas tendencias de comercialización también son conocidas como spin-off académicos (Vega et al., 2018) las cuales afectan de forma positiva a los procesos de innovación y emprendimiento tecnológico empresarial (Li & Nuccialleri, 2016).

Finalmente, la revisión de literatura reconoce que existe una relación directa entre la creación de empresas y el crecimiento económico, como lo mencionan Lupiáñez, López, Navarro, & Bergamini (2017) donde exponen que el factor fundamental del crecimiento económico es la innovación a través de la actividad económica digital, las cuales son impulsadas por el conocimiento especializado.

La discusión de literatura antes presentada permite plantear la primera hipótesis de esta investigación:

*H1: El conocimiento especializado tiene un efecto positivo en la actividad económica digital.*

## 2.5 Asentamiento humano

De acuerdo con un reporte de las Naciones Unidas la población urbana ha aumentado exponencialmente de 751 millones en 1950 a 4,200 millones en 2018.

Basados en este informe, el crecimiento previsto está altamente concentrado en áreas urbanas como es el ejemplo de América del Norte con el 82% de población urbana y América del Sur con el 81% (ONU-DAES, 2019).

Para Chowdhury, Bhaduri, & McKee (2018) este crecimiento requiere de esfuerzos que garantice la existencia del bienestar común entre la población, ya que las ciudades han integrado progresivamente servicios de tecnologías de la información y comunicaciones, así como otras infraestructuras y servicios urbanos. Esto ha permitido a los gobiernos locales, empresas e instituciones académicas explorar nuevos enfoques en el diseño y la prestación de servicios basados en innovación tecnológica e inversión de capital (Li & Nuccialleri, 2016), pernoctando como motores de crecimiento y desarrollo económico a nivel regional (Rahman et al., 2018).

Por consiguiente, esta integración fomenta el desarrollo de nuevos modelos de negocios a nivel regional (Meijer & Bolívar, 2016; Li, & Nuccialleri, 2016) en las ciudades vinculadas al desarrollo de innovación tecnológica, obteniendo de esta forma ventaja competitiva en comparación con las ciudades que no lo están (Ma & Zhao, 2019). Este hecho permite que las ciudades se transformen a futuro en unidades regionales con desarrollo económico (Nagy et al., 2018a).

No obstante, esta transformación trae consigo los efectos de la aglomeración que pueden ser subsanados mediante nuevas estrategias que vinculen a las ciudades con el desarrollo de procesos de innovación tecnológica (Kumar & Dahiya, 2016). Esta tendencia hacia la concentración en la actividad económica responde a varios factores como el tecnológico, ambiental, social, educativo, económico y espacial (Méndez, 2016;

Stern, Wares, & Epner, 2017). Estos factores influyen en el desarrollo de ciertas actividades económicas concentradas en determinados espacios geográficos.

A todo esto, se debe sumar la pobre gestión de datos espacio temporales a nivel regional, donde muchos de ellos son de baja calidad o no existen (Bruederle & Hodler, 2018). Por consiguiente, la búsqueda de alternativas en la recolección de datos espacio temporales se sustentan en la existencia de tres modelos que sirven para estudiar los efectos del asentamiento humano en la actividad económica (Lu et al., 2018). Primero: el análisis de datos socioeconómicos sobre la población y economía (mediante censo o datos secundarios); segundo: los datos de redes sociales basadas en geolocalización (como es el caso de Twitter); y tercero: las imágenes de teledetección satelital.

Aunque tradicionalmente los datos usados han sido los obtenidos mediante censo o datos secundarios, hoy en día se dispone de una nueva alternativa que es el uso de imágenes nocturnas satelitales adquiridas mediante teledetección (USGS, 2019). Estos datos se usan en ciertas investigaciones para determinar una relación de la estimación demográfica y crecimiento económico regional (Lu et al., 2018) mediante estudios de predicción del aumento del producto interno bruto, densidad de población, aglomeramiento y la contaminación lumínica (Ma & Zhao, 2019).

Las imágenes de teledetección satelital se han utilizado ampliamente para capturar características espacio temporales sobre la dinámica urbanística y el desarrollo económico regional (Zhao et al., 2018). Dentro de esta dinámica se puede mencionar el monitoreo de los procesos de expansión urbana a largo plazo (Lu et al., 2018) propiciados por observaciones y mediciones explícitas, continuas y uniformes del crecimiento y actividad económica (Lu et al., 2018) las cuales están relacionadas con los cambios en los asentamientos humanos (Gallaway et al., 2010).

Para concluir, aunque las ciudades son los motores del crecimiento económico, lo cual acelera la urbanización e inversión de capital, las innovaciones tecnológicas contribuyen al crecimiento económico y desarrollo social (Rahman et al., 2018), ya que las ciudades rezagadas en el desarrollo urbano tienden a ser excluidas dentro de los procesos de aglomeración lo cual fomenta las presiones urbanas de las ciudades más grandes, impidiendo el desarrollo de ciudades pequeñas (Manuel et al., 2016).

La discusión de literatura antes presentada permite plantear la segunda hipótesis de esta investigación:

*H2:* *La actividad económica digital impulsa significativamente al asentamiento humano.*

**2.6 Actividad económica digital**

La revolución digital mediante la transición a la economía de la información conocida también como economía digital (Li & Nuccialleri, 2016) ha traído efectos en la manera en que los mercados convencionales de productos y servicios compiten. Estos efectos de acuerdo con Gutiérrez, Heijs, Buesa, & Baumert (2016) se basan en la innovación, siendo este un proceso en el que las empresas no sólo optimizan una función de beneficios, sino también establecen modelos de búsqueda y selección de tecnologías.

La revisión de literatura establece que la actividad económica digital es la capacidad para generar, productividad, crecimiento económico y bienestar, mediante la construcción de economías regionales basadas en innovación y conocimiento (Brynjolfsson et al., 2018); (Mendez, 2016). Cónsono con este pensamiento el estudio de Neffati & Gouidar (2019) establece que la competitividad de las economías depende de la capacidad para aprovechar las nuevas tecnologías, las mismas que contribuyen al producto interno bruto.

Las empresas de tecnología de información (Giones & Brem, 2018) influyen en los procesos de producción y comercialización, así como en la creación de empleo, transformación de la fuerza laboral e innovación empresarial (Neffati & Gouidar, 2019). Esto debido al desarrollo de plataformas usadas por los nuevos modelos de negocios (Agrawal et al., 2019) las cuales maximizan la gestión en el número de participantes (Lu et al., 2018).

En conclusión, el impulso de la transferencia tecnológica mediante el conocimiento especializado por parte de las empresas de tecnología de la información resuelve la necesidad de construir economías regionales basadas en innovación y conocimiento (Mendez, 2016), ya que la innovación a través de la actividad económica digital (Lupiáñez et al., 2017) es una de las causas fundamentales del crecimiento económico (Kumar & Dahiya, 2016).

La discusión de literatura antes presentada permite plantear la tercera hipótesis de esta investigación:

*H3:* *La actividad económica digital impacta significativamente el crecimiento económico regional.*

## 2.7 El Producto interno bruto y el crecimiento económico

El análisis de crecimiento económico cobró notoriedad en la década de los cincuenta (Gómez, 2017), ya que su objetivo fue explicar el crecimiento de la renta per cápita mediante su relación con otras variables. El método de medir el crecimiento económico regional es una metodología propuesta por Smith en el siglo XVIII (Sierra, 2015), la cual hoy en día permite calcular el crecimiento a nivel regional, subregional, departamental y municipal.

El ejercicio de monitorear el desarrollo de la economía microrregional se fundamenta en la identificación de variables que sirvan para generar las estadísticas del cálculo del producto interno bruto municipal (Sierra, 2015). En Brasil el Instituto Brasileño de Geografía y Estadística, conocido como IBGE, en asociación con los organismos estadísticos estatales son los encargados de esta labor, este cálculo se inició por primera vez hace 24 años en 1996.

La información obtenida para realizar este cálculo de acuerdo con el IBGE (2016) debe satisfacer los siguientes requisitos básicos: adecuación (datos correlacionados), frecuencia (datos con encuestas sistemáticas), universalidad (disponibilidad de datos), puntualidad (uso oportuno y confiable) y consistencia (datos confiables). El IBGE (2016) establece que para obtener el (PIB) de los municipios, se debe ajustar el método de cálculo de los agregados relacionados con: a) la producción, b) el consumo intermedio y c) el valor agregado bruto de cada unidad de la federación a la especificidad de las actividades y la disponibilidad de información municipal.

Por lo tanto, el uso del producto interno bruto en estudios microrregionales ha permitido obtener resultados sobre los cambios espacio temporales en el crecimiento económico regional (Hernández, 2016). A lo largo de estos últimos setenta años muchos economistas como Solow (1956), Kuznets (1973), Lucas (1988), Barro (1991), han colocado la evolución tecnológica como el principal factor del crecimiento económico moderno (Ramfla, 2015).

Para concluir este capítulo, la tesis expuesta anteriormente será validada mediante un modelo empírico que mida el crecimiento económico regional, capturando los beneficios del conocimiento especializado, el asentamiento humano y la actividad economía digital en el incremento de la renta per cápita, por lo tanto, esta investigación busca determinar la existencia de este impacto en los 89 Municipios que componen la Mesoregión Norte de Minas Gerais.

## 2.8 Modelo de medida sugerido basado en las hipótesis de investigación

La revisión de literatura expuesta anteriormente sugiere el siguiente modelo de medida, el cual busca mostrar las relaciones entre las variables latentes y variables observables (Ávila & Moreno, 2018a). Evaluando la contribución de los indicadores a cada constructo y la fiabilidad de cada variable. Este modelo expone la relación entre las variables conocimiento especializado con actividad económica digital, luego actividad económica digital con asentamiento humano y actividad económica digital con crecimiento económico regional como variable dependiente.

## 2.9 Matriz de hipótesis

Tabla 2

*Descripción de las hipótesis de investigación (Elaboración propia)*

| Código | Dirección | Base Teórica | Descripción de Hipótesis | Referencia |
|---|---|---|---|---|
| **H1** | **Conocimiento Especializado**<br><br>**Afecta**<br><br>**Actividad Económica Digital** | *Nueva Teoría del Crecimiento* | *El conocimiento especializado tiene un efecto positivo en la actividad económica digital.* | (Giones & Brem, 2018; Ogundari & Awokuse, 2018; Bialetti, 2012; Cortright, 2001) |
| **H2** | **Actividad Económica Digital**<br><br>**Impulsa**<br><br>**Asentamiento Humano** | *Nueva Teoría del Crecimiento* | *La actividad económica digital impulsa significativamente el asentamiento humano* | (Ma & Zhao, 2019; Chowdhury, Bhaduri, & McKee, 2018; Rahman et al., 2018; Meijer & Bolívar, 2016; Li, & Nuccialleri, 2016) |
| **H3** | **Actividad Económica Digital**<br><br>**Impacta**<br><br>**Crecimiento Económico Regional** | *Nueva Teoría del Crecimiento* | *La actividad económica digital impacta significativamente el crecimiento económico regional.* | (Brynjolfsson et al., 2018; Gomez, 2017; dos Santos, 2017; Mendez, 2016; Manuel et al., 2016; Century et al., 2007) |

# CAPITULO III

# METODOLOGÍA

## 3.1 Diseño metodológico de la investigación

El enfoque de la presente investigación fue cuantitativo transversal ya que se estudió en el transcurso del tiempo las variables de interés de una determinada población utilizando estadísticas de análisis espacial multivariable (Gómez, 2017), así como ecuaciones estructurales con mínimos cuadrados parciales (Ajamieh, 2016). Esto se realizó implementando una matriz de pesos espaciales para determinar ciertas relaciones de crecimiento de un municipio anverso a su vecino. El proceso se llevó a cabo de manera secuencial y con el uso de razonamiento deductivo para el análisis de la realidad objetiva que nos propusimos observar.

La estructura de la metodología estuvo enmarcada en el diseño transeccional de corte correlacional-causal porque solo se midió el nivel de correlación entre las variables para luego identificar posibles causalidades en el fenómeno a estudiar (Orengo, 2008). La recolección de los datos de panel se realizó en un momento único sobre un conjunto de unidades transversales seguidas a lo largo de nueve años desde el 2009 hasta el 2018 en fuentes públicas gubernamentales del Gobierno de Brasil.

Los datos de panel permitieron identificar la presencia de diferencias sistemáticas y no observadas entre las unidades a correlacionarse con factores cuyos efectos deben medirse (Wooldridge, 2009). Las mismas permitieron generalizar los resultados toda vez que en este estudio se propuso obtener de dicha población los datos previamente organizados en tablas diseñadas metodológicamente para tales fines (IBGE, 2016).

## 3.2 Etapas de la investigación

Tabla 3

*Etapas para el proceso de esta investigación empírica*

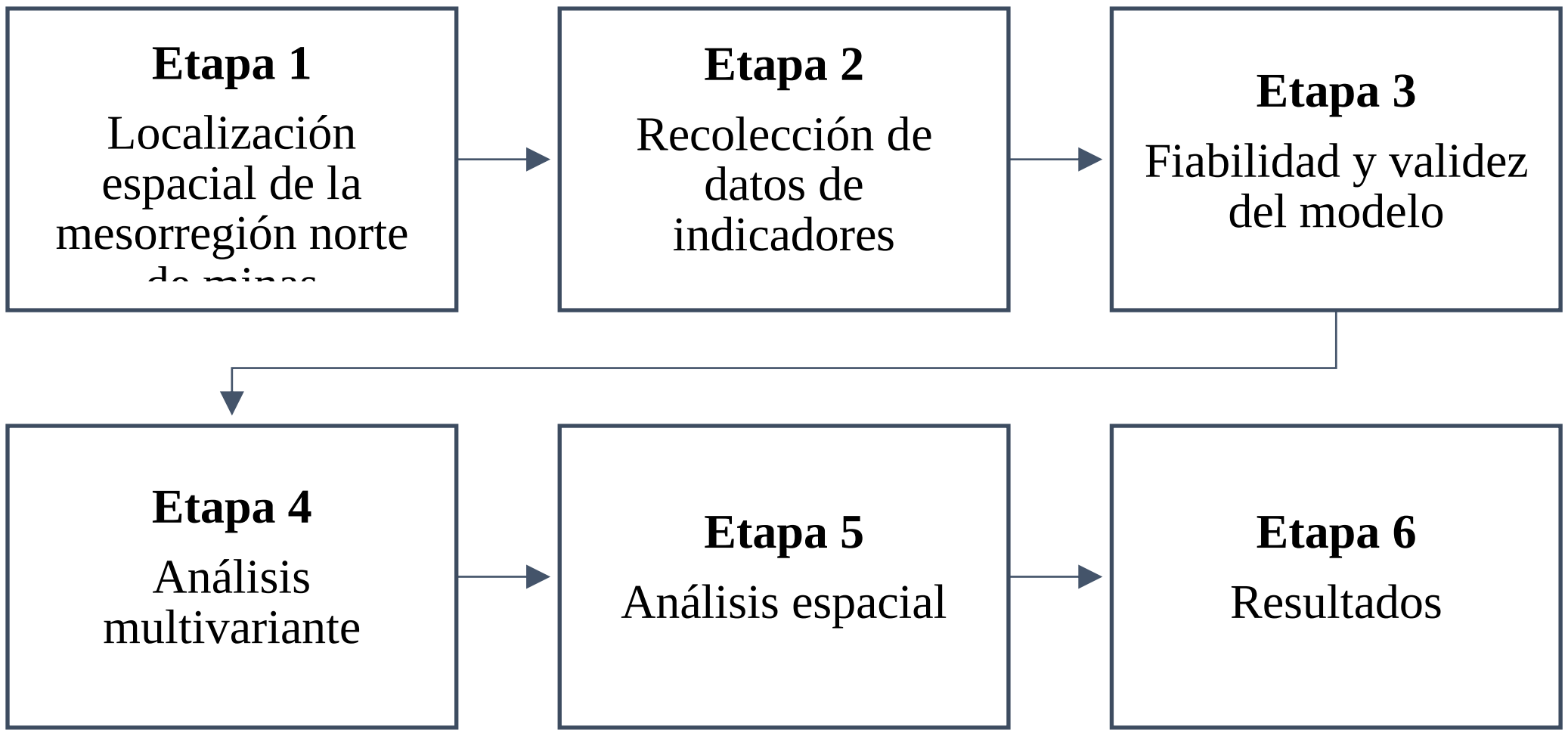

Nota: Estas etapas partieron de la premisa de establecer un proceso organizado y metodológico dentro de la investigación, el Capítulo 3 de esta Tesis cubrió la Etapa 1; el Capítulo 4 cubrió las Etapas 2, 3 y 4; y el Capítulo 5 cubrió la Etapa 5 juntamente con las limitaciones y conclusiones.

## 3.3 Población de estudio: mesorregión norte de Minas Gerais

La proyección geográfica de Brasil está compuesta por una distribución jerárquica (Bosco, 2009), donde todo el país representa el primer nivel de la jerarquía, seguido por el segundo nivel compuesto por la región norte, región noreste, región del medio oeste, región sureste y región sur. El tercer nivel está compuesto por veintisiete Estados o unidades administrativas. El cuarto nivel está representado por ciento treinta y siete mesorregiones, el quinto nivel corresponde a las microrregiones, que son subdivisiones de las mesorregiones el país cuenta con un total de 558 microrregiones (IBGE, 2019b).

Finalmente, el sexto nivel corresponde a los municipios, los cuales están compuesto por 5.565, pero en realidad dos tienen un estatus especial: el Distrito Federal el cual se considera una unidad en todos los niveles jerárquicos, es decir, es un Estado, una mesorregión, una microrregión y un municipio al mismo tiempo, y el archipiélago de Fernando de Noroña, que es un distrito estatal, perteneciente a Pernambuco, pero considerado como un municipio para fines de análisis. Este estudio no contó con muestreo ya que el mismo abarcó el total de la población compuesta por 7 microrregiones organizadas por 89 municipios que comprenden un área de 128.602 km², equivalente al 22% del estado de Minas Gerais, con una población de 1.610.413 habitantes, siendo el 69% urbana y el 31% rural (IBGE, 2017).

Mapa 1

*Distribución Georreferencial de los 89 Municipios de la Mesoregión Norte de Minas Gerais, compuesta de 7 Microrregiones (Elaboración Propia)*

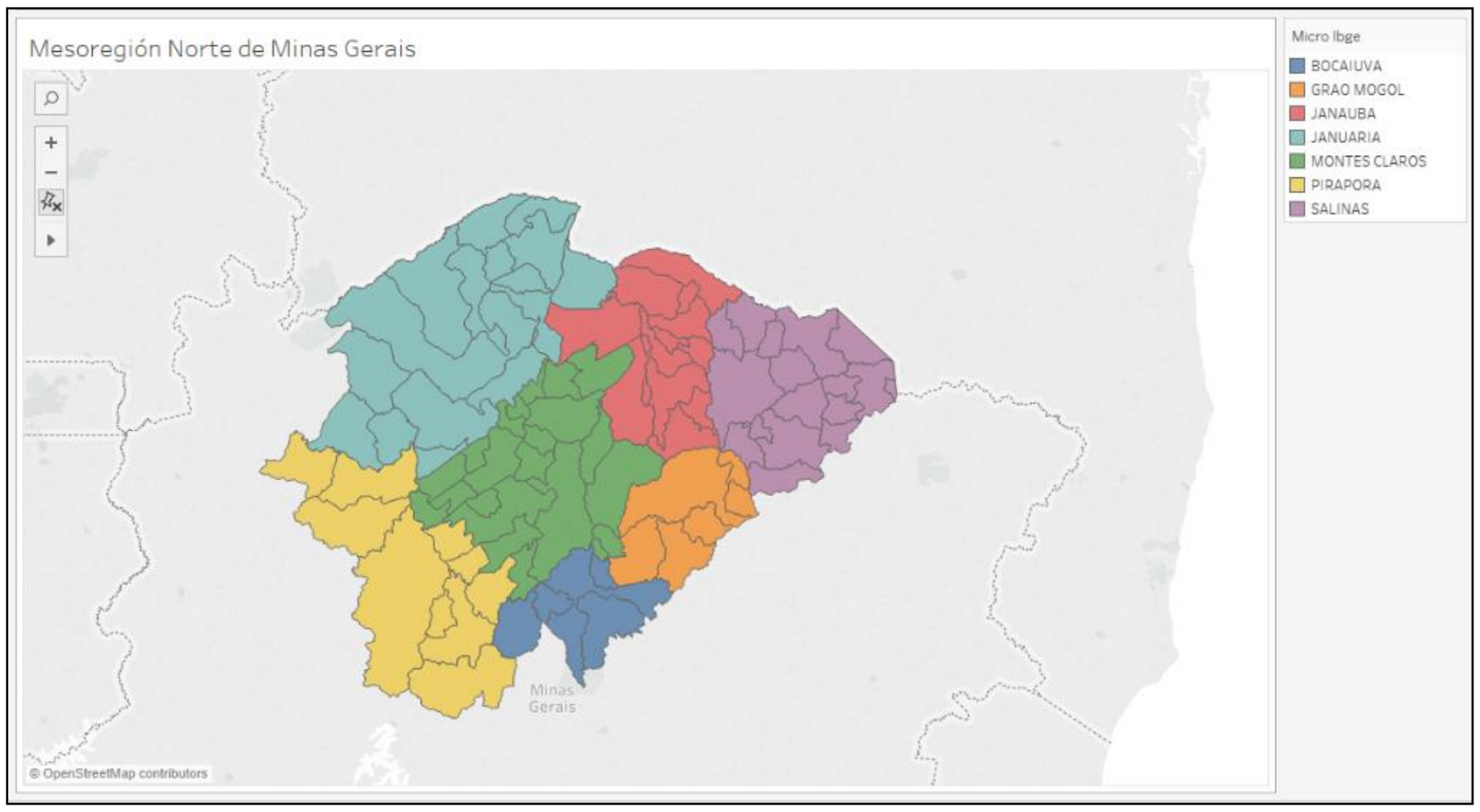

Tabla 4

*Población de estudio (Municipios de la Mesoregión norte de Minas Gerais)*

| **Microregión** | **Municipios** | **Municipios** | **Municipios** |
|---|---|---|---|
| **1. Bocaiúva** | **85 Bocaiúva *** | Engenheiro Navarro<br>Francisco Dumont | Guaraciama<br>Olhos de Água |
| **2. Grão-Mogol** | Botumirim<br>Cristália | **Grão-Mogol ***<br>Itacambira | Josenópolis<br>Padre Carvalho |
| **3. Janaúba** | Catuti<br>Espinosa<br>Gameleiras<br>Jaíba<br>**Janaúba *** | Mamonas<br>Mato Verde<br>Monte Azul<br>Nova Porteirinha<br>Pai Pedro | Porteirinha<br>Riacho dos Machados<br>Serranópolis de Minas |
| **4. Januária** | Bonito de Minas<br>Chapada Gaúcha<br>Cônego Marinho<br>Icaraí de Minas<br>Itacarambi<br>**Januária *** | São João das Missões<br>Urucuia<br>Juvenília<br>Manga<br>Matias Cardoso<br>Miravânia | Montalvânia<br>Pedras de Maria da Cruz<br>Pintópolis<br>São Francisco |
| **5. Montes Claros** | Brasília de Minas<br>Campo Azul<br>Capitão Enéias<br>Claro dos Poções<br>Coração de Jesus<br>Francisco Sá<br>Glaucilândia | Ibiracatu<br>Japonvar<br>Juramento<br>Lontra<br>Luislândia<br>Mirabela<br>**Montes Claros *** | Patis<br>Ponto Chique<br>São João da Lagoa<br>São João da Ponte<br>São João do Pacuí<br>Ubaí<br>Varzelândia<br>Verdelândia |
| **6. Salinas** | Águas Vermelhas<br>Berizal<br>Curral de Dentro<br>Divisa Alegre<br>Fruta de Leite<br>Indaiabira | Ninheira<br>Novorizonte<br>Rio Pardo de Minas<br>Rubelita<br>**Salinas ***<br>Santa Cruz de Salinas | Santo Antônio do Retiro<br>São João do Paraíso<br>Taiobeiras<br>Vargem Grande do Rio Pardo<br>Montezuma |
| **7. Pirapora** | Buritizeiro<br>Ibiaí<br>Jequitaí<br>Lagoa dos Patos | Lassance<br>**Pirapora ***<br>Riachinho | Santa Fé de Minas<br>São Romão<br>Várzea da Palma |

### 3.4 Recolección de datos

Los datos para realizar esta investigación fueron obtenidos en repositorios públicos de agencias gubernamentales de Brasil los cuales se componen de: a) la Clasificación Brasileña de Ocupaciones (CBO); b) la Clasificación Nacional de Actividades Económicas (CNAE); c) el Instituto Brasileño de Geografía y Estadísticas (IBGE); d) el Banco Central de Brasil (IBGE, 2019a); y en el Centro de Información Ambiental (NOAA) de los Estados Unidos (NOAA, 2019).

### 3.5 Aspectos éticos

#### 3.5.1 Consentimiento informado y confidencialidad de los datos

Por el diseño metodológico de este estudio y al no usarse datos primarios obtenidos a través de intervención o interacción con participantes humanos y cumpliendo con los requerimientos de la Junta de Revisión Institucional de la Universidad Ana G. Méndez, se obtuvo la certificación de estudios exentos. Por tal razón, tampoco se aplicó el protocolo de confidencialidad de datos ya que los datos usados son de base pública gubernamental.

#### 3.5.2 Riesgos potenciales de la investigación

Al no haberse aplicado los protocolos con sujetos humanos del comité de ética (IRB), ni tampoco el protocolo de confidencialidad de datos, esta investigación no tuvo ningún factor de riesgo humano ni animal.

### 3.6 Técnicas estadísticas y análisis de datos

#### 3.6.1 Datos de panel

El método de datos de panel son herramientas econométricas utilizadas para: a) estimar parámetros; b) calcular efectos parciales de interés en modelos no lineales; c) cuantificar enlaces dinámicos; y d) realizar inferencia válida cuando los datos están disponibles en secciones transversales repetidas (Wooldridge, 2009). Estos datos surgen de una misma observación cruzada (Sancho, 2005), repetidas a través del tiempo (Arellano, 1991).

#### 3.6.2 Tipos de datos de panel

A continuación, se describe los tipos de datos de panel, así como la importancia y relevancia que tuvo en esta investigación el uso de la mencionada metodología.

**Los micro paneles**: se basa en la información que se analiza mediante un número muy elevado de individuos y pocas observaciones temporales para cada uno.

**Los macro paneles:** son aquellas que disponen de muchas observaciones temporales correspondientes a pocos individuos.

A continuación, se presentan las razones para usar los datos de panel en esta investigación. De acuerdo con Sancho (2005) y Wooldridge (2009), los datos de panel tienen las siguientes características:

1. Permiten la presencia de diferencias sistemáticas y no observadas entre unidades que pueden correlacionarse con factores observados cuyos efectos deben medirse.

2. Explican mejor el fenómeno de cambio tecnológico.

3. Proporcionan información valiosa de una cierta población a través del tiempo.
4. Buscan interpretarse mejor la dinámica del cambio.
5. Eliminan el sesgo de agregación al trabajar con datos desagregados.
6. Eliminan el sesgo de especificación de los modelos de series temporales las cuales no tienen en cuenta las características inobservables de las regiones que podrían condicionar su comportamiento o bien efectos mismos de las regiones en distintos momentos del tiempo.
7. Permiten mitigar o reducir los problemas de multicolinealidad (VIF) ya que la misma ocurre cuando dos o más predictores en el modelo están correlacionados y proporcionan información redundante sobre la respuesta.

### 3.6.3 Estandarización de los datos

Considerando que las características individuales de los datos no son las mismas, se utilizó la técnica conocida como escalado estándar (Anselin, 2003), según el cual a cada observación se le restará la media de la variable y se le dividirá por la desviación típica (Kumar & Dahiya, 2016). La fórmula que se usó en la estandarización de los datos fue la siguiente:

Ecuación 1

*Ecuación para la estandarización de los datos (Kumar & Dahiya, 2016)*

$$X_{normalized}\frac{X - X_{mean}}{X_{stddev}}$$

### 3.6.4 Prueba de normalidad

Este estudio utilizó la prueba de normalidad de Ryan-Joiner, la cual evaluó la normalidad calculando la correlación entre los datos y las puntuaciones normales de los datos (Acuña, 2009). De acuerdo con Salgado (2015) si el coeficiente de correlación se encuentra cerca de 1, existirá normalidad en los datos, por lo tanto, el estadístico de Ryan-Joiner evalúa la fuerza de esta correlación; si esta es menor del valor apropiado entonces se rechaza la hipótesis nula de normalidad, es importante recalcar que esta prueba es similar a la prueba de normalidad de Shapiro-Wilk.

### 3.6.5 Modelo econométrico de crecimiento endógeno basado en la actividad económica digital y el conocimiento especializado

La función Cobb-Douglas permite desarrollar análisis teóricos y empíricos de crecimiento y productividad (Adams, 2005), siendo usados contemporáneamente en estudios que miden los efectos de cambio tecnológico y crecimiento económico (Hanclova et al., 2015). La función de Cobb-Douglas establece el supuesto de que si desaparece la mano de obra o el capital también lo hará la producción (McAuliffe, 2015).

La función de Cobb-Douglas descompone el aporte de capital en activos de las empresas de tecnología de información (Hanclova et al., 2015) mediante la contribución de los servicios de las empresas de tecnología y el conocimiento especializado como factores que aporta en el crecimiento económico (Neffati & Gouidar, 2019).

### 3.6.6 Diseño econométrico del modelo empírico

El modelo propuesto y sus hipótesis se probaron mediante un conjunto de datos de panel de nueve años entre el 2009 al 2018, los cuales sirvieron para determinar el impacto de la actividad económica digital en el crecimiento económico regional de los municipios seleccionados en esta investigación para ser estudiados.

Por lo tanto, para fines de este estudio se creó una formula exponencial usando como base la función Cobb-Douglas para establecer el modelo de análisis, el mismo está representado por el producto interno bruto per cápita municipal como indicador del crecimiento económico regional (Bosco, 2009; Camargos, 2017). La función está compuesta por:

Ecuación 2

*Modelo de análisis de crecimiento económico regional (elaboración propia)*

$$Y_{cer} = A_{aed}.K_{ce}$$

Donde ***Y*** representa el producto interno bruto per cápita municipal, la cual es la tasa que mide el crecimiento económico regional, representado por (***cer***). ***A*** representa la contribución del nivel de progreso tecnológico mediante la actividad económica digital (***aed***) al crecimiento del producto interno bruto per cápita municipal. ***K*** representa las habilidades de trabajo de capital humano mediante la contribución del conocimiento especializado (***ce***), al crecimiento del producto interno bruto per cápita municipal. Como parte de la aportación en este trabajo de investigación se elaboró un algoritmo en "Python" el cual es un lenguaje de programación orientado a objetos (Van, 2013), el mismo se encuentra incluido en el Anejo A.

### 3.6.7 Prueba con análisis de datos espaciales

La primera prueba de medición que se usó fue la de dependencia espacial donde su supuesto manifiesta que el valor observado en una ubicación depende de los valores observados en ubicaciones vecinas (Gómez, 2017). La dependencia espacial es el fenómeno que se presenta cuando los valores observados en una región dependen de los valores de las regiones vecinas (Padilla & Fabricio, 2014). Los mismos fueron analizados mediante el uso de datos de panel (Funke, 2005; Rey, 1999).

Para detectar la dependencia espacial se utilizó el Índice de Moran (Scheider & Huisjes, 2019; Tsutsumida, Rodríguez, Harris, Balzter, & Comber, 2019), mediante una variable X observada en N regiones (Franco, 2015), las regiones estudiadas no pueden tratarse como aisladas, por lo tanto, su crecimiento está vinculado con el de las áreas contiguas (Eduardo, Sanén, Romero, & Hernández, 2007; Franco, 2015).

Ecuación 3

*Índice de Moran (Moran, 1950)*

$$I = \frac{\sum_{j=1}^{N}\sum_{j=1}^{N} w_{ij}(x_j - \bar{x})(x_i - \bar{x})}{\sum_{i=1}^{N}\left(x_j - \bar{x}\right)^2}$$

Dentro de esta primera prueba de análisis de datos espaciales, se buscó medir el grado de asociación espacial mediante el gráfico de dispersión de Moran (Padilla & Fabricio, 2014). El cual consta de cuatro cuadrantes, los cuales presentan diferente tipo de asociación espacial, denotándose que los cuadrantes I y III corresponden a las formas positivas de autocorrelación espacial, mientras que los II y IV representan la autocorrelación espacial negativa.

Las cuatro áreas diferentes del gráfico corresponden a los cuatro tipos de asociación espacial local posibles entre una región y sus vecinas (Darmofal, 2006):

- Cuadrante I (AA) una región con alto valor (superior a la media) rodeada de regiones con alto valor (cuadrante I superior derecho).
- Cuadrante II (BA) una región con bajo valor rodeada de regiones con alto valor (cuadrante II superior izquierdo).
- Cuadrante III (BB) una región con bajo valor rodeada de regiones con bajo valor (cuadrante inferior izquierdo III).
- Cuadrante IV (AB) una región con alto valor, rodeada de regiones con bajo valor (cuadrante inferior derecho IV).

#### 3.6.7.1 Indicadores locales de asociación espacial (LISA)

El método de asociación espacial llamado LISA, descompone el índice I de Moran verificando de qué manera cada unidad espacial ayuda a la formación del valor general, permitiendo obtener un valor de significancia para cada clúster formado por los valores similares de cada unidad espacial y sus vecinos (Celemin, 2004). Además, permite relacionar los valores de una variable en una localización con los valores de esa misma u otra variable en las localizaciones vecinas (Pellegrini & Platino, 2013).

Estos agrupamientos de especial concentración de valores extremos de una variable se conocen también como zonas calientes y frías respectivamente (Villareal & Flores, 2015). Según se trate de una concentración de valores especialmente altos/bajos de una variable, correspondientemente. La autocorrelación espacial es usada frecuentemente en estudios de propagación epidémica, este método de análisis tuvo la capacidad de estudiar la forma en que un fenómeno se expande dentro de unidades espaciales (Celemín, 2009).

Para realizar esta prueba se usó el método de asociación espacial local LISA (Carro-Calvo et al., 2017) basada en el estadístico *I Moran* (Anselin, 2005). Esta prueba descompone el índice global de autocorrelación y verifica en cuánto contribuye cada unidad espacial a la formación del valor general. LISA es un coeficiente de correlación ponderado, donde los pesos reflejan la proximidad geográfica. En esta prueba los resultados se ajustan en valores mayores que *0* lo cual indica autocorrelación espacial positiva; no obstante, los valores menores que *0* indican autocorrelación espacial negativa (Yrigoyen, 2006).

### 3.6.8 Prueba de ecuaciones estructurales con mínimos cuadrados parciales.

Dentro del proceso de medición además del análisis espacial se usó ecuaciones estructurales con mínimos cuadrados (Ajamieh, 2016) mediante el modelo de medición reflexivo, ya que se ha demostrado que es útil para probar modelos que utilizan datos de panel (Rana et al., 2018). Por lo tanto, para medir la relación entre las variables de estudio es importante establecer unos conceptos básicos descritos a continuación:

De acuerdo con Martínez & Fierro (2018), las variables latentes o factores son aquellas no observadas directamente, esto significa que no pueden medirse en su forma, por lo tanto, es importante que existan indicadores para ser medidos. Las variables latentes pueden ser independientes “exógenas” o dependientes “endógenas” (Nascimento & Macedo, 2016; Bernal, Nieto, & Briones, 2016).

Las variables latentes exógenas son aquellas que no tienen variables que le antecedan, y las variables endógenas tienen variables que le anteceden (Taborga & Eduardo, 2013). De igual forma las variables observables o indicadores, son aquellas variables cuantitativas que en su naturaleza pueden ser medidas directamente (Martínez & Fierro, 2018).

#### 3.6.8.1 Modelo de medida con enfoque reflexivo

Las teorías de medición buscan explicar cómo las variables latentes son medidas mediante un enfoque de medición. Por consiguiente, el modelo de medida con enfoque reflexivo busca estimar el efecto o relación indirecta y total que existe entre una variable a otra (Manzano, 2017). Por lo tanto, si existe un cambio en la variable latente será reflejado en sus indicadores (Taborga & Eduardo, 2013; Martínez & Fierro, 2018).

Al buscar probar y estimar efectos o relaciones indirectas o totales, el modelo de ecuación estructural va a partir de la regresión, la cual de acuerdo con Cazorla (2008) busca analizar la relación o dependencia entre variables teniendo como interés conocer el efecto que una variable puede tener en otra. Una de las razones importantes para usar el modelo de análisis de mínimos cuadrados parciales (PLS), es por tratarse de una metodología de modelación más flexible al no exigir supuestos paramétricos rigurosos, principalmente en la distribución de los datos, lo que nos permitirá trabajar de una manera más eficiente con los datos a recopilarse.

Para concluir con este capítulo, la naturaleza de los datos de estudio basados en unidades regionales dinámicas y una organización espacial única compuesta por microrregiones (Ma & Zhao, 2019) llevo a la necesidad del uso de herramientas econométricas usadas con datos disponibles en secciones transversales que se utilizaron para realizar inferencias válidas dentro de este estudio (Wooldridge, 2009).

# CAPITULO IV

# RESULTADOS

## 4.1 Introducción

Este capítulo presenta el resultado obtenido del análisis estadístico de los datos secundarios recopilados en repositorios públicos compuesto por los resultados del análisis descriptivo, espacial y multivariante de las variables de estudio. Además, se detalla la evidencia estadística utilizadas para sustentar las hipótesis planteadas por el investigador.

Los datos fueron analizados en el programa para MS Windows 10, Minitab® versión 18.1, donde se realizaron estadísticas descriptivas e inferenciales. Para el análisis multivariante se usó el programa SmartPLS versión 3.2.9 para MS Windows 10. Para el análisis espacial se usó el programa GeoDa versión 1.14 para MS Windows. Para el desarrollo de las funciones de análisis se usó el programa Spyder Python versión 3.3.6.

## 4.2 Primera fase: análisis estadístico descriptivo

### 4.2.1 Constructo actividad económica digital

Se realizó el análisis estadístico descriptivo del constructo actividad económica digital (AED) la misma cuenta con los siguientes indicadores: **Primero:** cantidad de empresas relacionadas al rubro de las tecnologías de información (TIC) por municipio y clasificado en microrregiones con datos de panel desde el 2009 hasta el 2018. **Segundo:** indicador de la clasificación brasileña de ocupaciones (CBO) compuesta por: cantidad de Analista de Redes de Comunicación de Datos, Director de Servicios de Informática, Gerente de Producción de Tecnología de Información, Gerente de Proyectos de Tecnología de Información, Gerente de Suporte Técnico de Tecnología de información,

Administrador de Base de Datos, Electricista de Mantenimiento de Linas Eléctricas y Telefónicas de Comunicación de Datos, Instalador-Reparador de Redes Telefónicas y Comunicación de Datos, Técnico de Comunicación de Datos, Administrador en Seguridad de Información, Director de Servicios de Informática, Analista de Información (Investigador de Redes de Información), Gerente de Proyectos de Tecnología de Información, Gerente de Producción en Tecnología de Información, Gerente de Soporte Técnico de Tecnología de Información, Programador de Sistemas de Información, Técnico en Mantenimiento de Equipos de informática, Técnico de Apoyo al Usuario de informática (Helpdesk), Tecnólogo en Gestión de Tecnología de Información, Ingeniero en Programación de Computadoras, Ingeniero en Equipos de Computación, Ingeniero de Sistemas Operativos en Computación, Instalador de Equipos Electrónicos (Computadores y Equipos Auxiliares), Operador de Computador (Inclusive Microcomputador), Programador de Internet, organizados por municipio y clasificado en microrregiones con datos de panel desde el 2009 hasta el 2018.

Tabla 6

*Estadística descriptiva indicador TIC (N = 89)*

| **Año** | **Mun.** | **Micro Región** | ***M*** | ***SD*** | **Varianza** | **CoefVar** | **Mínimo** | ***Mdn*** | **Máximo** | **Rango** |
|---|---|---|---|---|---|---|---|---|---|---|
| 2009 | 89 | 7 | 285 | 284 | 80777 | 99.72 | 36 | 219 | 896 | 860 |
| 2010 | 89 | 7 | 289 | 310 | 96215 | 107.44 | 33 | 218 | 964 | 931 |
| 2011 | 89 | 7 | 288 | 300 | 90279 | 104.38 | 24 | 229 | 936 | 912 |
| 2012 | 89 | 7 | 385 | 380 | 144192 | 98.59 | 39 | 296 | 1192 | 1153 |
| 2013 | 89 | 7 | 398 | 430 | 184589 | 107.91 | 45 | 329 | 1331 | 1286 |
| 2014 | 89 | 7 | 418 | 468 | 219275 | 112.06 | 39 | 284 | 1441 | 1402 |
| 2015 | 89 | 7 | 310 | 329 | 107969 | 105.85 | 52 | 212 | 1023 | 971 |
| 2016 | 89 | 7 | 333 | 375 | 140329 | 112.35 | 42 | 194 | 1151 | 1109 |
| 2017 | 89 | 7 | 243 | 285 | 81096 | 117.33 | 38 | 168 | 873 | 835 |
| 2018 | 89 | 7 | 242 | 298 | 89070 | 123.11 | 29 | 136 | 903 | 874 |

Tabla 7

*Estadística descriptiva indicador CBO (N = 89)*

| Año | Mun. | Micro Región | *M* | *SD* | Varianza | CoefVar | Mínimo | *Mdn* | Máximo | Rango |
|---|---|---|---|---|---|---|---|---|---|---|
| 2009 | 89 | 7 | 52.6 | 79.2 | 6270.6 | 150.63 | 3.0 | 13.0 | 216.0 | 213.0 |
| 2010 | 89 | 7 | 71.9 | 99.7 | 9935.8 | 138.72 | 1.0 | 26.0 | 271.0 | 270.0 |
| 2011 | 89 | 7 | 82.1 | 147.5 | 21754.5 | 179.56 | 2.0 | 22.0 | 412.0 | 410.0 |
| 2012 | 89 | 7 | 82.9 | 150.4 | 22632.8 | 181.57 | 0.0 | 29.0 | 420.0 | 420.0 |
| 2013 | 89 | 7 | 92.1 | 175.4 | 30774.5 | 190.39 | 0.0 | 17.0 | 485.0 | 485.0 |
| 2014 | 89 | 7 | 85.7 | 144.5 | 20893.2 | 168.64 | 0.0 | 19.0 | 406.0 | 406.0 |
| 2015 | 89 | 7 | 88.1 | 157.8 | 24915.8 | 179.08 | 0.0 | 40.0 | 443.0 | 443.0 |
| 2016 | 89 | 7 | 90.0 | 172.4 | 29704.7 | 191.50 | 0.0 | 25.0 | 478.0 | 478.0 |
| 2017 | 89 | 7 | 73.9 | 129.9 | 16875.5 | 175.89 | 1.0 | 17.0 | 363.0 | 362.0 |
| 2018 | 89 | 7 | 72.1 | 111.8 | 12508.8 | 155.03 | 2.0 | 29.0 | 321.0 | 319.0 |

### 4.2.2 Constructo conocimiento especializado

Se realizó el análisis estadístico descriptivo del constructo conocimiento especializado (CE) el cual incluye los siguientes indicadores: **Primero:** cantidad de instituciones universitarias públicas y privadas (IUPP) que comprenden a instituciones que ofrecen educación técnica profesional, instituciones de educación tecnológica, instituciones educativas de educación subgraduada, instituciones educativas de educación graduada e instituciones educativas de educación extendida organizados por municipio y clasificado en microrregiones con datos de panel desde el 2009 hasta el 2018.

**Segundo:** cantidad de docentes (DCNT) con doctorado y maestría, instructores de estudios profesionales, profesores de informática, profesores de ciencias, profesores de ciencias económicas y administrativas, profesores de ciencias biológicas, profesores de ingeniería, profesores de matemática y estadística, profesores de ciencias humanas de instituciones universitarias públicas y privadas organizados por municipio y clasificado en microrregiones con datos de panel desde el 2009 hasta el 2018. **Tercero:** cantidad de aportación económica gubernamental en el sistema educativo (FUNDEB), organizados por municipio y clasificado en microrregiones con datos de panel desde el 2009 hasta el 2018.

Tabla 8

*Estadística descriptiva indicador IUPP (N = 89)*

| **Año** | **Mun.** | **Micro Región** | ***M*** | ***SD*** | **Varianza** | **CoefVar** | **Mínimo** | ***Mdn*** | **Máximo** | **Rango** |
|---|---|---|---|---|---|---|---|---|---|---|
| 2009 | 89 | 7 | 153 | 332 | 110266 | 217.64 | 0 | 5 | 897 | 897 |
| 2010 | 89 | 7 | 121 | 280 | 78521 | 231.04 | 0 | 14 | 756 | 756 |
| 2011 | 89 | 7 | 146 | 333 | 110566 | 227.30 | 0 | 23 | 899 | 899 |
| 2012 | 89 | 7 | 173 | 411 | 168709 | 237.42 | 0 | 18 | 1104 | 1104 |
| 2013 | 89 | 7 | 178 | 430 | 185115 | 242.10 | 0 | 16 | 1153 | 1153 |
| 2014 | 89 | 7 | 200 | 451 | 203462 | 226.02 | 0 | 48 | 1221 | 1221 |
| 2015 | 89 | 7 | 294 | 608 | 370071 | 206.92 | 0 | 60 | 1667 | 1667 |
| 2016 | 89 | 7 | 254 | 507 | 257204 | 200.00 | 0 | 48 | 1389 | 1389 |
| 2017 | 89 | 7 | 221 | 399 | 159284 | 180.82 | 0 | 24 | 1089 | 1089 |
| 2018 | 89 | 7 | 205 | 414 | 171808 | 202.48 | 0 | 33 | 1133 | 1133 |

Tabla 9

*Estadística descriptiva indicador DCNT (N = 89)*

| **Año** | **Mun.** | **Micro Región** | ***M*** | ***SD*** | **Varianza** | **CoefVar** | **Mínimo** | ***Mdn*** | **Máximo** | **Rango** |
|---|---|---|---|---|---|---|---|---|---|---|
| 2009 | 89 | 7 | 205 | 513 | 262732 | 250.21 | 0 | 11 | 1367 | 1367 |
| 2010 | 89 | 7 | 220 | 533 | 284400 | 242.88 | 0 | 13 | 1428 | 1428 |
| 2011 | 89 | 7 | 208 | 512 | 262463 | 246.81 | 0 | 13 | 1369 | 1369 |
| 2012 | 89 | 7 | 220 | 546 | 297701 | 247.69 | 0 | 14 | 1457 | 1457 |
| 2013 | 89 | 7 | 247 | 604 | 364712 | 244.08 | 0 | 19 | 1616 | 1616 |
| 2014 | 89 | 7 | 271 | 665 | 442433 | 245.32 | 0 | 19 | 1779 | 1779 |
| 2015 | 89 | 7 | 281 | 695 | 482529 | 246.95 | 0 | 21 | 1856 | 1856 |
| 2016 | 89 | 7 | 273 | 662 | 437811 | 242.62 | 0 | 17 | 1772 | 1772 |
| 2017 | 89 | 7 | 289 | 701 | 491548 | 242.72 | 0 | 17 | 1877 | 1877 |
| 2018 | 89 | 7 | 282 | 695 | 483148 | 246.74 | 0 | 17 | 1857 | 1857 |

Tabla 10

*Estadística descriptiva indicador FUNDEB (N = 89)*

| Año | Mun. | Micro Región | *M* | *SD* | Varianza | CoefVar | Mínimo | *Mdn* | Máximo |
|---|---|---|---|---|---|---|---|---|---|
| 2009 | 89 - 7 | 25730929 | 16818767 | 2.82871 | 65.36 | 6765283 | 25566365 | 56526729 | 49761446 |
| 2010 | 89 - 7 | 35095015 | 24238398 | 5.87500 | 69.07 | 8124034 | 34273665 | 81062139 | 72938105 |
| 2011 | 89 - 7 | 37762516 | 27115923 | 7.35273 | 71.81 | 8591606 | 36619002 | 89565927 | 80974321 |
| 2012 | 89 - 7 | 43208469 | 32209783 | 1.03747 | 74.55 | 10668510 | 40099069 | 106283978 | 95615468 |
| 2013 | 89 - 7 | 47311512 | 36793461 | 1.35376 | 77.77 | 12195978 | 47231633 | 119564436 | 107368458 |
| 2014 | 89 - 7 | 51377420 | 36746044 | 1.35027 | 71.52 | 12885206 | 49242707 | 121756289 | 108871083 |
| 2015 | 89 - 7 | 57713925 | 43296697 | 1.87460 | 75.02 | 13588765 | 53277843 | 140109234 | 126520469 |
| 2016 | 89 - 7 | 61867775 | 47011160 | 2.21005 | 75.99 | 14238850 | 55819072 | 152133419 | 137894569 |
| 2017 | 89 - 7 | 62696219 | 49077542 | 2.40861 | 78.28 | 14612288 | 56077798 | 157955390 | 143343102 |
| 2018 | 89 - 7 | 68854560 | 55142655 | 3.04071 | 80.09 | 15336456 | 59626609 | 178800518 | 163464062 |

### 4.2.3 Constructo asentamiento humano

Se realizó el análisis estadístico descriptivo del constructo asentamiento humano (AHU) el cual incluye el siguiente indicador: Cantidad de habitantes (POB) que están asentados en los 89 municipios que comprenden la mesoregión norte de Minas Gerais organizados por municipio y clasificado en microrregiones con datos de panel desde el 2009 hasta el 2018.

Tabla 11

*Estadística descriptiva indicador POB (N = 89)*

| Año | Mun. | Micro Región | *M* | *SD* | Varianza | CoefVar | Mínimo | *Mdn* | Máximo | Rango |
|---|---|---|---|---|---|---|---|---|---|---|
| 2009 | 89 | 7 | 233166 | 184370 | 33992125621 | 79.07 | 56508 | 216489 | 606401 | 549893 |
| 2010 | 89 | 7 | 235007 | 186413 | 34749856344 | 79.32 | 56761 | 218010 | 612786 | 556025 |
| 2011 | 89 | 7 | 231356 | 184437 | 34017151722 | 79.72 | 55545 | 211622 | 606698 | 551153 |
| 2012 | 89 | 7 | 232604 | 185926 | 34568400058 | 79.93 | 55680 | 212436 | 611361 | 555681 |
| 2013 | 89 | 7 | 240988 | 193178 | 37317630380 | 80.16 | 57503 | 219727 | 634942 | 577439 |
| 2014 | 89 | 7 | 242405 | 194796 | 37945386560 | 80.36 | 57680 | 220704 | 640028 | 582348 |
| 2015 | 89 | 7 | 243765 | 196348 | 38552600934 | 80.55 | 57852 | 221638 | 644907 | 587055 |
| 2016 | 89 | 7 | 245058 | 197824 | 39134468215 | 80.73 | 58016 | 222526 | 649544 | 591528 |
| 2017 | 89 | 7 | 246287 | 199229 | 39692168320 | 80.89 | 58171 | 223372 | 653953 | 595782 |
| 2018 | 89 | 7 | 244781 | 199251 | 39701126815 | 81.40 | 57407 | 221188 | 653411 | 596004 |

### 4.2.4 Constructo crecimiento económico regional

Se realizó el análisis estadístico descriptivo del constructo crecimiento económico regional (CER) el cual incluye el indicador compuesto por el producto interno bruto municipal (PIB), organizado por municipio y clasificado en microrregiones con datos de panel desde el 2009 hasta el 2018.

Tabla 12

*Estadística descriptiva indicador PIB (N = 89)*

| **Año** | **Mun.** | **Micro Región** | ***M*** | ***SD*** | **Varianza** | **CoefVar** | **Mínimo** | ***Mdn*** | **Máximo** |
|---|---|---|---|---|---|---|---|---|---|
| 2009 | 89 - 7 | 1636452 | 1673073 | 2.79917E+12 | 102.24 | 339840 | 1213096 | 5252848 | 4913008 |
| 2010 | 89 - 7 | 1894516 | 1953628 | 3.81666E+12 | 103.12 | 346253 | 1354459 | 6085195 | 5738942 |
| 2011 | 89 - 7 | 2172268 | 2201567 | 4.84690E+12 | 101.35 | 459529 | 1511966 | 6897900 | 6438371 |
| 2012 | 89 - 7 | 2706891 | 2477673 | 6.13886E+12 | 91.53 | 1330650 | 1727078 | 8227748 | 6897098 |
| 2013 | 89 - 7 | 2732578 | 2872492 | 8.25121E+12 | 105.12 | 541482 | 1964128 | 8998814 | 8457332 |
| 2014 | 89 - 7 | 2980221 | 3137514 | 9.84399E+12 | 105.28 | 548843 | 2183035 | 9811017 | 9262174 |
| 2015 | 89 - 7 | 3039433 | 3207401 | 1.02874E+13 | 105.53 | 549813 | 2330617 | 10068372 | 9518559 |
| 2016 | 89 - 7 | 3383862 | 3523930 | 1.24181E+13 | 104.14 | 524082 | 2568738 | 11057938 | 10533856 |
| 2017 | 89 - 7 | 3485352 | 3565467 | 1.27126E+13 | 102.30 | 522026 | 2628430 | 11204841 | 10682815 |

## 4.3 Segunda fase: estandarización regional de indicadores

Con el fin de reducir la multicolinealidad y comparar el tamaño de los coeficientes en una escala comparable, se realizó la estandarización de datos usando la metodología expuesta en el capítulo anterior, la estandarización busco centrar los datos y cambiar las unidades a desviaciones estándar.

Tabla 13

*Estandarización indicador TIC (N = 89)*

| Región | 2009 | 2010 | 2011 | 2012 | 2013 | 2014 | 2015 | 2016 | 2017 | 2018 |
|---|---|---|---|---|---|---|---|---|---|---|
| Januária | 0.288 | -0.227 | 0.195 | -0.234 | -0.389 | -0.309 | -0.314 | -0.388 | -0.409 | 0.423 |
| Janauba | 0.098 | 0.0009 | 0.070 | 0.178 | -0.028 | 0.023 | 0.099 | 0.076 | -0.072 | 0.008 |
| Salinas | 0.232 | -0.263 | 0.195 | -0.305 | -0.160 | -0.285 | -0.299 | -0.163 | -0.199 | 0.266 |
| Pirapora | 0.200 | -0.221 | 0.345 | -0.189 | -0.126 | -0.215 | -0.205 | -0.372 | -0.262 | 0.356 |
| M Claros | 2.149 | 2.177 | 2.157 | 2.124 | 2.171 | 2.184 | 2.168 | 2.182 | 2.213 | 2.213 |
| Bocaiúva | 0.650 | -0.640 | 0.611 | -0.661 | -0.645 | -0.589 | -0.661 | -0.556 | -0.550 | 0.460 |
| G Mongol | 0.876 | -0.824 | 0.878 | -0.911 | -0.821 | -0.809 | -0.786 | -0.777 | -0.718 | 0.715 |

Tabla 14

*Estandarización del indicador CBO (N = 89)*

| Región | 2009 | 2010 | 2011 | 2012 | 2013 | 2014 | 2015 | 2016 | 2017 | 2018 |
|---|---|---|---|---|---|---|---|---|---|---|
| Januária | 0.512 | -0.460 | -0.407 | -0.357 | -0.428 | -0.496 | -0.476 | -0.435 | -0.437 | -0.421 |
| Janauba | 0.499 | -0.299 | -0.373 | -0.331 | -0.331 | -0.198 | -0.203 | -0.319 | -0.252 | -0.260 |
| Salinas | 0.436 | -0.580 | -0.462 | -0.457 | -0.428 | -0.461 | -0.305 | -0.377 | -0.491 | -0.510 |
| Pirapora | 0.573 | 0.683 | -0.021 | -0.072 | -0.046 | 0.022 | -0.241 | -0.145 | 0.001 | -0.019 |
| M Claros | 2.063 | 1.997 | 2.236 | 2.241 | 2.239 | 2.215 | 2.248 | 2.251 | 2.225 | 2.225 |
| Bocaiúva | 0.562 | -0.630 | -0.428 | -0.470 | -0.479 | -0.489 | -0.463 | -0.452 | -0.483 | -0.385 |
| G Mongol | 0.626 | -0.710 | -0.543 | -0.550 | -0.525 | -0.592 | -0.558 | -0.522 | -0.560 | -0.627 |

Tabla 15

*Estandarización del indicador POB (N = 89)*

| Región | 2009 | 2010 | 2011 | 2012 | 2013 | 2014 | 2015 | 2016 | 2017 | 2018 |
|---|---|---|---|---|---|---|---|---|---|---|
| Januária | 0.184 | 0.184 | 0.169 | 0.168 | 0.167 | 0.166 | 0.165 | 0.164 | 0.163 | 0.160 |
| Janauba | 0.098 | 0.093 | 0.091 | 0.087 | 0.083 | 0.080 | 0.077 | 0.074 | 0.071 | 0.062 |
| Salinas | 0.090 | -0.091 | -0.106 | -0.108 | -0.110 | -0.111 | -0.112 | -0.113 | -0.115 | -0.118 |
| Pirapora | 0.366 | -0.366 | -0.356 | -0.355 | -0.355 | -0.355 | -0.355 | -0.355 | -0.355 | -0.354 |
| M Claros | 2.024 | 2.026 | 2.035 | 2.037 | 2.039 | 2.041 | 2.043 | 2.044 | 2.046 | 2.050 |
| Bocaiúva | 0.893 | -0.890 | -0.880 | -0.877 | -0.874 | -0.872 | -0.870 | -0.868 | -0.866 | -0.860 |
| G Mongol | 0.958 | -0.956 | -0.953 | -0.951 | -0.949 | -0.948 | -0.946 | -0.945 | -0.944 | -0.940 |

Tabla 16

*Estandarización del indicador IUPP (N = 89)*

| Región | 2009 | 2010 | 2011 | 2012 | 2013 | 2014 | 2015 | 2016 | 2017 | 2018 |
|---|---|---|---|---|---|---|---|---|---|---|
| Januária | 0.387 | -0.368 | -0.325 | -0.353 | -0.375 | -0.311 | -0.355 | -0.432 | -0.522 | -0.414 |
| Janauba | 0.040 | -0.279 | -0.277 | -0.326 | -0.350 | -0.324 | -0.182 | -0.405 | -0.299 | -0.334 |
| Salinas | 0.450 | -0.382 | -0.409 | -0.379 | -0.396 | -0.420 | -0.461 | -0.486 | -0.545 | -0.474 |
| Pirapora | 0.444 | -0.414 | -0.439 | -0.409 | -0.401 | -0.429 | -0.384 | -0.028 | 0.238 | -0.042 |
| M Claros | 2.241 | 2.265 | 2.263 | 2.266 | 2.266 | 2.264 | 2.256 | 2.238 | 2.175 | 2.239 |
| Bocaiúva | 0.459 | -0.386 | -0.370 | -0.377 | -0.329 | -0.336 | -0.389 | -0.385 | -0.492 | -0.479 |
| G Mongol | 0.459 | -0.432 | -0.439 | -0.421 | -0.413 | -0.442 | -0.483 | -0.499 | -0.553 | -0.493 |

Tabla 17

*Estandarización del indicador DCNT (N = 89)*

| Región | 2009 | 2010 | 2011 | 2012 | 2013 | 2014 | 2015 | 2016 | 2017 | 2018 |
|---|---|---|---|---|---|---|---|---|---|---|
| Januária | 0.378 | -0.387 | -0.379 | -0.376 | -0.378 | -0.379 | -0.374 | -0.386 | -0.387 | -0.379 |
| Janauba | 0.335 | -0.293 | -0.327 | -0.310 | -0.300 | -0.327 | -0.325 | -0.292 | -0.273 | -0.300 |
| Salinas | 0.384 | -0.406 | -0.387 | -0.396 | -0.404 | -0.391 | -0.397 | -0.395 | -0.407 | -0.403 |
| Pirapora | 0.372 | -0.370 | -0.368 | -0.378 | -0.368 | -0.358 | -0.364 | -0.368 | -0.380 | -0.380 |
| M Claros | 2.267 | 2.265 | 2.267 | 2.266 | 2.266 | 2.266 | 2.266 | 2.265 | 2.265 | 2.266 |
| Bocaiúva | 0.397 | -0.396 | -0.399 | -0.401 | -0.404 | -0.403 | -0.399 | -0.410 | -0.403 | -0.396 |
| G Mongol | 0.399 | -0.411 | -0.405 | -0.403 | -0.409 | -0.407 | -0.404 | -0.412 | -0.412 | -0.405 |

Tabla 18

*Estandarización del indicador FUNDEB (N = 89)*

| Región | 2009 | 2010 | 2011 | 2012 | 2013 | 2014 | 2015 | 2016 | 2017 | 2018 |
|---|---|---|---|---|---|---|---|---|---|---|
| Januária | 0.567 | 0.478 | 0.493 | 0.431 | 0.327 | 0.411 | 0.502 | 0.474 | 0.457 | 0.415 |
| Janauba | 0.009 | 0.015 | -0.035 | -0.096 | 0.127 | 0.053 | 0.018 | 0.052 | 0.027 | -0.043 |
| Salinas | 0.053 | -0.033 | -0.042 | -0.015 | -0.002 | -0.058 | -0.102 | -0.128 | -0.134 | -0.167 |
| Pirapora | 0.351 | -0.336 | -0.337 | -0.364 | -0.587 | -0.285 | -0.312 | -0.344 | -0.352 | -0.362 |
| M Claros | 1.831 | 1.896 | 1.910 | 1.958 | 1.963 | 1.915 | 1.903 | 1.920 | 1.940 | 1.993 |
| Bocaiúva | 0.962 | -0.906 | -0.912 | -0.903 | -0.874 | -0.988 | -0.989 | -0.960 | -0.959 | -0.865 |
| G Mongol | 1.127 | -1.112 | -1.075 | -1.010 | -0.954 | -1.047 | -1.019 | -1.013 | -0.979 | -0.970 |

Tabla 19

*Estandarización del indicador FUNDEB (N = 89)*

| Región | 2009 | 2010 | 2011 | 2012 | 2013 | 2014 | 2015 | 2016 | 2017 |
|---|---|---|---|---|---|---|---|---|---|
| Januária | 0.253 | -0.276 | -0.299 | -0.395 | -0.267 | -0.254 | -0.220 | -0.231 | -0.240 |
| Janauba | 0.206 | -0.247 | -0.230 | -0.341 | -0.200 | -0.167 | -0.112 | -0.133 | -0.111 |
| Salinas | 0.326 | -0.329 | -0.337 | -0.416 | -0.336 | -0.327 | -0.287 | -0.332 | -0.314 |
| Pirapora | 0.108 | 0.197 | 0.199 | 0.009 | 0.043 | 0.024 | -0.137 | -0.023 | 0.005 |
| M Claros | 2.161 | 2.145 | 2.146 | 2.228 | 2.181 | 2.177 | 2.191 | 2.177 | 2.165 |
| Bocaiúva | 0.708 | -0.696 | -0.700 | -0.529 | -0.657 | -0.677 | -0.657 | -0.645 | -0.672 |
| G Mongol | 0.774 | -0.792 | -0.777 | -0.555 | -0.762 | -0.774 | -0.776 | -0.811 | -0.831 |

## 4.4 Tercera fase: prueba de hipótesis para búsqueda de normalidad

Para cumplir con los supuestos de normalidad y determinar si los datos no siguen una distribución normal se realizó la respectiva prueba de Ryan-Joiner (Shapiro-Wilk) (Horta, 2015), con el propósito de establecer dentro del diseño de esta investigación las pruebas paramétricas o no paramétricas a usarse. Para confirmar el supuesto de normalidad de los datos recolectados se realizó dos pruebas de hipótesis de normalidad donde:

$\boldsymbol{H_o}$ *establece que los datos siguen una distribución normal*

$\boldsymbol{H}_1$ establece que los *datos no siguen una distribución normal*

Dado los resultados de la prueba con valores menores a $p = .05$ en las variables (TIC, CBO, IUPP, DCNT y PIB) se rechaza la hipótesis nula y se apoya la alterna, por lo tanto, los datos que componen esas variables no siguen una distribución normal. No obstante, los resultados de la prueba con valores mayores a $p = .05$ en las variables (POB y FUNDEB) confirma la hipótesis nula, por lo tanto, los datos que componen esas variables siguen una distribución normal como lo menciona la literatura para la naturaleza de este tipo de datos (María Russell, 2002).

Tabla 20

*Prueba de normalidad variable TIC de las Regiones de Januária, Janauba, Salinas, Pirapora, Montes Claros, Bocaiúva y Grao Mongol.*

| Año | Shapiro-Wilk | *p* |
|---|---|---|
| T2009 | .853 | .011 |
| T2010 | .829 | .010 |
| T2011 | .855 | .013 |
| T2012 | .870 | .026 |
| T2013 | .838 | .010 |
| T2014 | .827 | .010 |
| T2015 | .838 | .010 |
| T2016 | .829 | .010 |
| T2017 | .800 | .010 |
| T2018 | .059 | .010 |

Tabla 21

*Prueba de normalidad variable CBO de las Regiones de Januária, Janauba, Salinas, Pirapora, Montes Claros, Bocaiúva y Grao Mongol.*

| Año | Shapiro-Wilk | *p* |
|---|---|---|
| C2009 | .822 | .010 |
| C2010 | .864 | .021 |
| C2011 | .748 | .010 |
| C2012 | .750 | .010 |
| C2013 | .750 | .010 |
| C2014 | .781 | .010 |
| C2015 | .742 | .010 |
| C2016 | .732 | .010 |
| C2017 | .768 | .010 |
| C2018 | .779 | .010 |

Tabla 22

*Prueba de normalidad variable POB de las Regiones de Januária, Janauba, Salinas, Pirapora, Montes Claros, Bocaiúva y Grao Mongol.*

| Año | Shapiro-Wilk | *p* |
|---|---|---|
| PO2009 | .911 | .082* |
| PO2010 | .910 | .080* |
| PO2011 | .908 | .074* |
| PO2012 | .907 | .072* |
| PO2013 | .906 | .070* |
| PO2014 | .906 | .069* |
| PO2015 | .905 | .067* |
| PO2016 | .906 | .066* |
| PO2017 | .904 | .065* |
| PO2018 | .902 | .061* |

Tabla 23

*Prueba de normalidad variable IUPP de las Regiones de Januária, Janauba, Salinas, Pirapora, Montes Claros, Bocaiúva y Grao Mongol.*

| Año | Shapiro-Wilk | *p* |
|---|---|---|
| IU2009 | .747 | .010 |
| IU2010 | .684 | .010 |
| IU2011 | .712 | .010 |
| IU2012 | .675 | .010 |
| IU2013 | .672 | .010 |
| IU2014 | .688 | .010 |
| IU2015 | .715 | .010 |
| IU2016 | .740 | .010 |
| IU2017 | .793 | .010 |
| IU2018 | .741 | .010 |

Tabla 24

*Prueba de normalidad variable DCNT de las Regiones de Januária, Janauba, Salinas, Pirapora, Montes Claros, Bocaiúva y Grao Mongol.*

| Año | Shapiro-Wilk | *p* |
|---|---|---|
| D2009 | .666 | .010 |
| D2010 | .676 | .010 |
| D2011 | .669 | .010 |
| D2012 | .671 | .010 |
| D2013 | .681 | .010 |
| D2014 | .670 | .010 |
| D2015 | .670 | .010 |
| D2016 | .677 | .010 |
| D2017 | .679 | .010 |
| D2018 | .673 | .010 |

Tabla 25

*Prueba de normalidad variable FUNDEB de las Regiones de Januária, Janauba, Salinas, Pirapora, Montes Claros, Bocaiúva y Grao Mongol.*

| Año | Shapiro-Wilk | *p* |
|---|---|---|
| F2009 | .963 | .100* |
| F2010 | .952 | .100* |
| F2011 | .946 | .100* |
| F2012 | .935 | .100* |
| F2013 | .928 | .100* |
| F2014 | .942 | .100* |
| F2015 | .944 | .100* |
| F2016 | .941 | .100* |
| F2017 | .937 | .100* |
| F2018 | .925 | .100* |

Tabla 26

*Prueba de normalidad variable PIB de las Regiones de Januária, Janauba, Salinas, Pirapora, Montes Claros, Bocaiúva y Grao Mongol.*

| Año | Shapiro-Wilk | *p* |
|---|---|---|
| PI2009 | .842 | .010 |
| PI2010 | .851 | .010 |
| PI2011 | .850 | .010 |
| PI2012 | .764 | .010 |
| PI2013 | .829 | .010 |
| PI2014 | .833 | .010 |
| PI2015 | .815 | .010 |
| PI2016 | .833 | .010 |
| PI2017 | .843 | .010 |

## 4.5 Modelo de análisis multivariante

### 4.5.1 Descripción del modelo

Los datos para este análisis se componen de 6,141 observaciones obtenidas de $N$ = 89 municipios que componen la mesoregión norte de Minas Gerais. Este modelo reflectivo se compone por 4 constructos y 69 indicadores debidamente estandarizados.

Ilustración 3

*Modelo empírico final de la investigación*

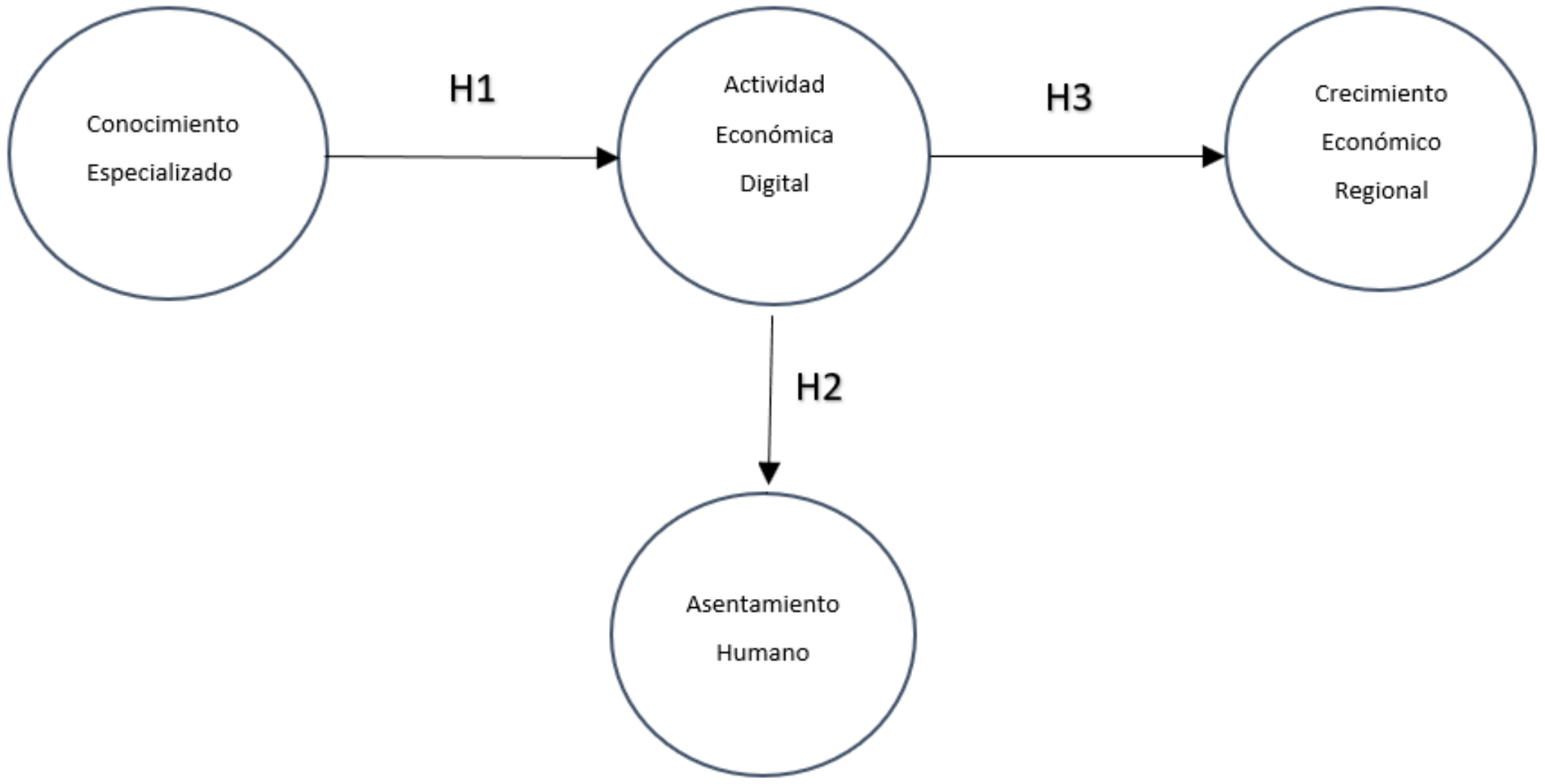

### 4.5.1 Evaluación del modelo

Antes de iniciar con el respectivo análisis multivariante Hair et al. (2012); Martínez & Fierro (2018b) establecen la importancia primero de la evaluación del modelo, lo cual implica examinar la confiabilidad de su consistencia interna, así como la validación convergente y discriminante. Para detectar la ausencia de multicolinealidad se realizó la prueba de factor de inflación de la varianza o (FIV), el cual establece la existencia de multicolinealidad para valores superiores a 5 en la escala de este índice (Mulero & Revuelta, 2010). Para cumplir con dichas evaluaciones se realizó las siguientes pruebas que arrojaron los siguientes resultados.

Tabla 27

*Prueba de Consistencia Interna*

| **Código** | **Constructos** | **Cronbach's α** |
|---|---|---|
| CE | Conocimiento Especializado | .998 |
| AED | Actividad Económica Digital | .999 |
| AHU | Asentamiento Humano | 1.000 |
| CER | Crecimiento Económico Regional | 1.000 |

**Tabla 28**

*Prueba de Varianza Extraída Media (AVE)*

| **Código** | **Constructos** | **AVE** |
|---|---|---|
| CE | Conocimiento Especializado | .945 |
| AED | Actividad Económica Digital | .987 |
| AHU | Asentamiento Humano | 1.000 |
| CER | Crecimiento Económico Regional | .998 |

Tabla 29

*Prueba de Fiabilidad Compuesta*

| **Código** | **Constructos** | **PFC** |
|---|---|---|
| CE | Conocimiento Especializado | .998 |
| AED | Actividad Económica Digital | .999 |
| AHU | Asentamiento Humano | 1.000 |
| CER | Crecimiento Económico Regional | 1.000 |

Tabla 30

*Prueba de Colinealidad Modelo*

| **Código** | **Constructos** | **VIF** |
|---|---|---|
| CE--AED | Conocimiento Especializado vs. Actividad Económica Digital | 1.000 |
| AED--AHU | Actividad Económica Digital vs. Asentamiento Humano | 1.000 |
| AED--CER | Actividad Económica Digital vs. Crecimiento Económico Regional | 1.000 |

Tabla 31

*Prueba de Validez Discriminante (Heterotrait-Monotrait HTMT)*

| | **AED** | **AHU** | **CE** |
|---|---|---|---|
| AHU | .731 | | |
| CE | .731 | .779 | |
| CER | .872 | .801 | .746 |

Luego de obtenidos los resultados se puede inferir que existe fiabilidad de constructo en el modelo, ya que en las pruebas obtuvieron valores superiores a valores $p$ = .7 en relación con la validación convergente mediante la prueba (AVE) se puede llegar a la conclusión que el conjunto de indicadores representa a un único constructo subyacente, ya que se obtuvo valores superiores a valores $p$ = .50 (Martínez & Fierro, 2018b). Esto significa que cada constructo explica al menos el 50% de la varianza de los indicadores. En la evaluación del nivel de colinealidad la prueba (VIF) no encontró problemas relacionados con la colinealidad, ya que sus valores fluctuaron en un valor $p$ = 1.00.

### 4.5.2 Calidad predictiva y precisión del modelo

Esta prueba permitió medir la calidad predictiva del modelo, la cual se realizó mediante la prueba Stone-Geisser de redundancia de validación cruzada del constructo o $Q^2$, la misma, evalúa la relevancia predictiva del modelo estructural y teórico, permitiendo examinar la pertinencia de predicción. Con los resultados alcanzados con un valor superior a cero *0* se concluye con la existencia de valides predictiva y relevancia del modelo (Thaisaiyi, 2020).

Tabla 32

*Prueba $Q^2$*

| **Código** | **Constructos** | **$Q^2$** |
|---|---|---|
| AED | Actividad Económica Digital | .323 |
| AHU | Asentamiento Humano | .534 |
| CER | Crecimiento Económico Regional | .473 |

A continuación, se presenta la visualización de la relación entre variables de estudio mediante el uso de los gráficos de los efectos totales de la fuerza de la relación entre constructos.

Gráfico Estadístico 1

*Efectos totales relación CE - AED*

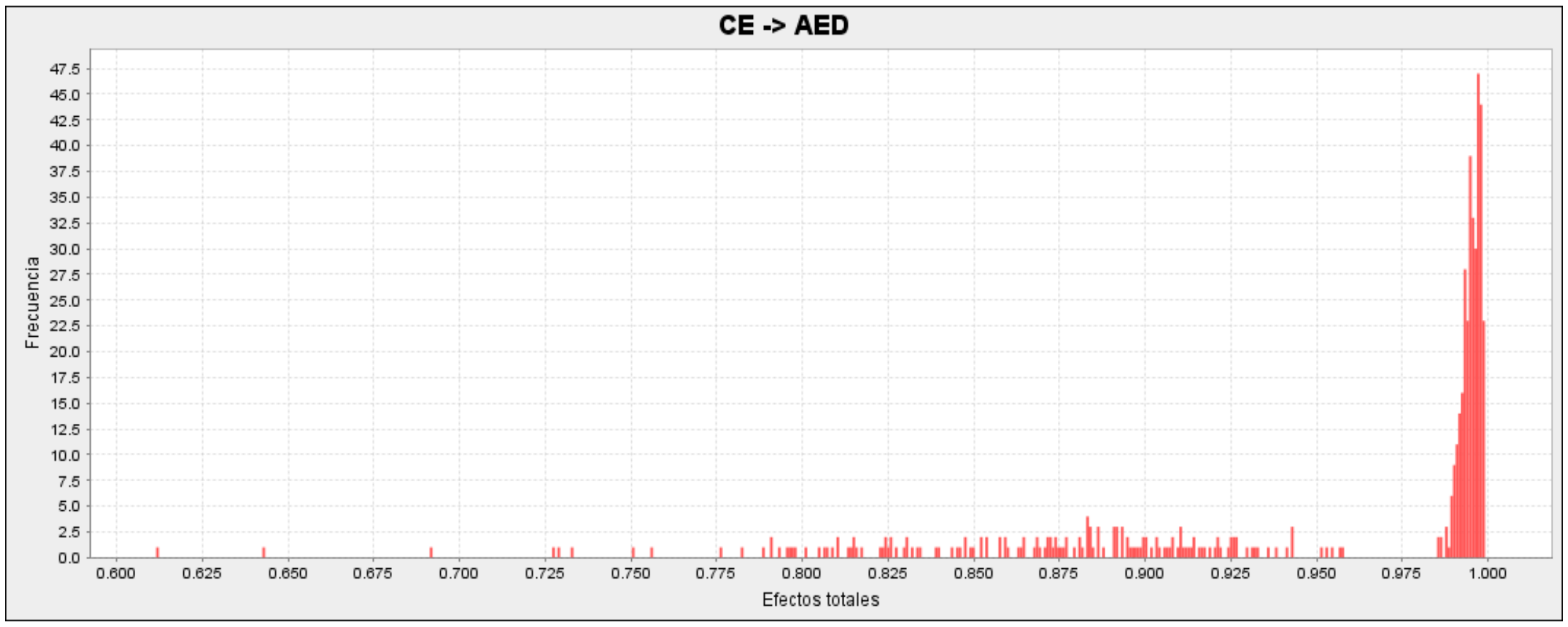

Gráfico Estadístico 2

*Efectos totales relación AED - AHU*

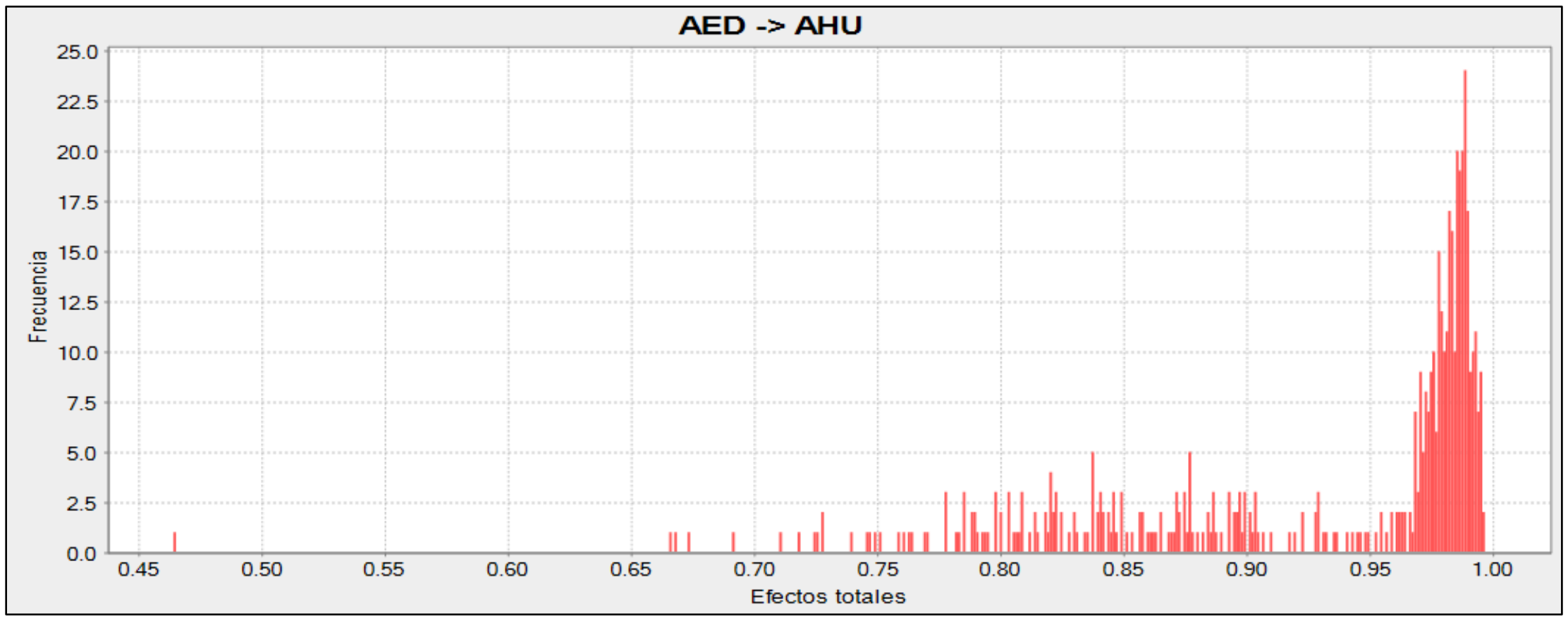

Gráfico Estadístico 3

*Efectos totales relación AED - CER*

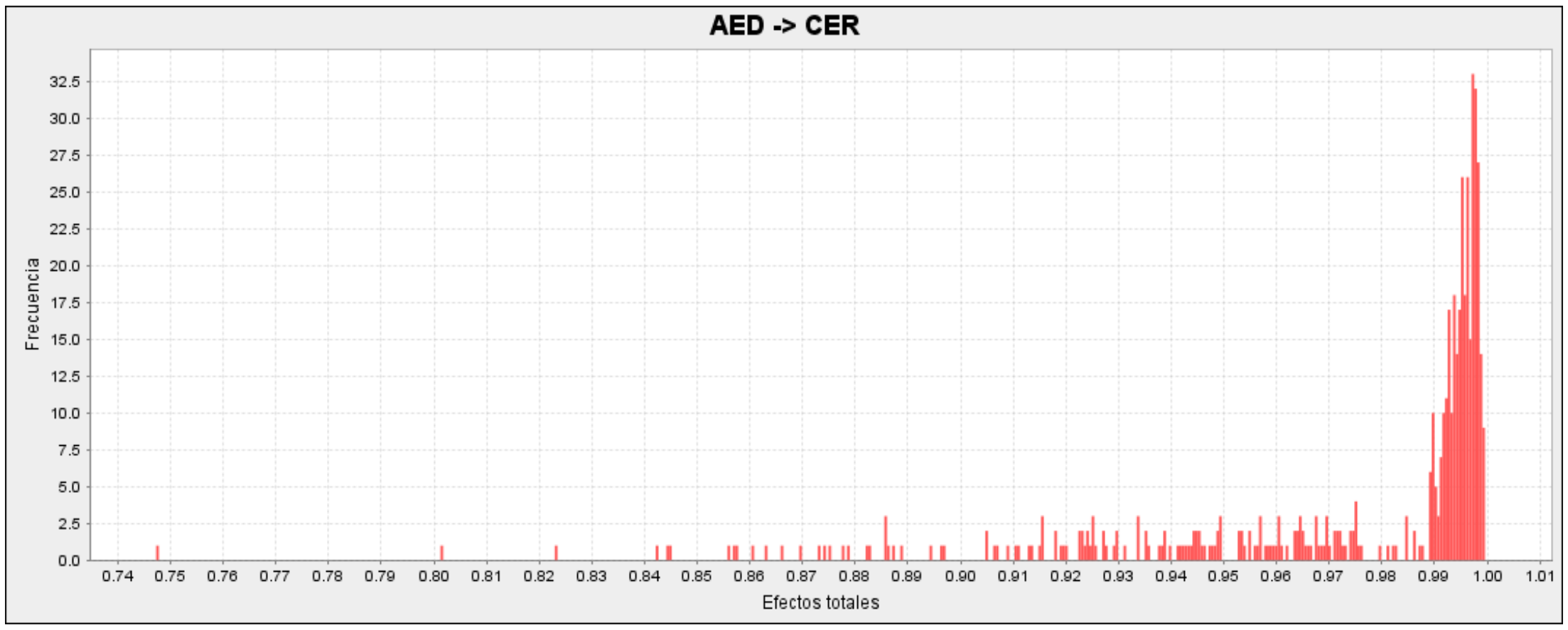

### 4.5.1 Magnitud y significancia estadísticas del modelo

Gráfico Estadístico 4

*Primer análisis mediante PLS-SEM del modelo de investigación: Resultados del coeficiente de ruta (β) y valores (p).*

El primer análisis de magnitud y significancia permitirá medir y probar las respectivas relaciones de hipótesis de este modelo de investigación. La magnitud se observa en el análisis de coeficiente de regresión estandarizado (β), y la significancia observada con los resultados de los valores (*p*). Para cumplir con dichas evaluaciones se usó el algoritmo PLS, las mismas arrojaron los siguientes resultados.

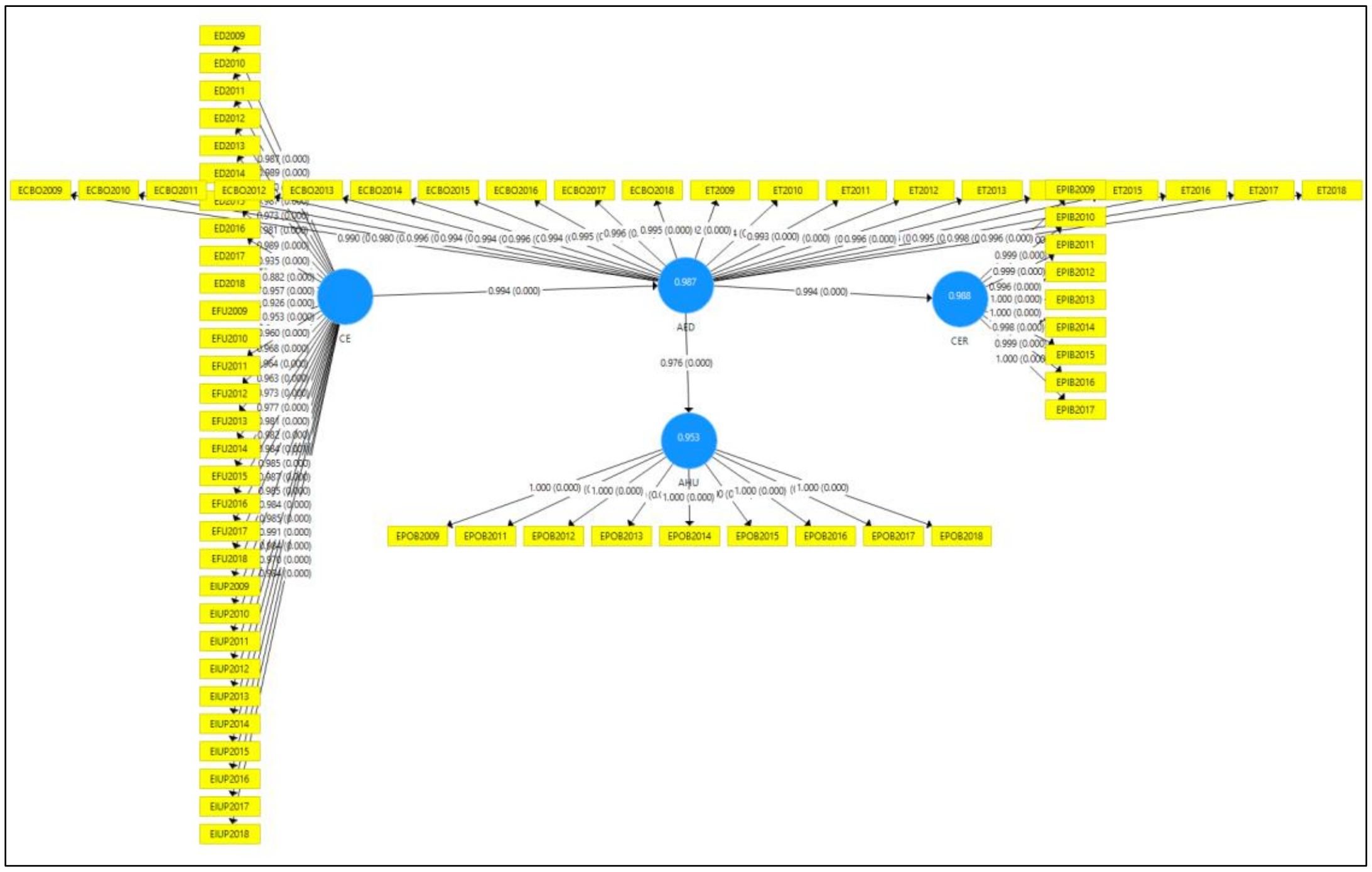

Gráfico Estadístico 5

*Segundo análisis mediante PLS-SEM del modelo de investigación: Resultados del coeficiente de ruta (β) y valores (t).*

El segundo análisis de magnitud y significancia permitirá medir y probar las respectivas relaciones de hipótesis de este modelo de investigación. La magnitud se observa en el análisis de coeficiente de regresión estandarizado (β), y la significancia observada con los resultados de los valores (*t*). Para cumplir con dichas evaluaciones se usó el análisis de remuestreo (5000 submuestras), las mismas arrojaron los siguientes resultados.

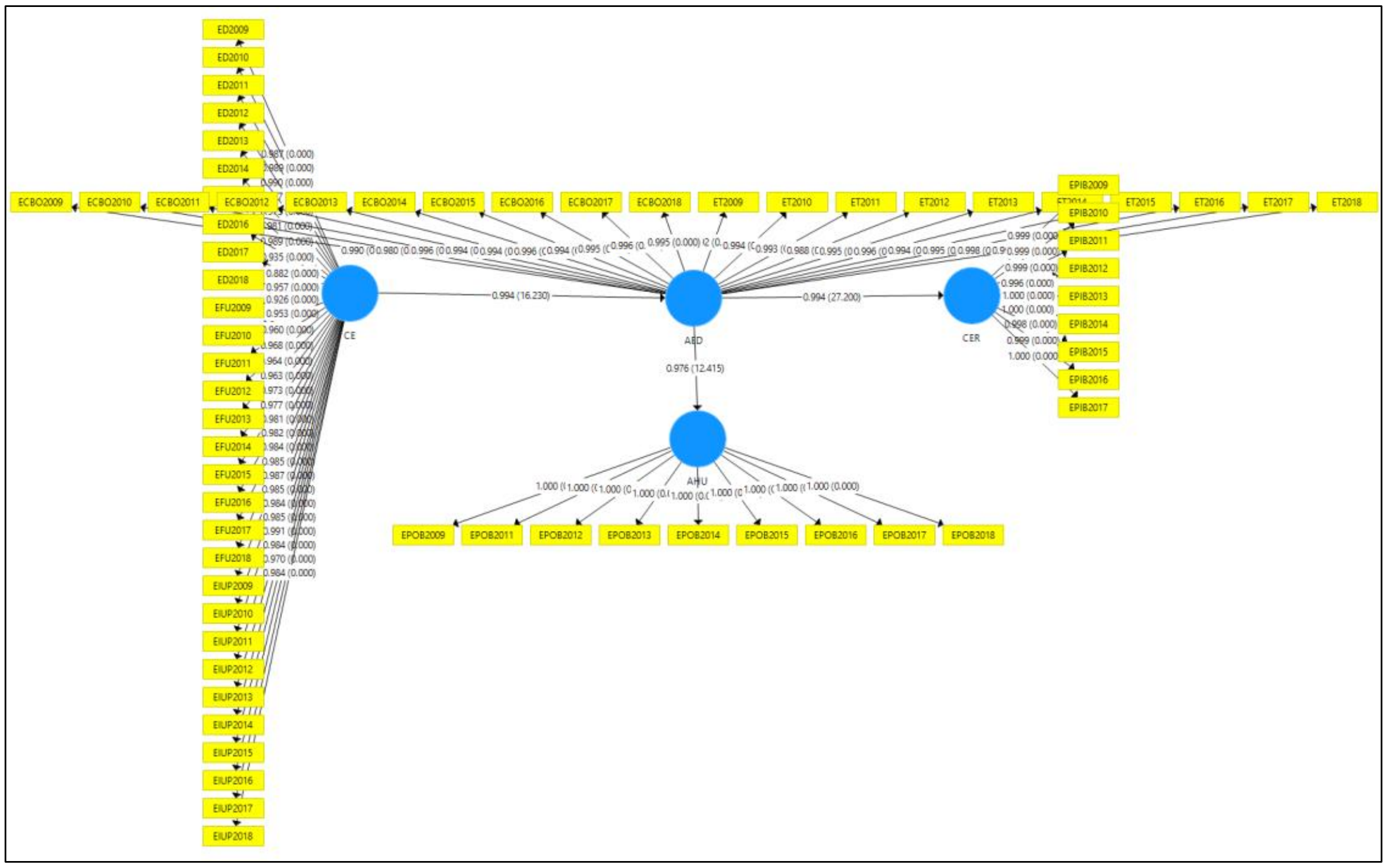

Tabla 33

*Validación de Hipótesis: Coeficientes de Regresión Estandarizado (β), valores (t), ($R^2$) y (p).*

| **Relación hipotética** | ***Coeficiente* β** | ***t* Student *Boostrapping*** | ***$R^2$*** | ***p*** | ***Hipótesis Apoyada*** |
|---|---|---|---|---|---|
| CE-----AED | .994 | 16.230 | .987 | .000 | SI |
| AED---AHU | .976 | 12.415 | .953 | .000 | SI |
| AED---CER | .994 | 27.200 | .988 | .000 | SI |

El nivel de significancia se determinó a partir de los valores proporcionados por los resultados de la prueba *t Student* por medio del algoritmo de remuestreo llamado (Boostrapping) con 5000 submuestras, los resultados están basados en una distribución t (4900) de dos colas; donde el valor critico *t* (0.05;4999) = 1,960 (Martínez & Fierro, 2018). El análisis arrojó los siguientes resultados empíricos dentro de la relación CE-AED valor *t* = 16.230 (*p* = .000); AED-AHU valor *t* = 12.415 (*p* = .000); y AED-CER valor *t* = 27.200 (*p* = .000). Estos resultados sobrepasan el valor umbral o critico (*t* y *p*) (Martínez & Fierro, 2018). Por lo tanto, existe una relación estadísticamente significativa entre los mencionados constructos.

Estos resultados apoyan la hipótesis H3 la cual confirma la relación predictiva entre la actividad económica digital (AED) y el crecimiento económico regional (CER) con un valor predictivo alto de $R^2$ = .988, β = .994, p = .000 (valor t = 27.200), donde por cada unidad de aumento en (AED) el crecimiento económico regional (CER) aumenta un 98.8%. En el caso de actividad económica digital (AED) con asentamiento humano (AHU) se halló un valor predictivo alto $R^2$ = .953, β = .976, p = .000 (valor t = 12.415), donde por cada unidad de aumento en (AED) el asentamiento humano (AHU) aumenta un 95.3%.; por último, conocimiento especializado CE con actividad económica digital AED se halló un valor predictivo alto de $R^2$ = .987, β = .994, p = .000 (valor t = 16.230), donde por cada unidad de aumento en (CE) la actividad económica digital (AED) aumenta un 98.7%.

## 4.6 Modelo de análisis espacial

### 4.6.1 Análisis exploratorio de datos espaciales

#### 4.6.1.1 Mapeo y geovisualización

Dentro del análisis cuantitativo espacial, el uso de mapas coropléticos es de suma relevancia para la representación de fenómenos asociados al análisis de unidades espaciales como es el caso en este estudio de los 89 municipios de la región norte de Minas Gerais (Scheider & Huisjes, 2019). Dentro del proceso de Geovisualización las tramas de colores de los mapas coropléticos se utilizan dentro del proceso de representación de los datos en las respectivas unidades espaciales (Buckley, 2019).

#### 4.6.1.2 Descripción del modelo de análisis espacial

Este análisis se realizó con el uso de datos de panel compuesta por observaciones de un mismo corte seccional para varios periodos de tiempo, los mismos están formadas por 89 municipios que componen la mesoregión norte de Minas Gerais estudiadas en un periodo de 10 años desde el 2009 hasta el 2018 (Toledo, 2012). Estas unidades espaciales componen un conjunto de datos debidamente estandarizados y organizados por los respectivos constructos e indicadores del estudio.

#### 4.6.1.3 Construcción de pesos espaciales basados en contigüidad

Este análisis espacial inicia con la construcción de pesos espaciales de contigüidad o conexión (Rubio, 2012) para determinar la relación de vecindad de primer orden. De esta forma se diseñó la matriz de pesos espaciales de tipo “Queen” (Anselin et al., 2006), la cual busca expresar la existencia de una relación vecinal. A continuación, se presenta la matriz de contigüidad espacial:

Mapa 2

*Mapa de conectividad espacial por municipio*

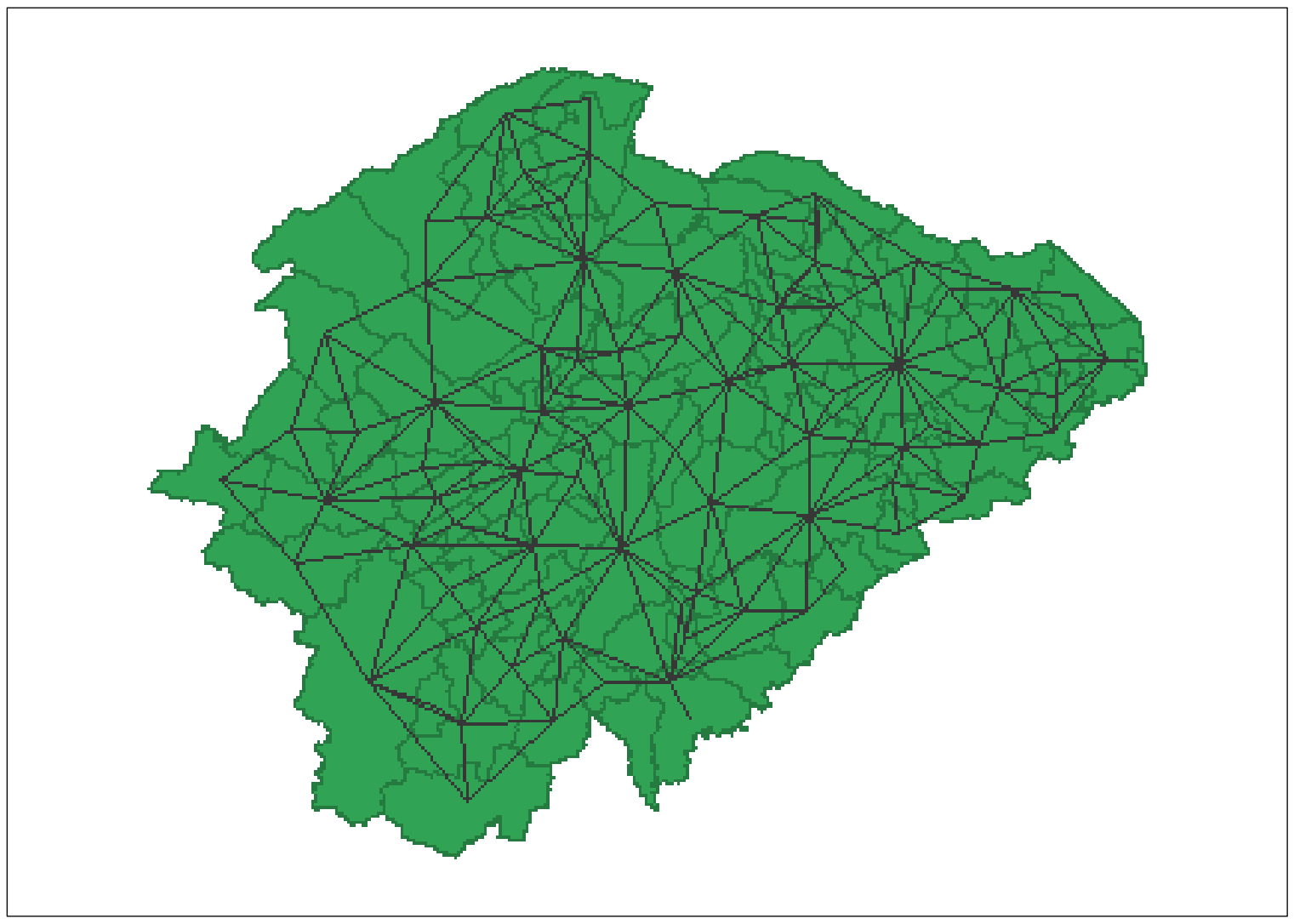

Tabla 34

*Matriz de contigüidad espaciales*

| **Property** | **Value** |
|---:|---|
| Tipo | Reina |
| Simetría | Simétrico |
| Archivo | TESIS-Q.gal |
| Id variable | Munic_PIO |
| No. Observaciones | 89 |
| Min. De Vecinos | 1 |
| Max. De Vecinos | 13 |
| Media Vecinos | 5,28 |
| Mediana Vecinos | 5,00 |
| % Distinto de cero | 5,93% |

Gráfico Estadístico 6

*Gráfico de relación vecinal por municipio*

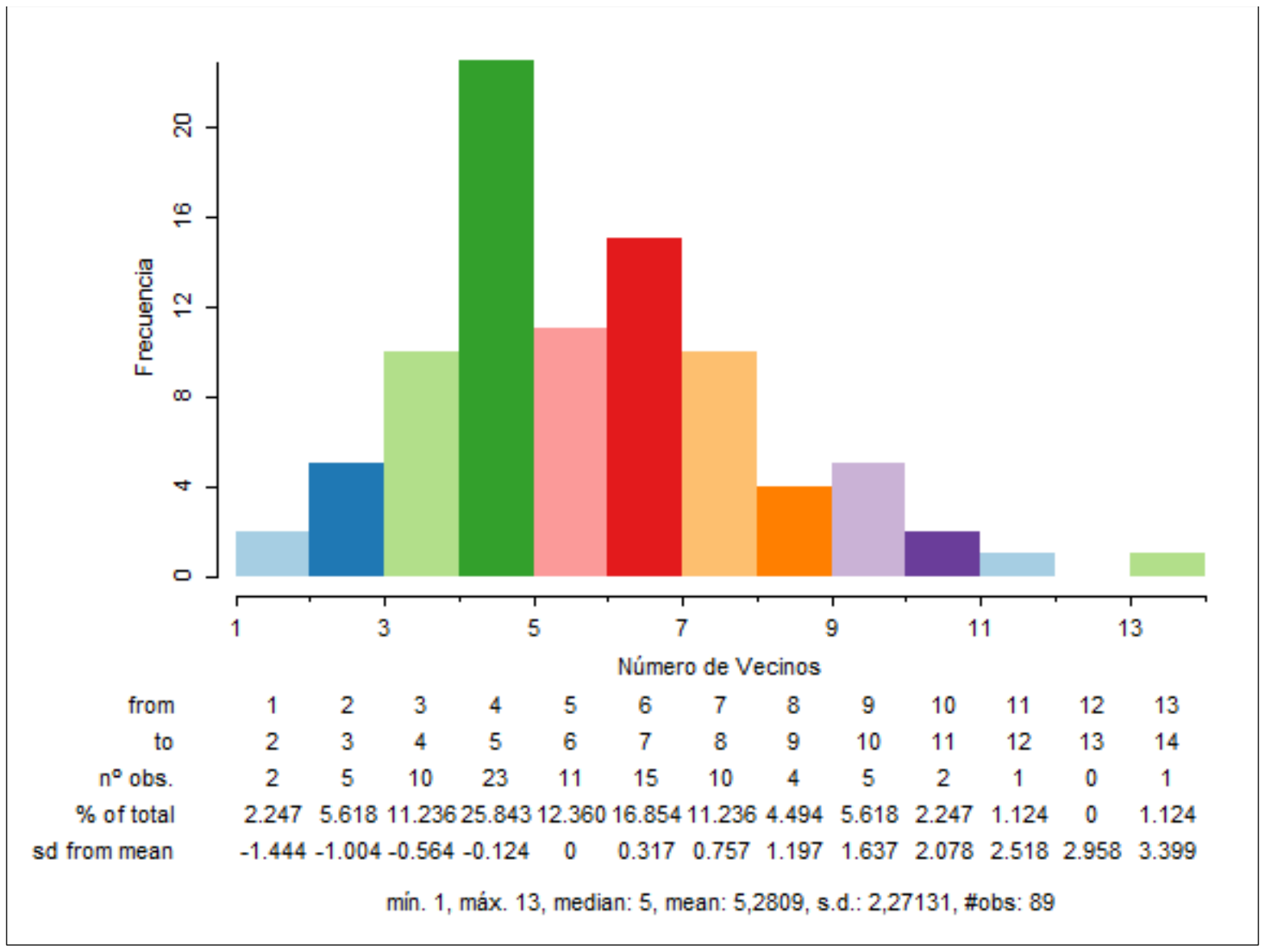

Donde 23 de los municipios que representa el 25.84% tienen 4 vecinos, 15 de los municipios que representa el 16.85% tienen 6 vecinos, 11 de los municipios que representa el 12.36% tienen 5 vecinos, 10 de los municipios que representa el 11.23% tienen 7 vecinos, 5 de los municipios que representa el 5.61% tienen 9 vecinos, 5 de los municipios que representa el 5.61% tienen 2 vecinos, 4 de los municipios que representa el 4.94% tienen 8 vecinos, 2 de los municipios que representa el 2.24% tienen 10 vecinos, 2 de los municipios que representa el 2.24% tienen 1 vecino, 1 de los municipios que representan al 1.12% tienen 13 vecinos, 1 de los municipios que representan al 1.12% tienen 11 vecinos y 12 municipios no tienen relación de contigüidad mediante el uso de pesos espaciales. Dentro de este análisis se encontró dos patrones extraños basados en conectividad vecinal, donde los Municipios de Divisa Alegre y Olhos de Agua solamente tienen un solo vecino.

#### 4.6.1.4 Relación vecinal de microrregiones del norte de Minas Gerais

Mediante el mapa de conectividad con el uso de pesos espaciales se establece la distribución regional de la relación vecinal con los municipios capitales de las microrregiones (ver proyección geográfica del Capítulo 3).

Mapa 3

*Proyección geográfica vecinal de Pirapora*

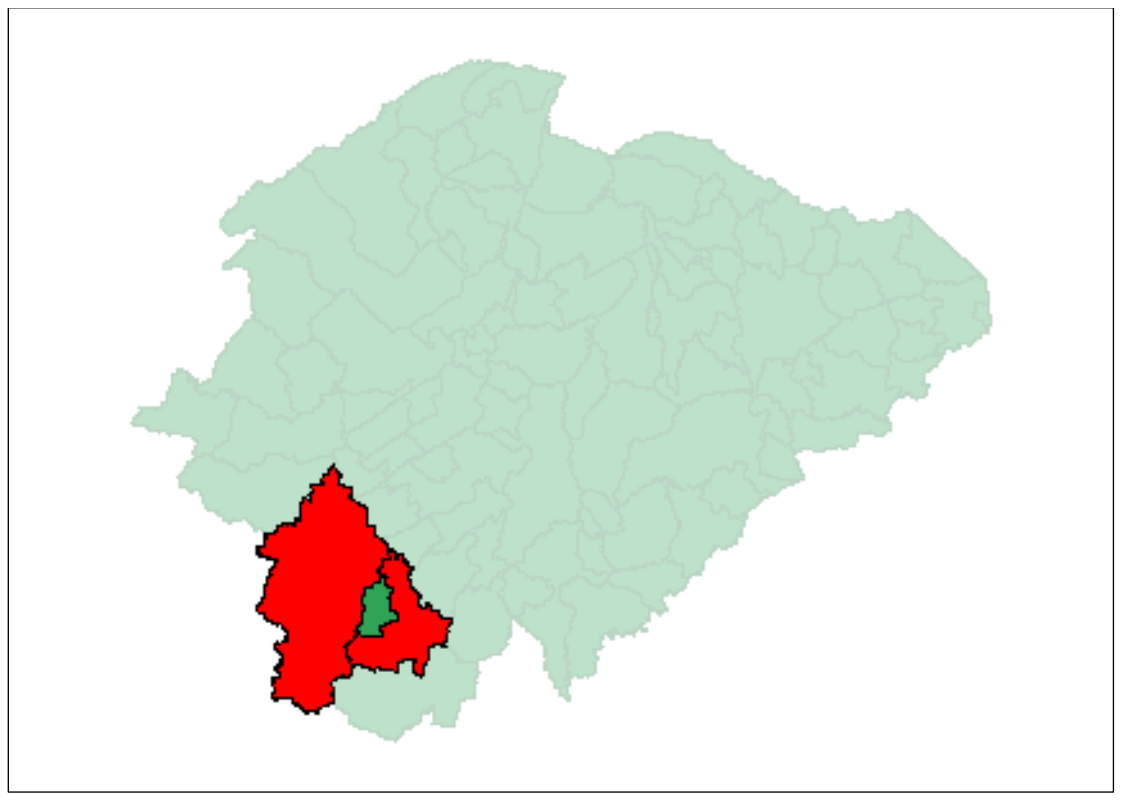

Mapa 4

*Proyección geográfica vecinal de Montes Claros*

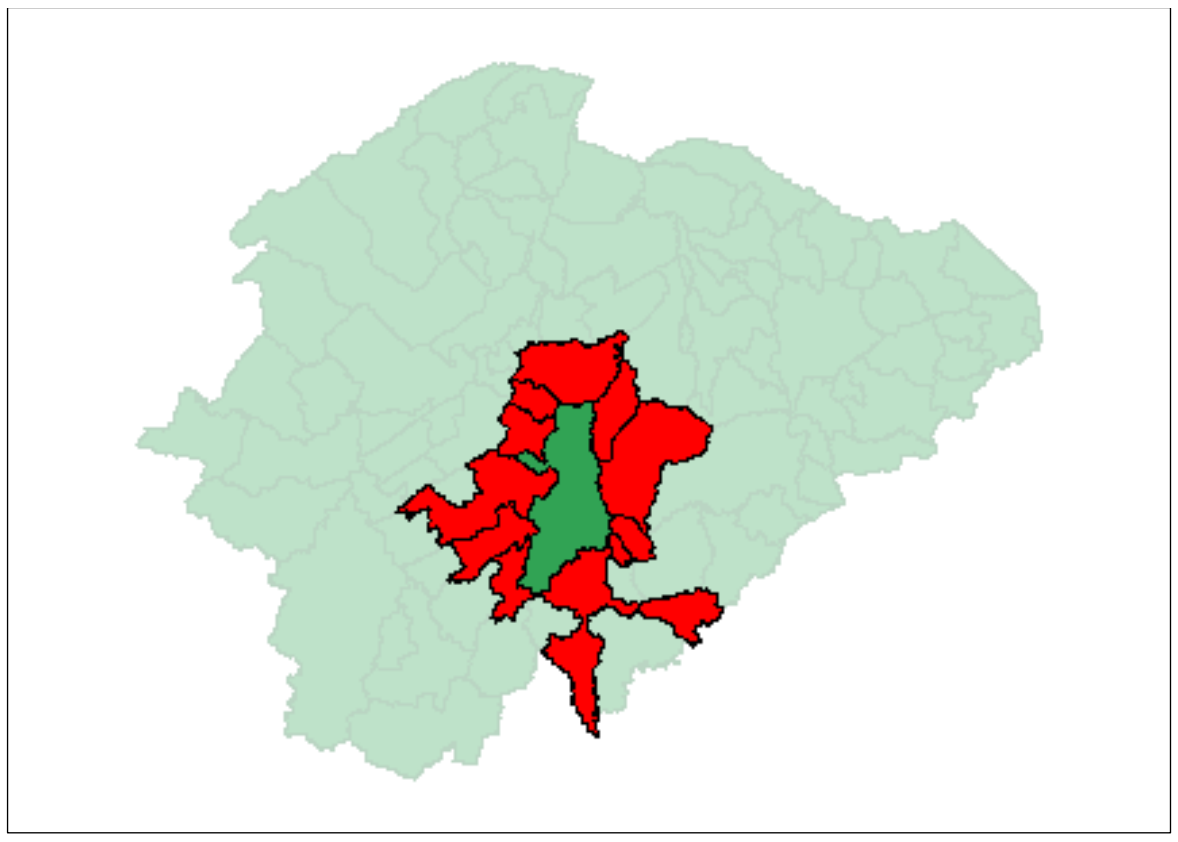

Mapa 5

*Proyección geográfica vecinal de Bocaiúva*

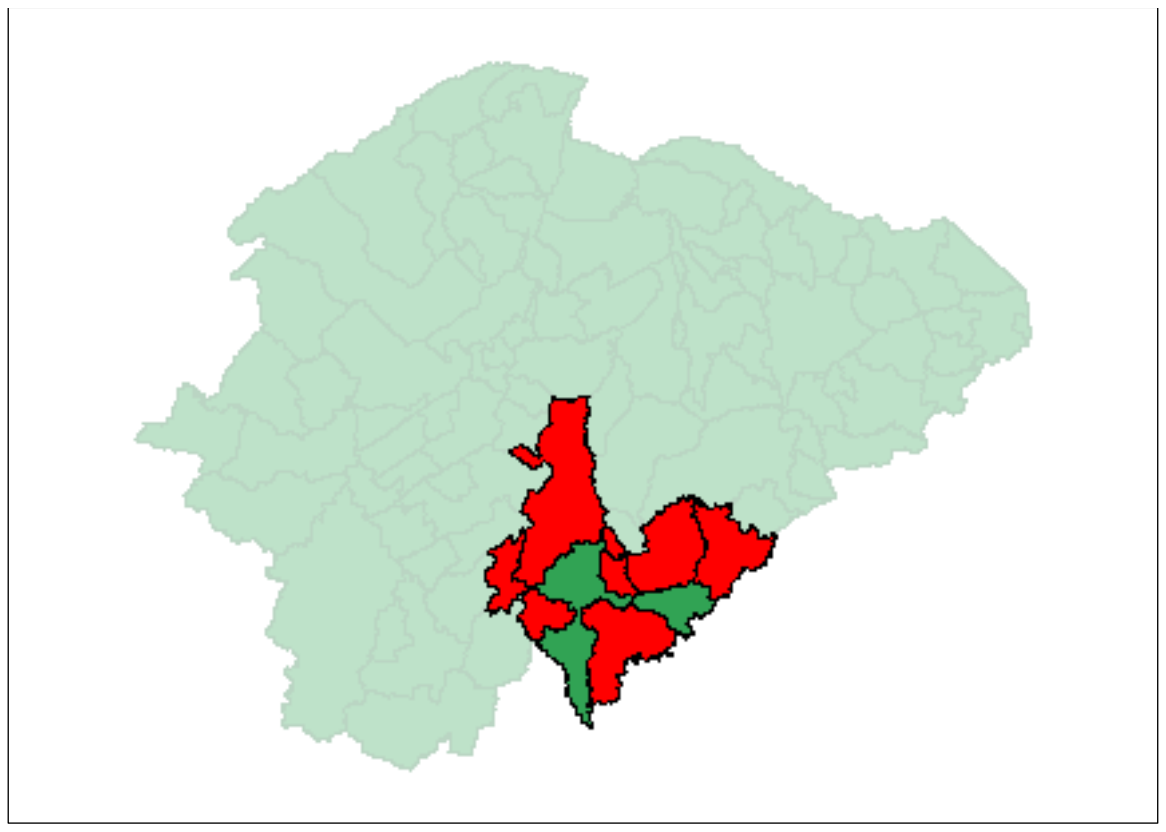

Mapa 6

*Proyección geográfica vecinal de Janauba*

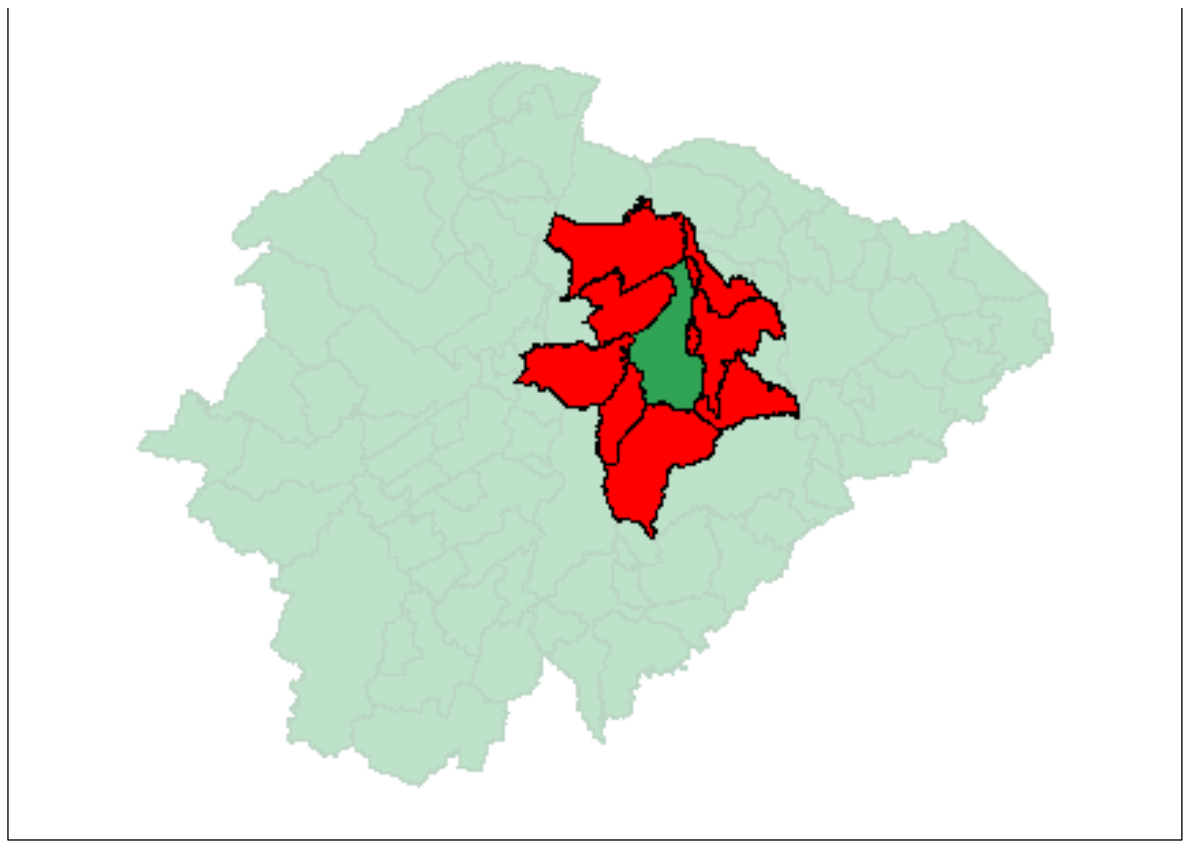

Mapa 7

*Proyección geográfica vecinal de Salinas*

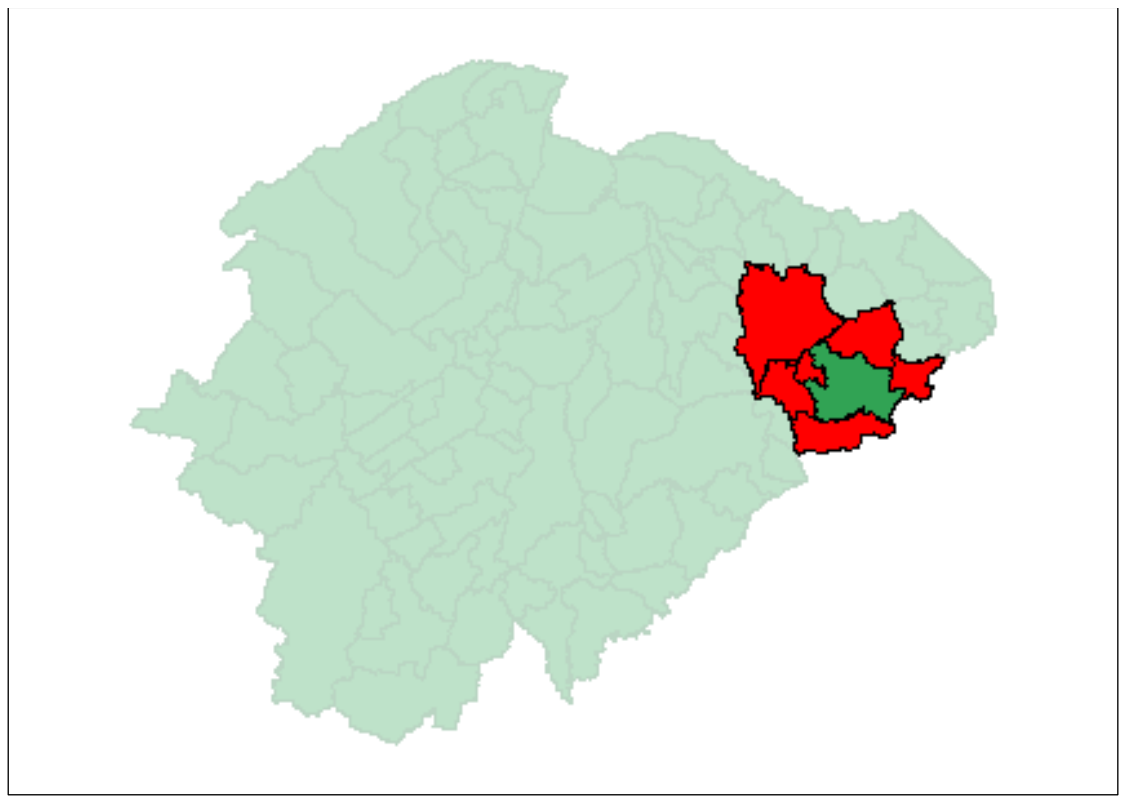

Mapa 8

*Proyección geográfica vecinal de Grao Mogol*

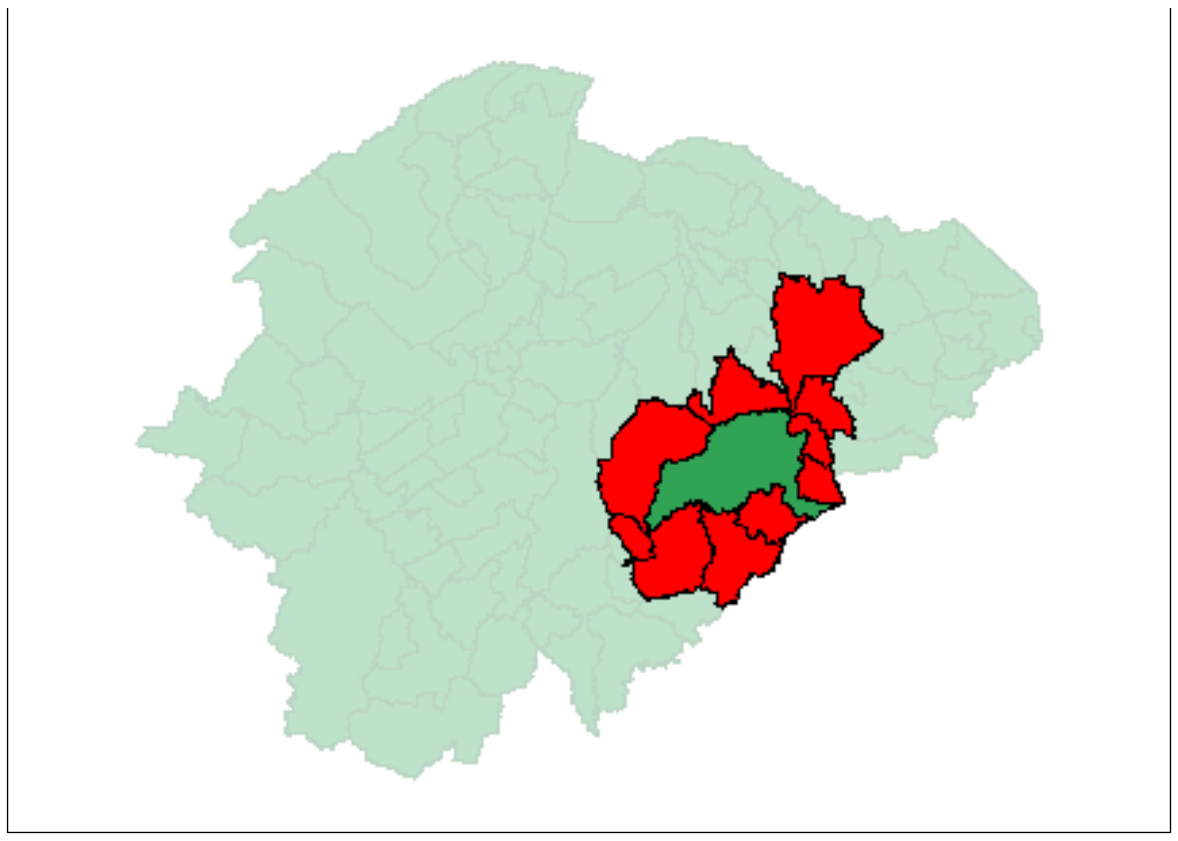

Mapa 9

*Proyección geográfica vecinal de Januária*

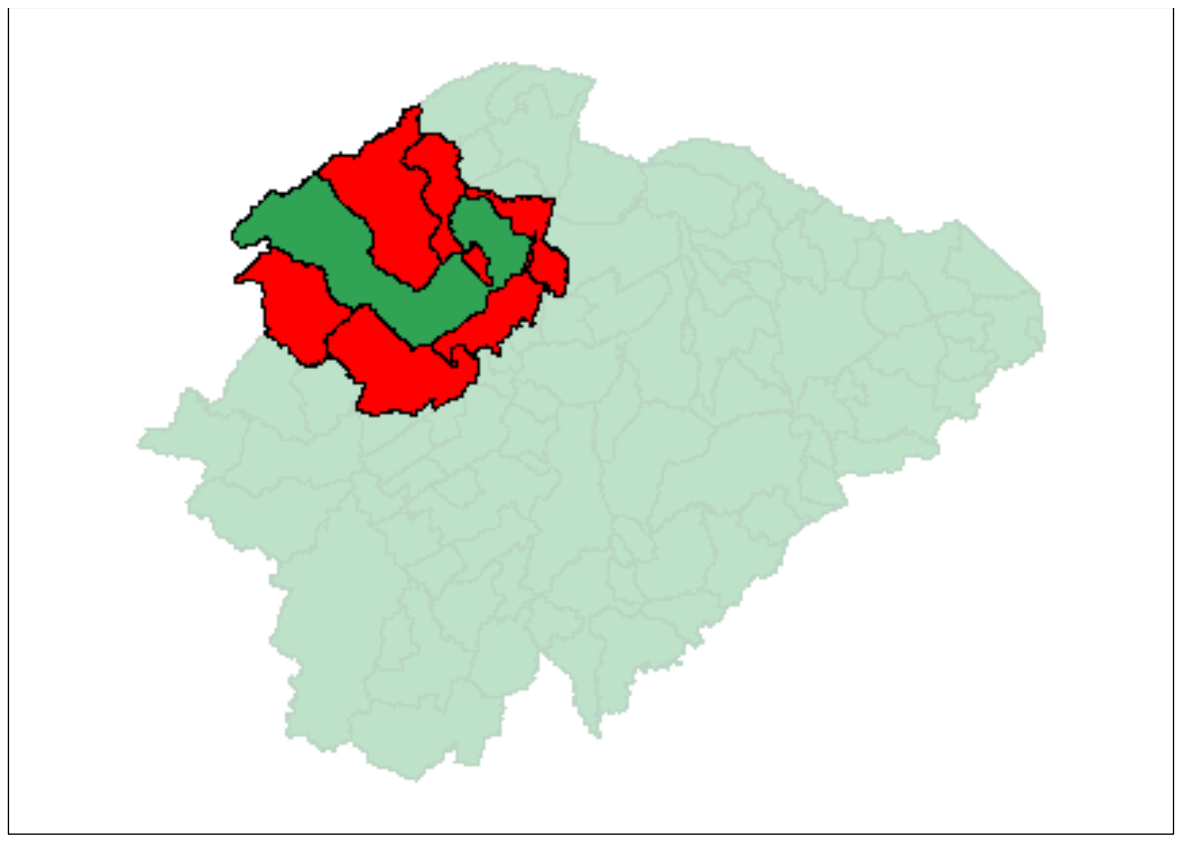

Estos mapas coropléticos fueron de mucha relevancia en este estudio pues permitieron observar la relación vecinal que tiene cada una de las regiones que componen este estudio, ya que en el análisis posterior de autocorrelación espacial se consideran las relaciones de vecinos próximos (Celemín, 2009).

Este análisis de contigüidad espacial de primer orden se realizó identificando cada una de las regiones que comprenden la mesorregión norte de minas y sus respectivos municipios, con el propósito de seguir un proceso metodológico comúnmente usado en este tipo de análisis de datos espaciales (Siabato & Guzmán, 2019). Esto con el propósito de cuantificar la autocorrelación espacial y poder definir si es positiva o negativa la misma.

### 4.6.2 Identificación de patrones espaciales en los municipios de la mesoregión norte de Minas Gerais

Este análisis exploratorio de datos espaciales busca identificar los patrones de crecimiento por variable, dentro de la distribución espacial de las regiones de estudio.

A continuación, se presentan el análisis exploratorio espacial para el año 2009 y 2018. Prat & Cànoves (2014) establecen la necesidad de distinguir la concentración geográfica causada por fenómenos que impactan el marco geográfico, como una de las características más relevantes de la distribución espacial. Es importante resaltar que, la actividad económica no es consecuencia de un incidente aleatorio, sino que responde a varios factores, ambientales, sociales, económicos y espaciales, que influyen a que ciertas actividades económicas que se concentran en determinados espacios.

El proceso de gestión de datos en el análisis de concentración se realiza para identificar municipios similares según las variables seleccionadas, esto se logra mediante el uso de una métrica apropiada[1] y un criterio de vinculación que especifique la disimilitud de los conjuntos en función a la distancia.

Por lo tanto, el objetivo principal en este análisis fue identificar grupos similares, teniendo como punto base la matriz de similitudes, distancia entre municipios y las variables que buscamos observar. Esto permitió enfocar ciertos fenómenos que se pudieron observar en segmentos espaciotemporales.

[1] *Una medida de la distancia entre pares de observaciones*

Mapa 10

*Agrupamiento espacial regional basado en patrones de indicador IUPP (elaboración propia)*

| EIUP 2009 | EIUP 2018 |
|---|---|
| 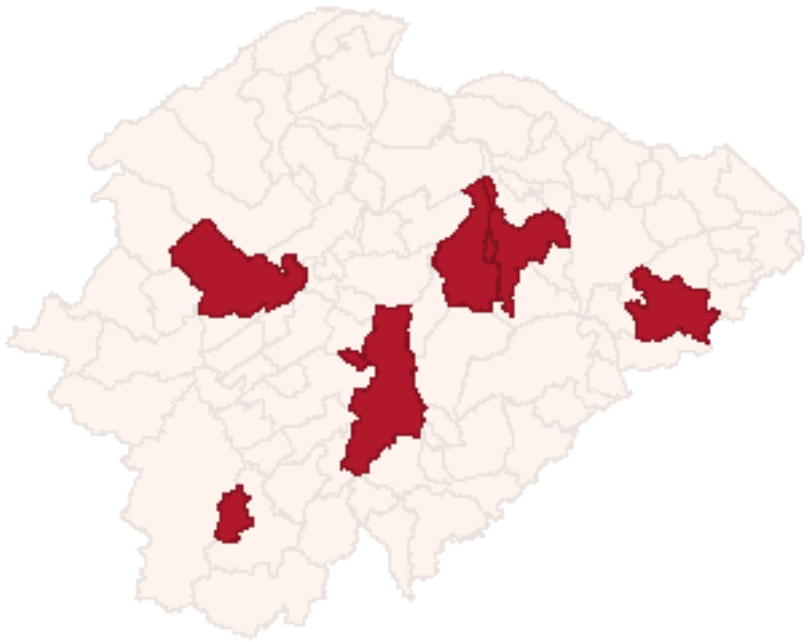 | 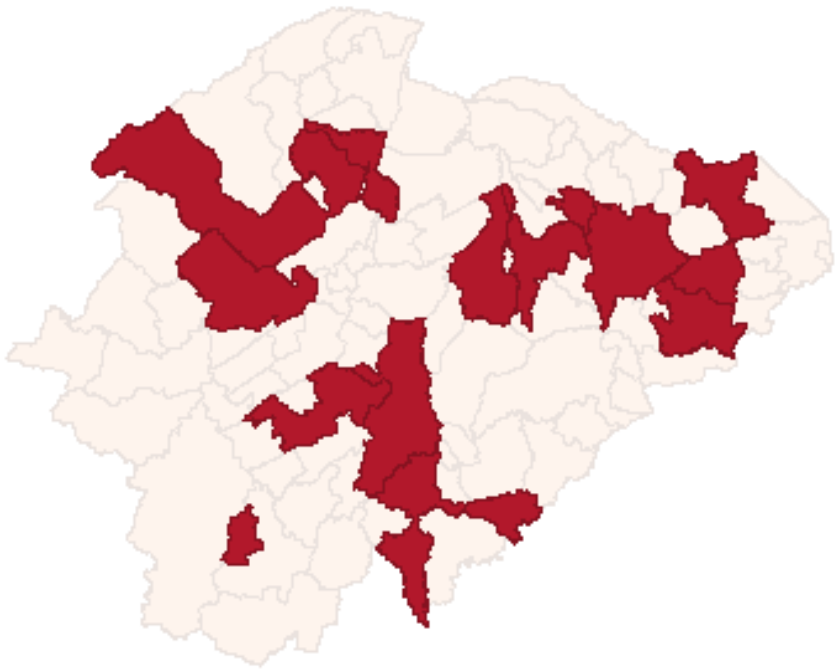 |
| Valores Superiores al 75% = 7 Municipios<br>1) San Francisco; 2) Pirapora; 3) Montes Claros; 4) Janauba; 5) Nova Porterinha; 6) Porterinha; 7) Salinas | Valores Superiores al 75% = 14 Municipios<br>1) Januária; 2) San Francisco; 3) Pirapora; 4) Montes Claros; 5) Corazón de Jesús; 6) Bocaiúva 7) Janauba; 8) Mato Verde; 9) Porterinha; 10) Salinas; 11) Rio Pardo de Minas; 12) Taiobeiras; 13) Sao Joao do Paraíso; 14) Itacarambi |

Mapa 11

*Agrupamiento espacial regional basado en patrones de indicador DNCT (elaboración propia)*

| ED 2009 | ED 2018 |
|---|---|
| 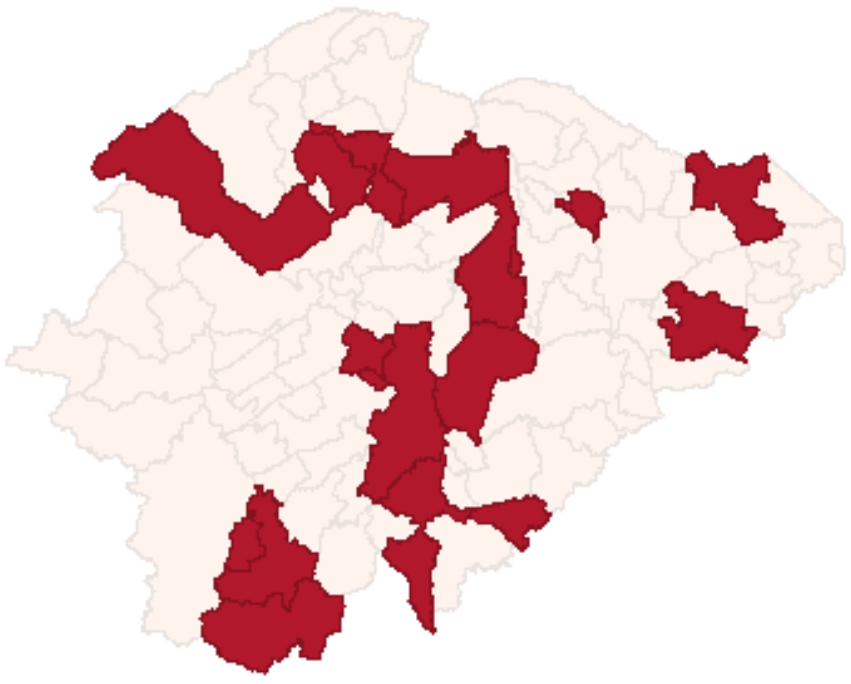 | 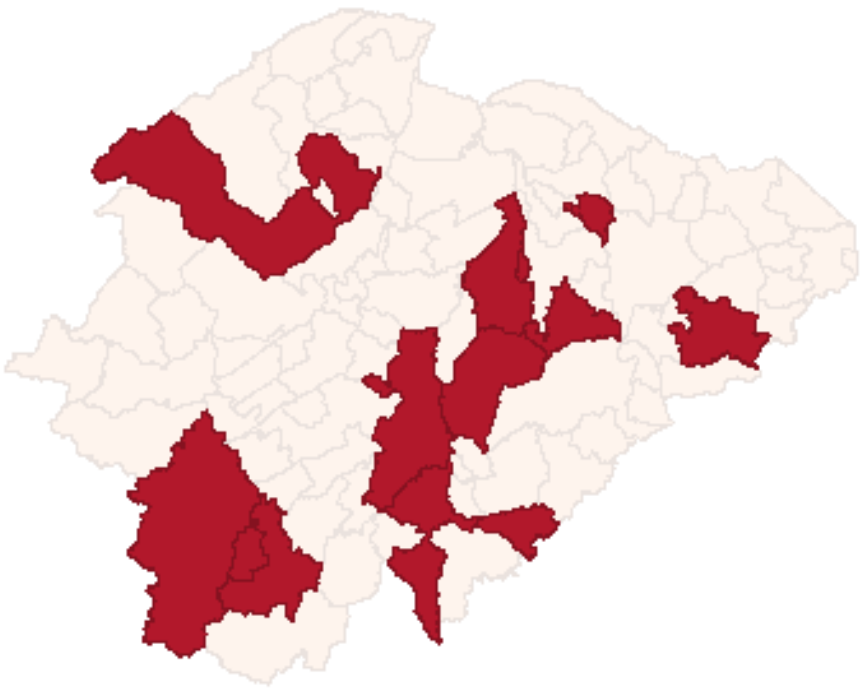 |
| Valores Superiores al 75% = 15 Municipios<br>1) Januária; 2) Pirapora; 3) Mirabela; 4) Montes Claros; 5) Bocaiúva; 6) Jaiba; 7) Francisco Sa; 8) Janauba; 9) Nova Porterinha; 10) Mato Verde; 11) Salinas; 12) Sao Joao do Paraíso; 13) Itacarambi; 14) Lassance; 15) Varzea da Palma | Valores Superiores al 75% = 12 Municipios<br>1) Januária; 2) Pirapora; 3) Montes Claros; 4) Bocaiúva; 5) Francisco Sa; 6) Janauba; 7) Nova Porterinha; 8) Riacho dos Machados; 9) Mato Verde; 10) Salinas; 11) Buritizeiro; 12) Varzea da Palma |

Mapa 12

*Agrupamiento espacial regional basado en patrones de indicador FUDEM (elaboración propia)*

| **EFU 2009** | **EFU 2018** |
|---|---|
| 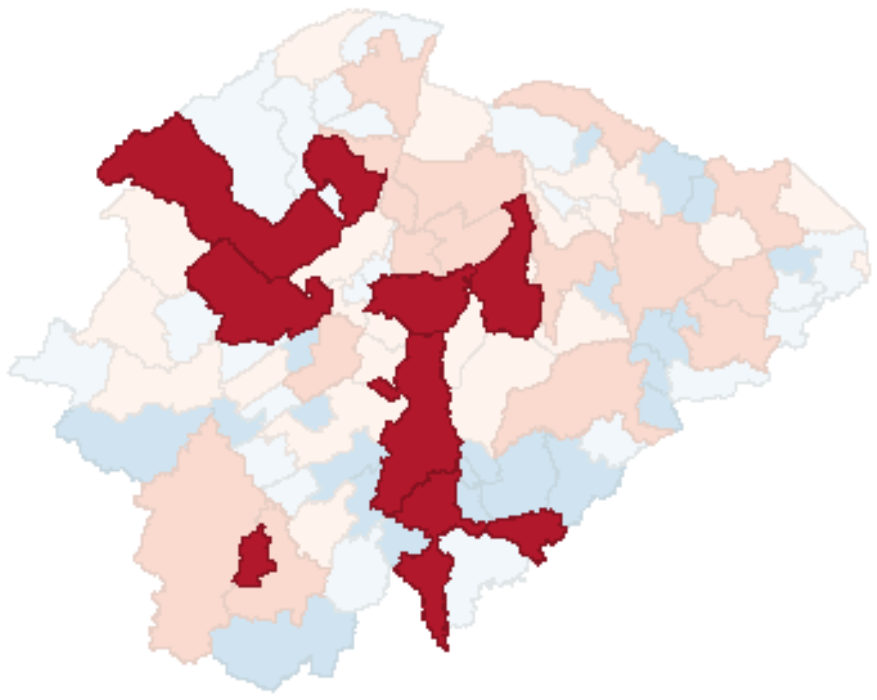 | 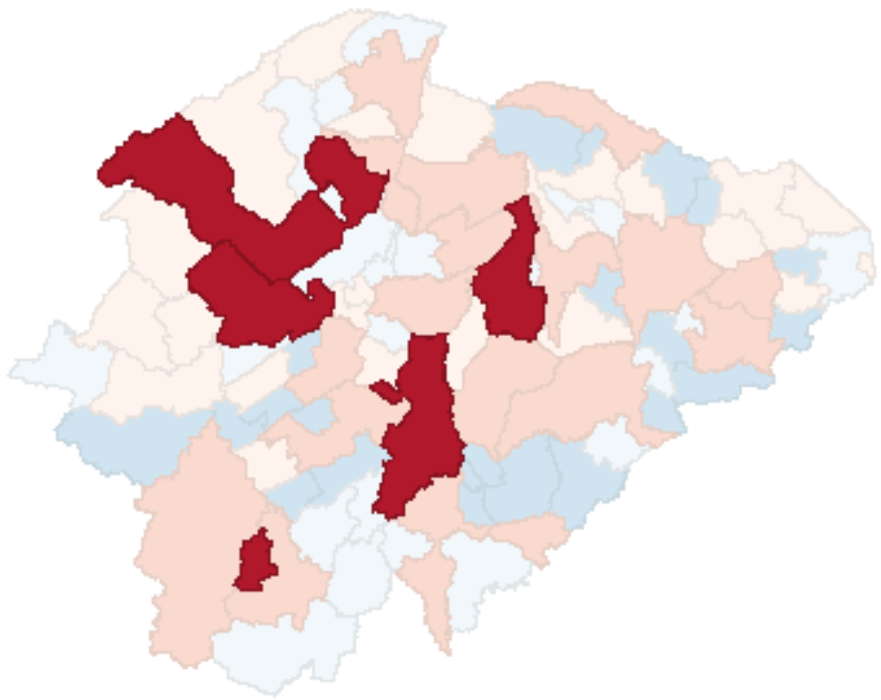 |
| Valores Superiores al 75% = 7 Municipios<br>1) Januária; 2) San Francisco; 3) Pirapora; 4) Sao Joao da Ponte; 5) Montes Claros; 6) Bocaiuva; 7) Janauba | Valores Superiores al 75% = 5 Municipios<br>1) Januária; 2) San Francisco; 3) Pirapora; 4) Montes Claros; 5) Janauba |

Mapa 13

*Agrupamiento espacial regional basado en patrones de indicador CBO (elaboración propia)*

| ECBO 2009 | ECBO 2018 |
|---|---|
| 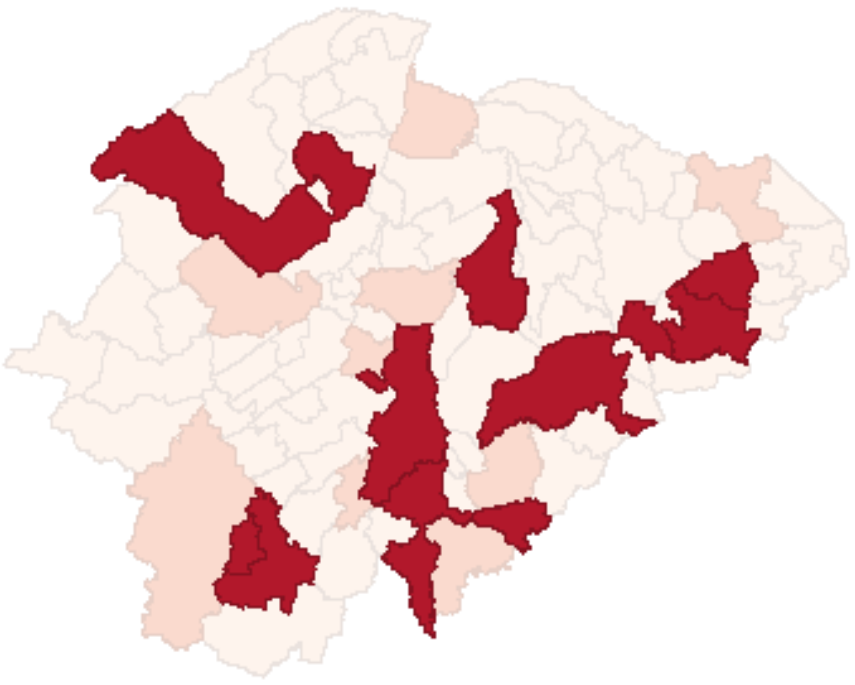 | 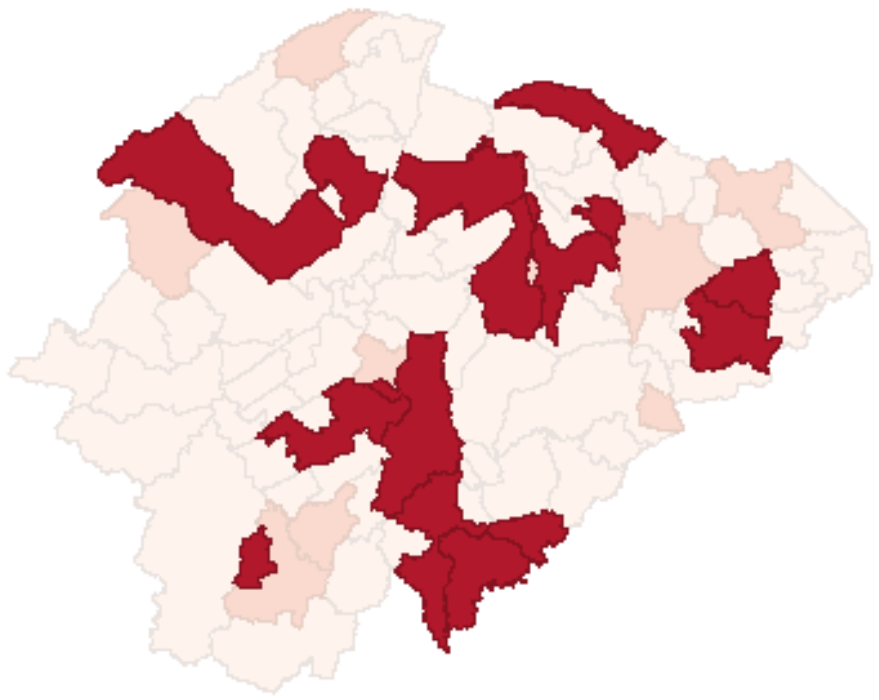 |
| Valores Superiores al 75% = 10 Municipios<br>1) Janauria; 2) Pirapora; 3) Montes Claros; 4) Bocaiuva; 5) Janauba; 6) Grao Mogol; 7) Fruta de Leite; 8) Salinas; 9) Taiobeiras; 10) Varzea da Palma | Valores Superiores al 75% = 13 Municipios<br>1) Janauria; 2) Pirapora; 3) Corazon de Jesus; 4) Montes Claros; 5) Bocaiuva; 6) Jaiba; 7) Olhos de Agua; 8) Janauba; 9) Espinosa; 10) Porteirinha; 11) Mato Verde; 12) Salinas; 13) Taiobeiras |

Mapa 14

*Agrupamiento espacial regional basado en patrones de indicador TIC (elaboración propia)*

| **ETIC 2009** | **ETIC 2018** |
|---|---|
| 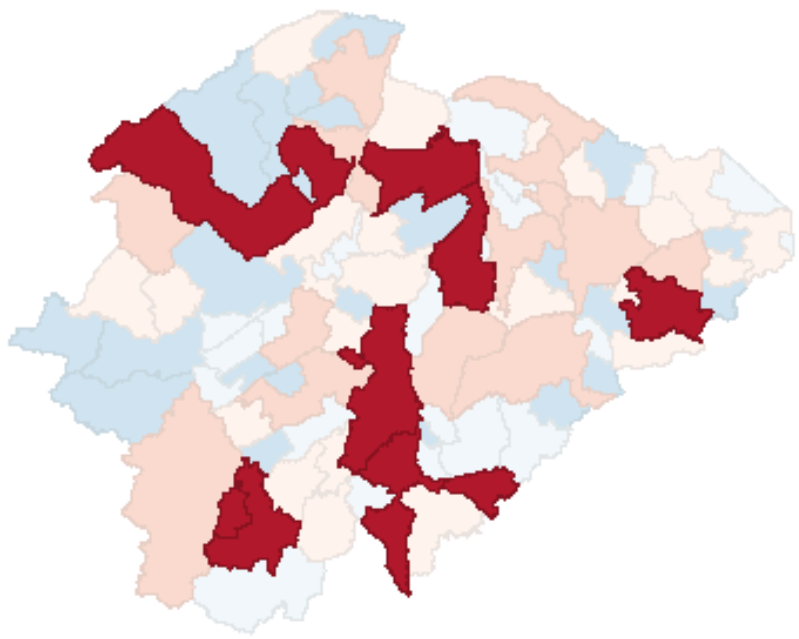 | 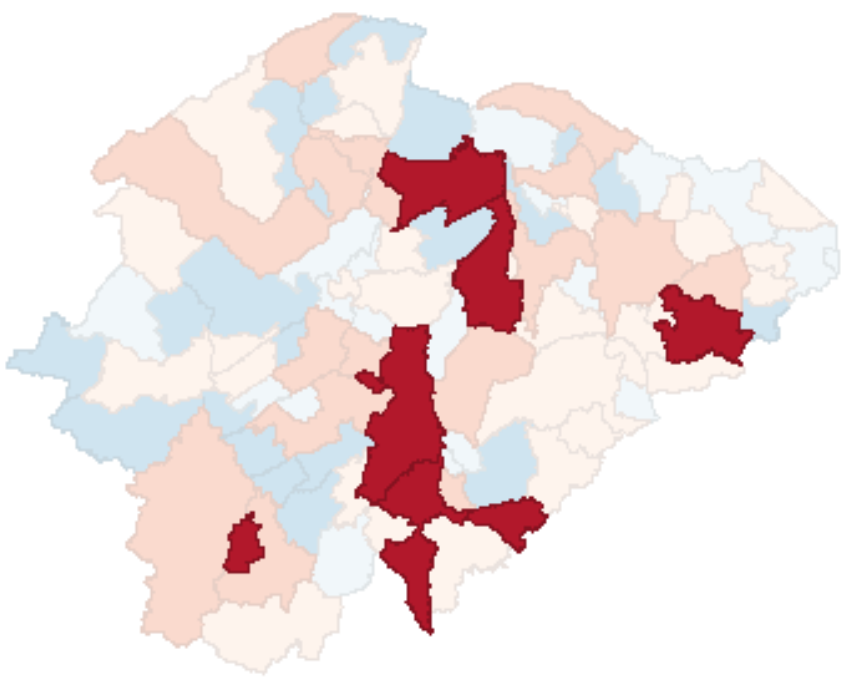 |
| Valores Superiores al 75% = 8 Municipios<br>1) Janauria; 2) Pirapora; 3) Montes Claros; 4) Bocaiuva; 5) Jaiba; 6) Janauba; 7) Salinas; 8) Varzea da Palma | Valores Superiores al 75% = 6 Municipios<br>1) Janauba; 2) Pirapora; 3) Montes Claros; 4) Bocaiuva; 5) Jaiba; 6) Salinas |

Mapa 15

*Agrupamiento espacial regional basado en patrones de indicador POB (elaboración propia)*

| **EPOB 2009** | **EPOB 2018** |
|---|---|
| 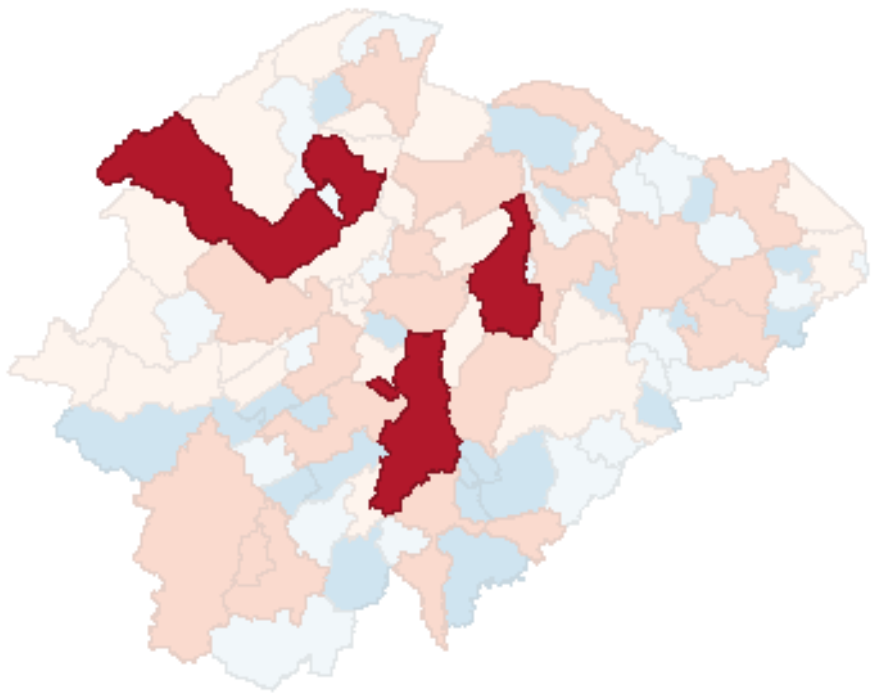 | 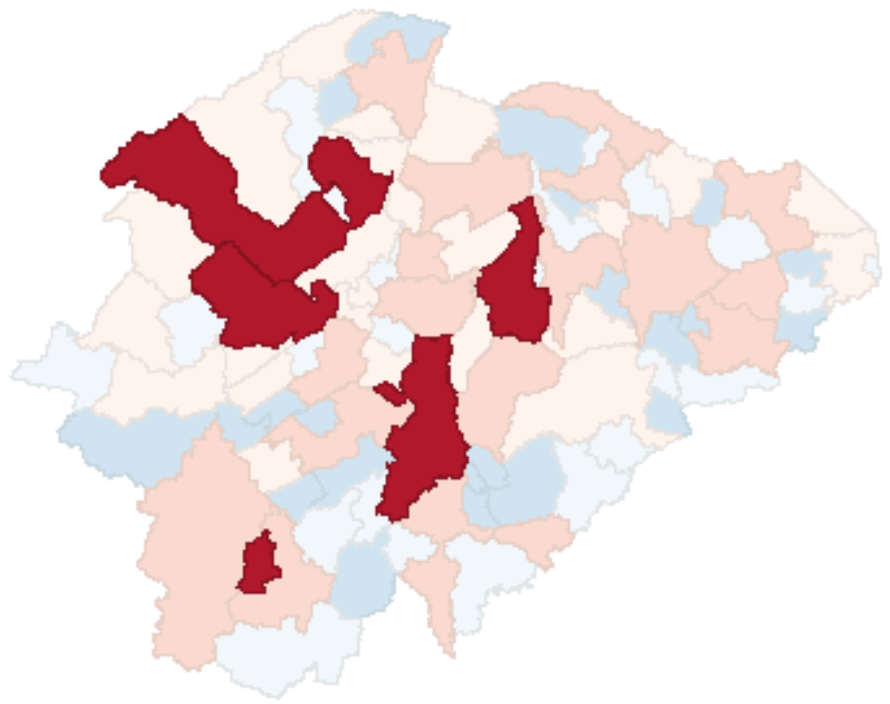 |
| Valores Superiores al 75% = 3 Municipios<br>1) Januária; 2) Montes Claros; 3) Janauba | Valores Superiores al 75% = 5 Municipios<br>1) Januária; 2) San Francisco; 3) Pirapora; 4) Montes Claros; 5) Janauba |

### 4.6.3 Evaluación de la relación hipotética mediante el análisis de regresión espacial en la mesoregión norte de Minas Gerais

El siguiente análisis busca realizar un modelo de regresión ponderada geográficamente, la misma es una forma local de regresión lineal que se utiliza para modelar las relaciones entre la actividad económica digital y el asentamiento humano con el propósito de observar si varían espacialmente, buscando identificar si los hallazgos son constantes en toda la región (Sabogal, 2013).

Tabla 35

*Análisis de Regresión Regional de AED con AHU*

| **Región** | **Relación Variables** | **2009** $R^2$ | $p$ | **2018** $R^2$ | $p$ |
|---|---|---|---|---|---|
| Norte Minas | ET----EPOB | .962 | .000 | .961 | .000 |
| N=89 | ECBO---EPOB | .902 | .000 | .927 | .000 |

Estos resultados antes presentados llevan a confirma la relación predictiva y validación de hipótesis entre la actividad económica digital AED y asentamiento humano AHU donde se halló un valor predictivo alto de $R^2$ = .962, valor $p$ =.000; para el indicador cantidad de empresas relacionadas al rubro de las tecnologías de información ET en el año 2009; y un valor predictivo alto de $R^2$ = .961, valor $p$ = .000 para el año 2018.

En relación con el Indicador de la clasificación brasileña de ocupaciones CBO en el rubro de las tecnologías de información se encontró un valor predictivo alto de $R^2$ = .902, valor $p$ = .000 para el año 2009; y un valor predictivo alto de $R^2$ = .927, valor $p$ = .000 para el año 2018. Estos resultados muestran una fuerte relación estadística entre las variables predictoras ET la cual explica el 96% de la variable de respuesta EPOB para el año 2009 y 2018. En cambio, la variable predictora CBO explica el 90 y 92% de la variable de respuesta EPOB para el año 2009 y 2018. Confirmándose y validándose así la hipótesis H2 la cual establece que la actividad económica digital impulsa significativamente el asentamiento humano.

### 4.6.4 Análisis de regresión espacial regional en la mesoregión norte de Minas Gerais

Este análisis busca evaluar y modelar las relaciones espaciales que varían entre el conocimiento especializado, actividad económica digital y crecimiento económico regional a nivel regional, comparándolo con las demás unidades espaciales (Yrigoyen, 2006). Para realizar este estudio se usó las estructuras de división regional de Brasil conocida como "Microrregiones" (Camargos, 2017; Bosco, 2009).

Mapa 16

*División regional mesoregión norte de Minas Gerais*

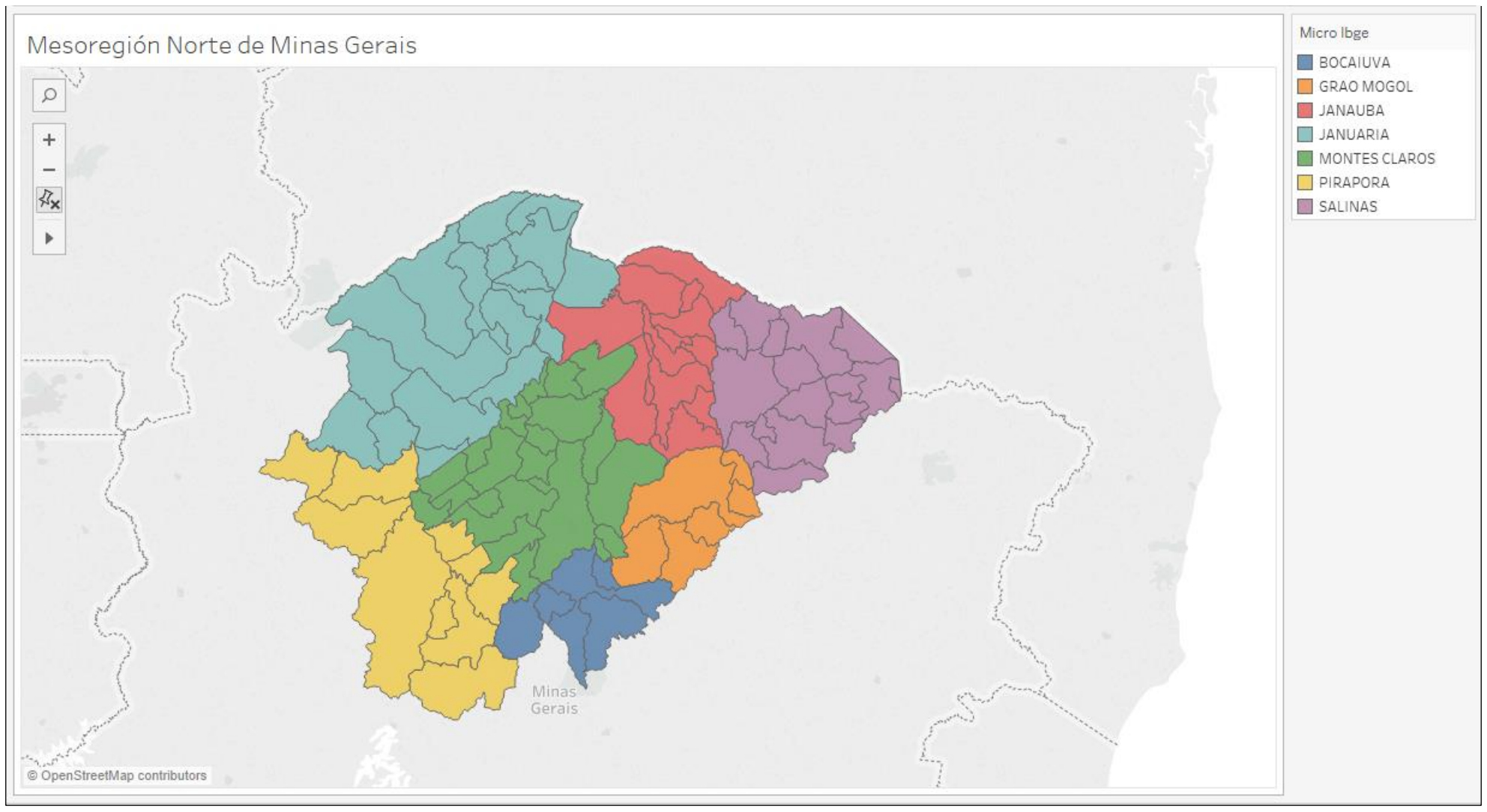

### 4.6.4.1 Análisis comparativo regresión regional microrregión salinas

Mapa 17

*Microrregión de Salinas (elaboración propia)*

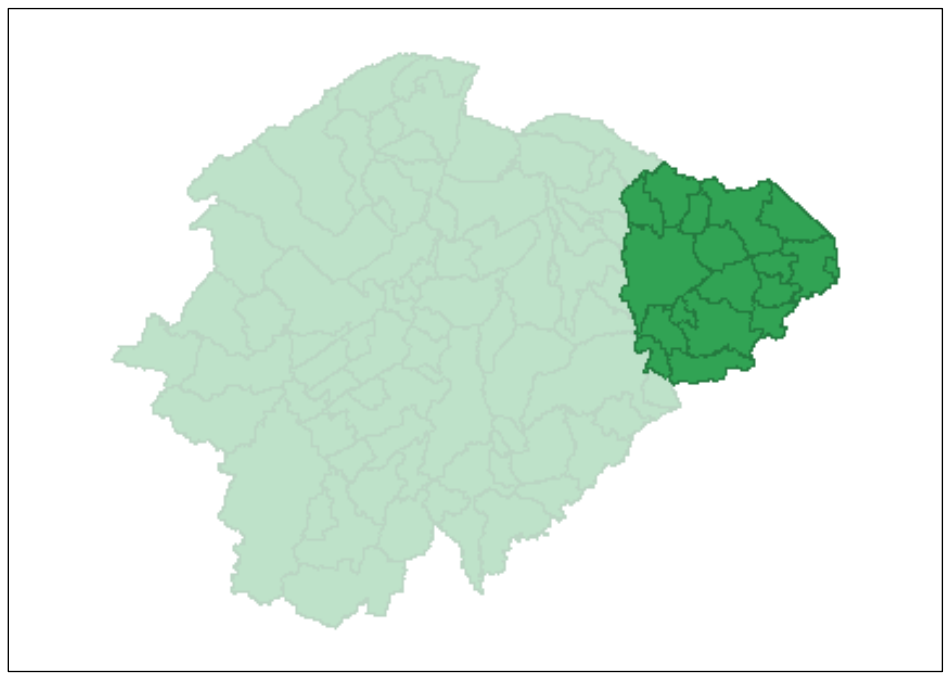

| | | | |
|---|---|---|---|
| **Microrregión de Salinas** | Águas Vermelhas<br>Berizal<br>Curral de Dentro<br>Divisa Alegre<br>Fruta de Leite<br>Indaiabira | Ninheira<br>Novorizonte<br>Rio Pardo de Minas<br>Rubelita<br>**Salinas ***<br>Santa Cruz de Salinas | Santo Antônio do Retiro<br>São João do Paraíso<br>Taiobeiras<br>Vargem Grande do Rio Pardo<br>Montezuma |

Tabla 36

*Matriz de Regresión Regional de Salinas*

| **Región** | **Relación Variables** | **2009** $R^2$ | $p$ | **2018** $R^2$ | $p$ |
|---|---|---|---|---|---|
| Salinas | EIUP----ET | .538 | .034 | .712 | .000 |
| n=18 | EIUP----ECBO | .522 | .002 | .916 | .000 |
| 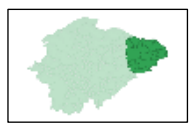 | ED------ECBO | .318 | .292 | .772 | .000 |
| | EFU----ET | .350 | .163 | .633 | .983 |
| | ET------EPIB | .835 | .000 | .857 | .000 |
| | ECBO—EPIB | .658 | .728 | .706 | .000 |

En los 18 Municipios de la Microrregión de Salinas los resultados encontrados llevan a confirma la relación predictiva y validación de hipótesis entre el conocimiento especializado CE y la actividad económica digital AED, donde se halló un valor predictivo alto entre los indicadores EIUP - ECBO de $R^2$ = .916, valor $p$ = .000 (2018). Para los indicadores EIUP - ET un valor predictivo sustancial de $R^2$ = .712, valor $p$ = .000 (2018). Con relación a la actividad económica digital AED y el crecimiento económico regional CER, se halló un valor predictivo alto entre los indicadores ET - EPIB de $R^2$ = .857, valor $p$ = .000 (2018). Para los indicadores ECBO - EPIB un valor predictivo sustancial de $R^2$ = .706, valor $p$ = .000 (2018).

#### 4.6.4.2 Análisis comparativo regresión regional microrregión Pirapora

Mapa 18

*Microrregión de Pirapora (elaboración propia)*

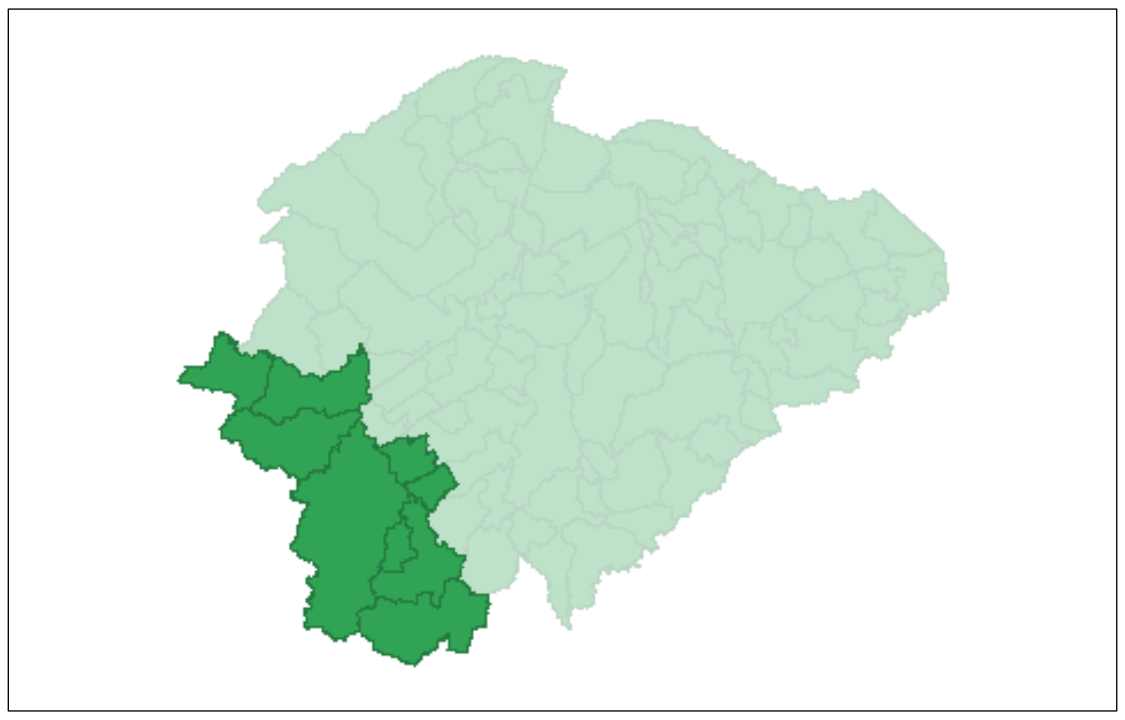

| | | | |
|---|---|---|---|
| **Microregion de Pirapora** | Buritizeiro<br>Ibiaí<br>Jequitaí<br>Lagoa dos Patos | Lassance<br>**Pirapora ***<br>Riachinho | Santa Fé de Minas<br>São Romão<br>Várzea da Palma |

Tabla 37

*Matriz de Regresión Regional de Pirapora*

| **Región** | **Relación Variables** | **2009** $R^2$ | $p$ | **2018** $R^2$ | $p$ |
|---|---|---|---|---|---|
| Pirapora | EIUP----ET | .580 | .017 | .737 | .208 |
| n=9 | EIUP----ECBO | .989 | .000 | .993 | .000 |
| 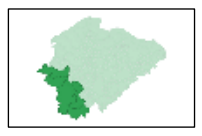 | ED------ECBO | .619 | .0.14 | .995 | .000 |
| | EFU----ET | .902 | .810 | .973 | .044 |
| | ET------EPIB | .972 | .013 | .979 | .004 |
| | ECBO—EPIB | .789 | .568 | .892 | .091 |

En los 9 Municipios de la Microrregión de Pirapora los resultados encontrados llevan a confirmar la relación predictiva y validación de hipótesis entre el conocimiento especializado CE y la actividad económica digital AED, donde se halló un valor predictivo alto entre los indicadores EIUP - ECBO de $R^2$ = .993, valor $p$ = .000 (2018); $R^2$ = .989, valor $p$ = .000 (2009). Para los indicadores EIUP - ET un valor predictivo moderado de $R^2$ = .580, valor $p$ = .017 (2009). En relación con los indicadores ED - ECBO se encontró un valor predictivo alto de $R^2$ = .995, valor $p$ = .000 (2018), de la misma forma para los indicadores EFU - ET un valor predictivo alto de $R^2$ = .973, valor p = .044 (2018). Con relación a la actividad económica digital AED y el crecimiento económico regional CER, se halló un valor predictivo alto entre los indicadores ET - EPIB de $R^2$ = .979, valor $p$ = .004 (2018) y $R^2$ = .972, valor $p$ = .013 (2009). Para los indicadores ECBO - EPIB un valor predictivo alto de $R^2$ = .892, valor $p$ = .091 (2018).

#### 4.6.4.3 Análisis comparativo regresión regional microrregión Bocaiúva

Mapa 19

*Microrregión de Bocaiúva (elaboración propia)*

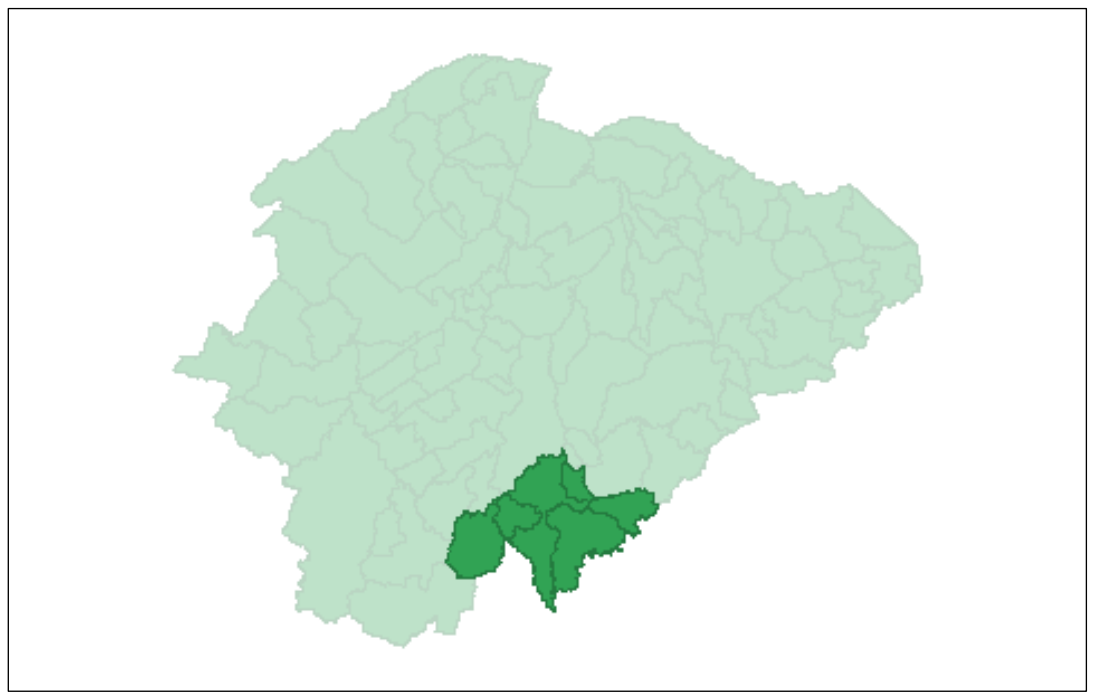

| | | | |
|---|---|---|---|
| **Microregion de Bocaiúva** | **85 Bocaiúva *** | Engenheiro Navarro<br>Francisco Dumont | Guaraciama<br>Olhos de Água |

Tabla 38

*Matriz de Regresión Regional de Bocaiúva*

| **Región** | **Relación Variables** | **2009** $R^2$ | $p$ | **2018** $R^2$ | $p$ |
|---|---|---|---|---|---|
| Bocaiúva | EIUP----ET | 0 | 0 | .978 | .001 |
| n=5 | EIUP----ECBO | 0 | 0 | .940 | .006 |
| 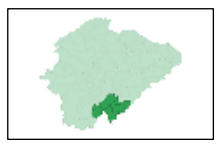 | ED------ECBO | .889 | .023 | .940 | .014 |
| | EFU----ET | .998 | .304 | .972 | .042 |
| | ET------EPIB | .997 | .006 | .989 | .134 |
| | ECBO—EPIB | .899 | .038 | .997 | .065 |

En los 5 Municipios de la Microrregión de Bocaiúva los resultados encontrados llevan a confirmar la relación predictiva y validación de hipótesis entre el conocimiento especializado CE y la actividad económica digital AED, donde se halló un valor predictivo alto entre los indicadores EIUP - ECBO de $R^2$ = .940, valor $p$ = .006 (2018). Para los indicadores EIUP - ET un valor predictivo alto de $R^2$ = .978, valor $p$ =.001 (2018). En relación con los indicadores ED - ECBO se encontró un valor predictivo alto de $R^2$ = .940, valor $p$ = .014 (2018) y $R^2$ = .889, valor p = .023 (2009), de la misma forma para los indicadores EFU - ET un valor predictivo alto de $R^2$ = .972, valor $p$ = .042 (2018). Con relación a la actividad económica digital AED y el crecimiento económico regional CER, se halló un valor predictivo alto entre los indicadores ET - EPIB de $R^2$ = .997, valor $p$ = .006 (2009). Para los indicadores ECBO - EPIB un valor predictivo alto de $R^2$ = .997, valor $p$ = .065 (2018) y un valor predictivo alto de $R^2$ = .899, valor $p$ = .038 (2009).

#### 4.6.4.4 Análisis comparativo regresión regional microrregión Grao Mogol

Mapa 1

*Microrregión de Grao Mogol (elaboración propia)*

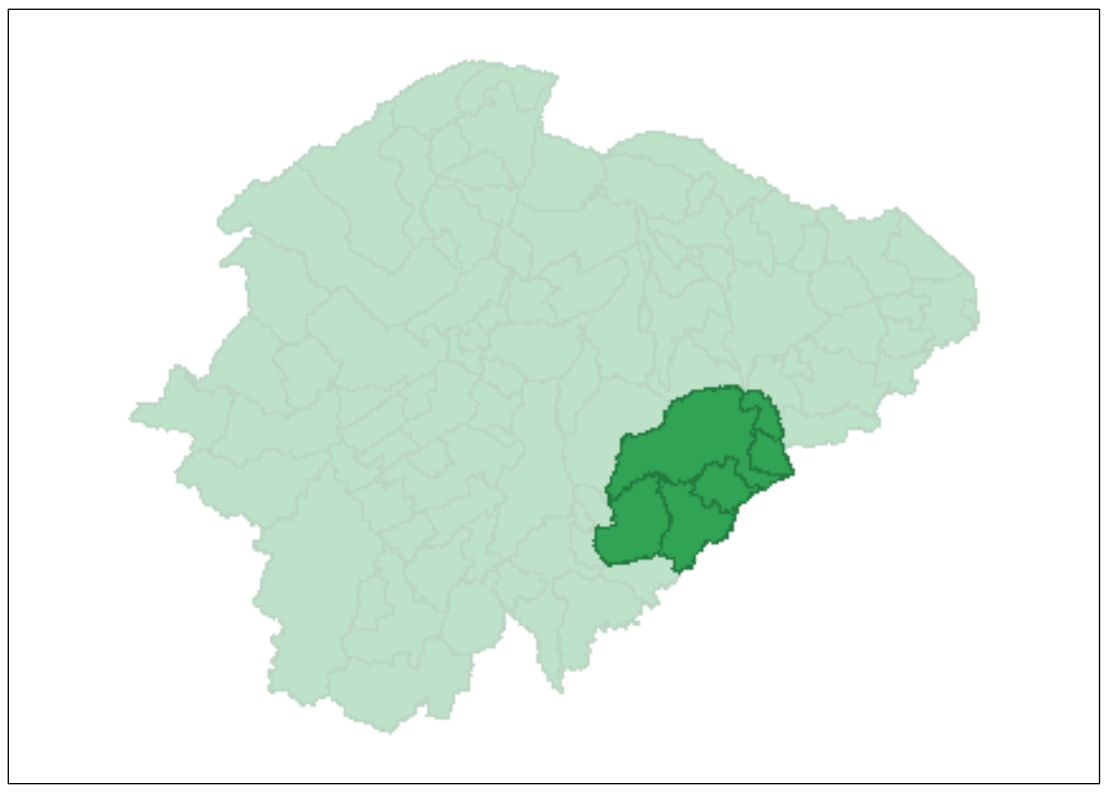

| | | | |
|---|---|---|---|
| **Microregion de Grão-Mogol** | Botumirim<br>Cristália | **Grão-Mogol *** <br>Itacambira | Josenópolis<br>Padre Carvalho |

Tabla 39

*Matriz de Regresión Regional de Grao Mogol*

| **Región** | **Relación** | **2009** | | **2018** | |
|---|---|---|---|---|---|
| | **Variables** | $R^2$ | $p$ | $R^2$ | $p$ |
| Grao Mogol | EIUP----ET | 0 | 0 | 0 | 0 |
| n=6 | EIUP----ECBO | 0 | 0 | 0 | 0 |
| 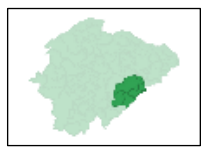 | ED------ECBO | 0 | 0 | 0 | 0 |
| | EFU----ET | .921 | .025 | .645 | .002 |
| | ET------EPIB | .986 | .001 | .108 | .501 |
| | ECBO—EPIB | .770 | .093 | .065 | .506 |

En los 6 Municipios de la Microrregión de Grao Mogol los resultados encontrados nos llevan a confirmar que en esta región no existió ninguna relación predictiva entre el conocimiento especializado CE y la actividad económica digital AED, con la única excepción de valores predictivos altos entre los indicadores EFU - ET con un $R^2 = .921$, valor $p = .025$ (2009) y un valor predictivo moderado de $R^2 = .645$, valor $p = .002$ (2018). Con relación a la actividad económica digital AED y el crecimiento económico regional CER, se halló un valor predictivo alto entre los indicadores ET - EPIB de $R^2 = .986$, valor $p = .001$ (2009). Para los indicadores ECBO - EPIB se halló un valor predictivo sustancial de $R^2 = .770$, valor $p = .093$ (2009).

### 4.6.4.5 Análisis comparativo regresión regional microrregión Januária

Mapa 21 *Microrregión de Januária (elaboración propia)*

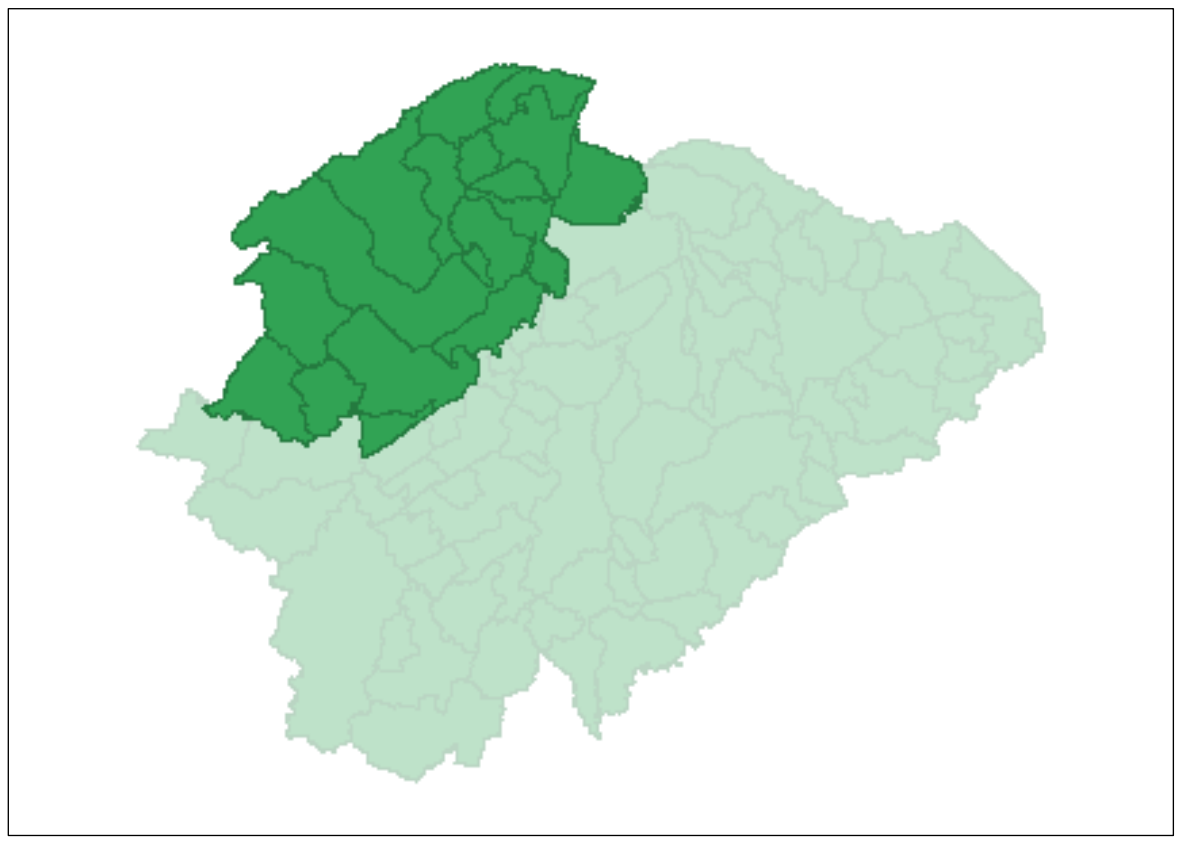

| | | | |
|---|---|---|---|
| **Microregión de Januária** | Bonito de Minas<br>Chapada Gaúcha<br>Cônego Marinho<br>Icaraí de Minas<br>Itacarambi<br>**Januária *** | São João das Missões<br>Urucuia<br>Juvenília<br>Manga<br>Matias Cardoso<br>Miravânia | Montalvânia<br>Pedras de Maria da Cruz<br>Pintópolis<br>São Francisco |

Tabla 40

*Matriz de Regresión Regional de Januária*

| **Región** | **Relación Variables** | **2009 $R^2$** | ***p*** | **2018 $R^2$** | ***p*** |
|---|---|---|---|---|---|
| Januária | EIUP----ET | .039 | .089 | .198 | .937 |
| n=16 | EIUP----ECBO | .014 | .005 | .380 | .379 |
| 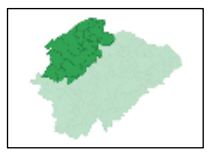 | ED------ECBO | .901 | .000 | .981 | .000 |
| | EFU----ET | .402 | .005 | .260 | .000 |
| | ET------EPIB | .525 | .342 | .498 | .615 |
| | ECBO—EPIB | .785 | .001 | .662 | .012 |

En los 16 Municipios de la Microrregión de Januária los resultados encontrados llevan a confirmar que en esta región no existió ninguna relación predictiva entre el conocimiento especializado CE y la actividad económica digital AED, con la única excepción de valores predictivos altos entre los indicadores ED - ECBO con un $R^2$ = .901, valor $p$ = .000 (2009); valor predictivo alto de $R^2$ = .981, valor $p$ = .000 (2018) y valores predictivos bajos entre los indicadores EFU - ET con un $R^2$ = .402, valor $p$ = .005 (2009). Con relación a la actividad económica digital AED y el crecimiento económico regional CER, se halló un valor predictivo sustancial entre los indicadores ECBO - EPIB con un $R^2$ = .785, valor $p$ = .001 (2009) y un valor predictivo moderado de $R^2$ = .662, valor $p$ = .012 (2018).

#### 4.6.4.6 Análisis comparativo regresión regional microrregión Janauba

Mapa 2

Microrregión de Janauba (elaboración propia)

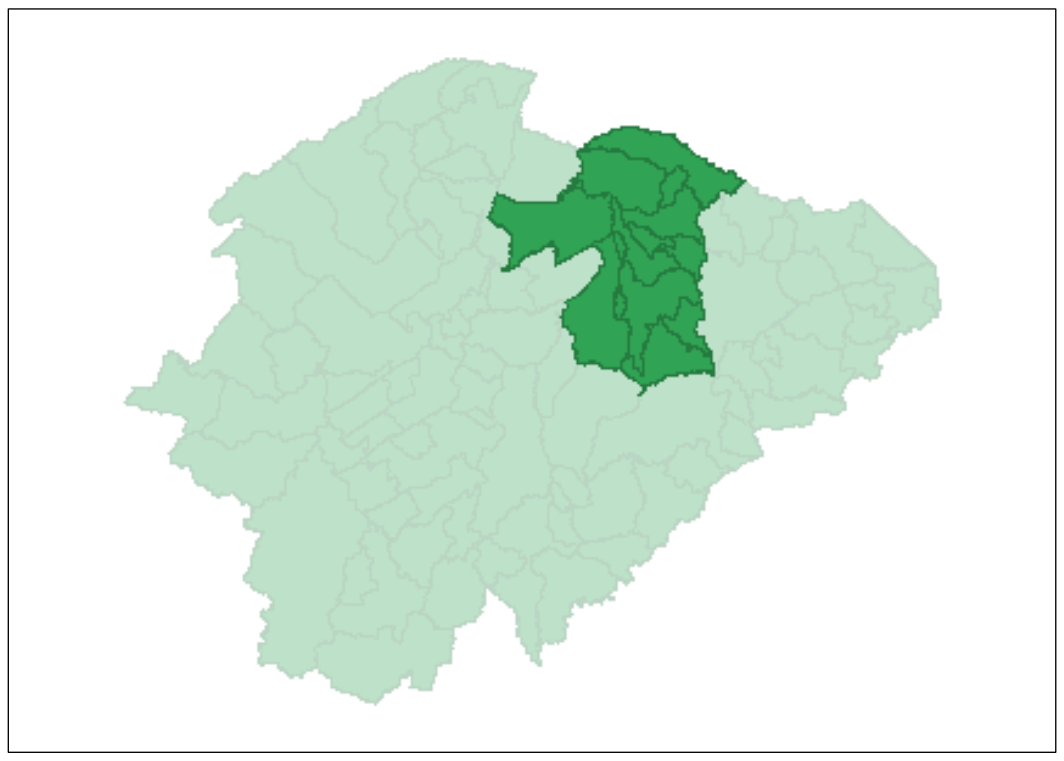

| | | | |
|---|---|---|---|
| **Microregion de Janaúba** | Catuti<br>Espinosa<br>Gameleiras<br>Jaíba<br>**Janaúba *** | Mamonas<br>Mato Verde<br>Monte Azul<br>Nova Porteirinha<br>Pai Pedro | Porteirinha<br>Riacho dos Machados<br>Serranópolis de Minas |

Tabla 41

*Matriz de Regresión Regional de Janauba*

| **Región** | **Relación Variables** | **2009** $R^2$ | $p$ | **2018** $R^2$ | $p$ |
|---|---|---|---|---|---|
| Janauba | EIUP----ET | .065 | .797 | .003 | .931 |
| n=13 | EIUP----ECBO | .029 | .009 | .136 | .007 |
| 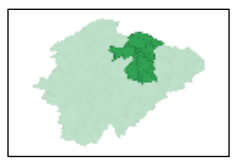 | ED------ECBO | .016 | .022 | .045 | .000 |
| | EFU----ET | .781 | .499 | .921 | .489 |
| | ET------EPIB | .902 | .008 | .913 | .681 |
| | ECBO—EPIB | .791 | .009 | .650 | .133 |

En los 13 Municipios de la Microrregión de Janauba los resultados encontrados llevan a confirmar que en esta región no existió ninguna relación predictiva entre el conocimiento especializado CE y la actividad económica digital AED. Con relación a la actividad económica digital AED y el crecimiento económico regional CER, se halló un valor predictivo alto entre los indicadores ET - EPIB de $R^2$ = .902, valor $p$ = .008 (2009). Para los indicadores ECBO - EPIB un valor predictivo sustancial de $R^2$ = .791, valor p =.009 (2009).

### 4.6.4.7 Análisis comparativo regresión regional microrregión Montes Claros

Mapa 3

*Microrregión de Montes Claros (elaboración propia)*

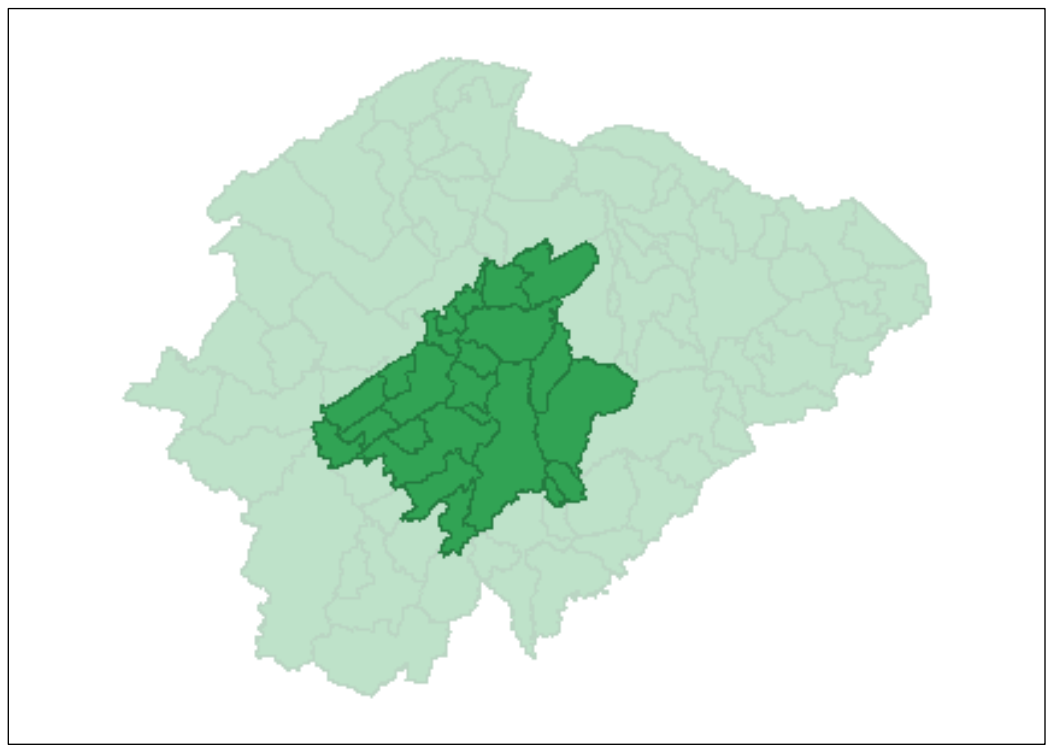

| | | | |
|---|---|---|---|
| **Microregión de Montes Claros** | Brasília de Minas | Ibiracatu | Patis |
| | Campo Azul | Japonvar | Ponto Chique |
| | Capitão Enéias | Juramento | São João da Lagoa |
| | Claro dos Poções | Lontra | São João da Ponte |
| | Coração de Jesus | Luislândia | São João do Pacuí |
| | Francisco Sá | Mirabela | Ubaí |
| | Glaucilândia | **Montes Claros *** | Varzelândia |
| | | | Verdelândia |

Tabla 42

*Matriz de Regresión Regional de Montes Claros*

| **Región** | **Relación Variables** | **2009** $R^2$ | $p$ | **2018** $R^2$ | $p$ |
|---|---|---|---|---|---|
| M. Claros | EIUP----ET | .994 | .154 | .995 | .854 |
| n=22 | EIUP----ECBO | 1.000 | .000 | 1.000 | .000 |
| 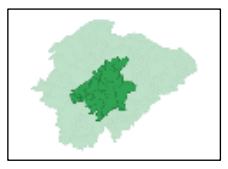 | ED------ECBO | 1.000 | .000 | 1.000 | .000 |
| | EFU----ET | .954 | .688 | .993 | .083 |
| | ET------EPIB | .998 | .021 | .999 | .010 |
| | ECBO—EPIB | .997 | .012 | .998 | .035 |

En los 22 Municipios de la Microrregión de Montes Claros los resultados encontrados llevan a confirmar la relación predictiva y validación de hipótesis entre el conocimiento especializado CE y la actividad económica digital AED, donde se halló un valor predictivo alto entre los indicadores EIUP - ECBO de $R^2 = 1.000$, valor $p = .000$ (2018) y $R^2 = 1.000$, valor $p = .000$ (2009). En relación con los indicadores ED - ECBO se encontró un valor predictivo alto de $R^2 = 1.000$, valor $p = .000$ (2018) y $R^2 = 1.000$, valor $p = .000$ (2009). De la misma forma para los indicadores EFU - ET un valor predictivo alto de $R^2 = .993$, valor $p = .083$ (2018). Con relación a la actividad económica digital AED y el crecimiento económico regional CER, se halló un valor predictivo alto entre los indicadores ET - EPIB de $R^2 = .999$, valor $p = .010$ (2018). Para los indicadores ECBO - EPIB un valor predictivo alto de $R^2 = .998$, valor $p = .038$ (2018) y un valor predictivo alto de $R^2 = .997$, valor $p = .012$ (2009).

### 4.6.5 Prueba de correlación espacial de los municipios de la mesoregión norte de Minas Gerais

La prueba de correlación espacial permitió mediante el mapa de significancia identificar la manera en que un fenómeno se transformó y afectó a una segunda variable a través del marco geográfico (Pellegrini & Platino, 2013). Este método sigue la función de la ubicación del valor mediante I de Moran (Celemín, 2009), buscando comprender la distribución y el comportamiento del fenómeno estudiado, para de esta manera determinar las áreas locales de impacto significativo sobre las tendencias generales de otras unidades espacialmente próximas.

#### 4.6.5.1 Correlación espacial y significancia estadística entre actividad económica digital y crecimiento económico regional (9999 permutaciones[2])

##### 4.6.5.1.1 Impacto de las empresas registradas en el rubro de informática, tecnología y comunicación

Mapa 24

*Análisis de correlación espacial local entre ET y EPIB 2009*

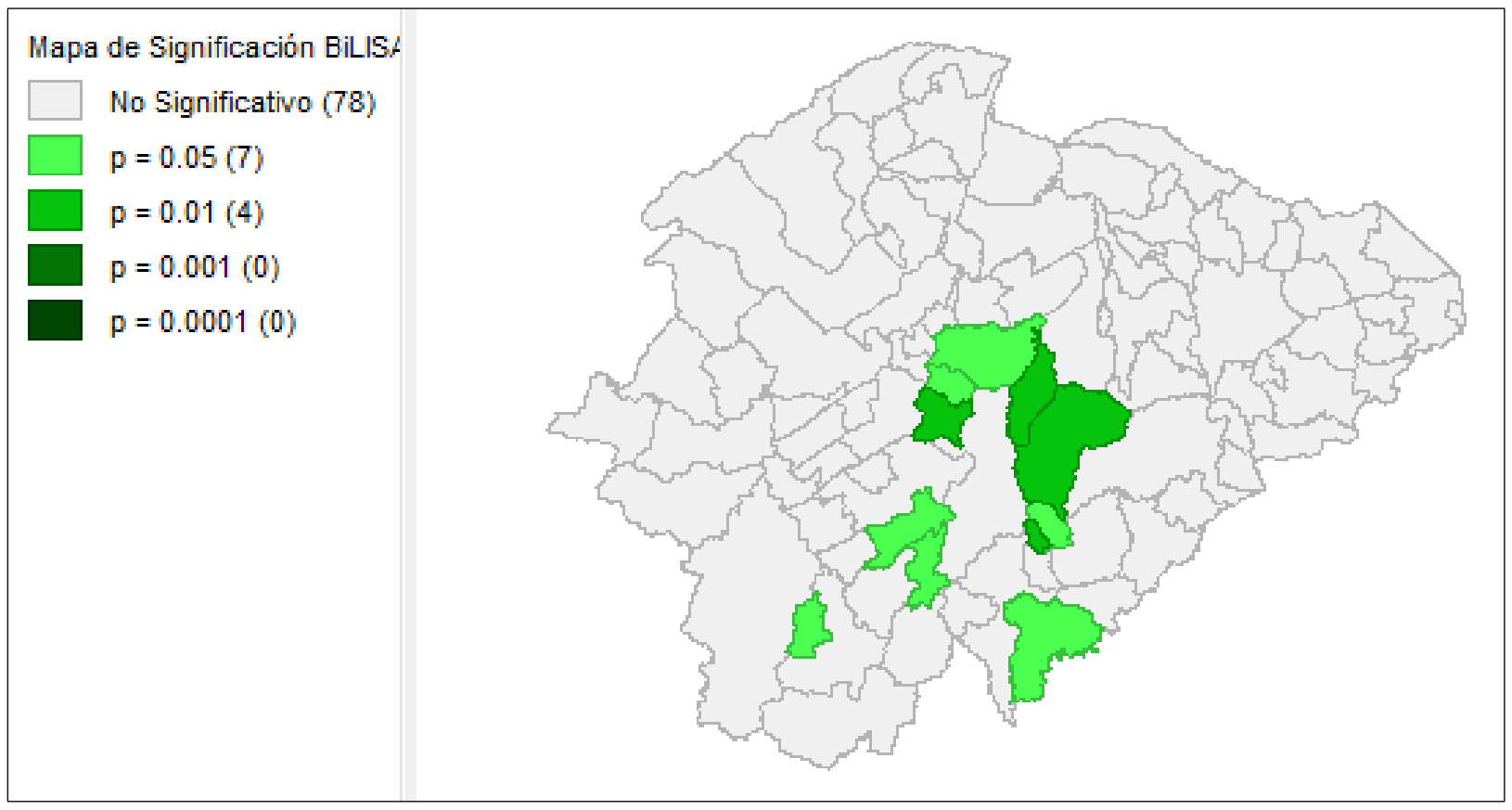

Para el 2009, las áreas de impacto significativo con un valor $p < .01$ que ejercieron las empresas con el rubro de informática, tecnología y comunicaciones en relación con el producto interno bruto municipal se focalizó en los Municipios de Mirabela (MOC[3]) con un valor $p = .009$, Capitán Eneas (MOC) con un valor $p = .001$, Francisco Sa (MOC) con un valor $p = .007$, y Glaucilandia (MOC) con un valor $p = .009$. Confirmando de esta manera la evidencia empírica del estudio, ya que los Municipios marcados con la abreviatura (MOC) pertenecen a la región de Montes Claros el cual alcanzo un valor predictivo alto entre los indicadores ET - EPIB de $R^2 = .998$, valor $p = .021$. Lo cual confirma y valida la hipótesis H3 estableciendo que la actividad económica digital impacta significativamente el crecimiento económico regional.

---

[2] La *prueba de permutación* se basa en la idea de que si existe [*n*] observaciones sobre una región en particular, [*n*] permutaciones de los datos son posibles (Celemín, 2009)

[3] MOC representa a la región de Montes Claros.

De la misma forma, se logró identificar las áreas de impacto significativo con un valor $p < .05$ que ejercieron las empresas de tecnología en relación con el producto interno bruto municipal, la misma se focalizó en los Municipios de Sao Joao de Lagoa (MOC) con un valor $p = .03$, Claro dos Pecoes (MOC) con un valor $p = .014$, Patis (MOC) con un valor $p = .012$, Sao Joao da Ponte (MOC) con un valor $p = .026$, Olhos de Agua (MOC) con un valor $p = .036$ y Juramento (BC[4]) con un valor $p = .029$, las mismas están rodeados de regiones con alta influencia como es el Municipio de Montes Claros[5].

Confirmando de esta manera la evidencia empírica del estudio, ya que los Municipios marcados con la abreviatura (MOC) pertenecen a la región de Montes Claros el cual alcanzo un valor predictivo alto entre los indicadores ET - EPIB de $R^2 = .998$, valor $p = .021$; de la misma manera el Municipio marcado con la abreviatura (BC) perteneciente a la región de Bocaiúva el cual alcanzo un valor predictivo alto entre los indicadores ET - EPIB de $R^2 = .997$, valor $p = .006$. Lo cual confirma y valida la hipótesis H3 estableciendo que la actividad económica digital impacta significativamente el crecimiento económico regional.

---

[4] BC representa a la región de Bocaiúva.

[5] El Municipio de *Montes Claros* es el centro económico y académico de gran impacto en la Región Norte del Estado de Minas Gerais.

Mapa 25

*Análisis de correlación espacial local entre ET y EPIB 2017 (elaboración propia)*

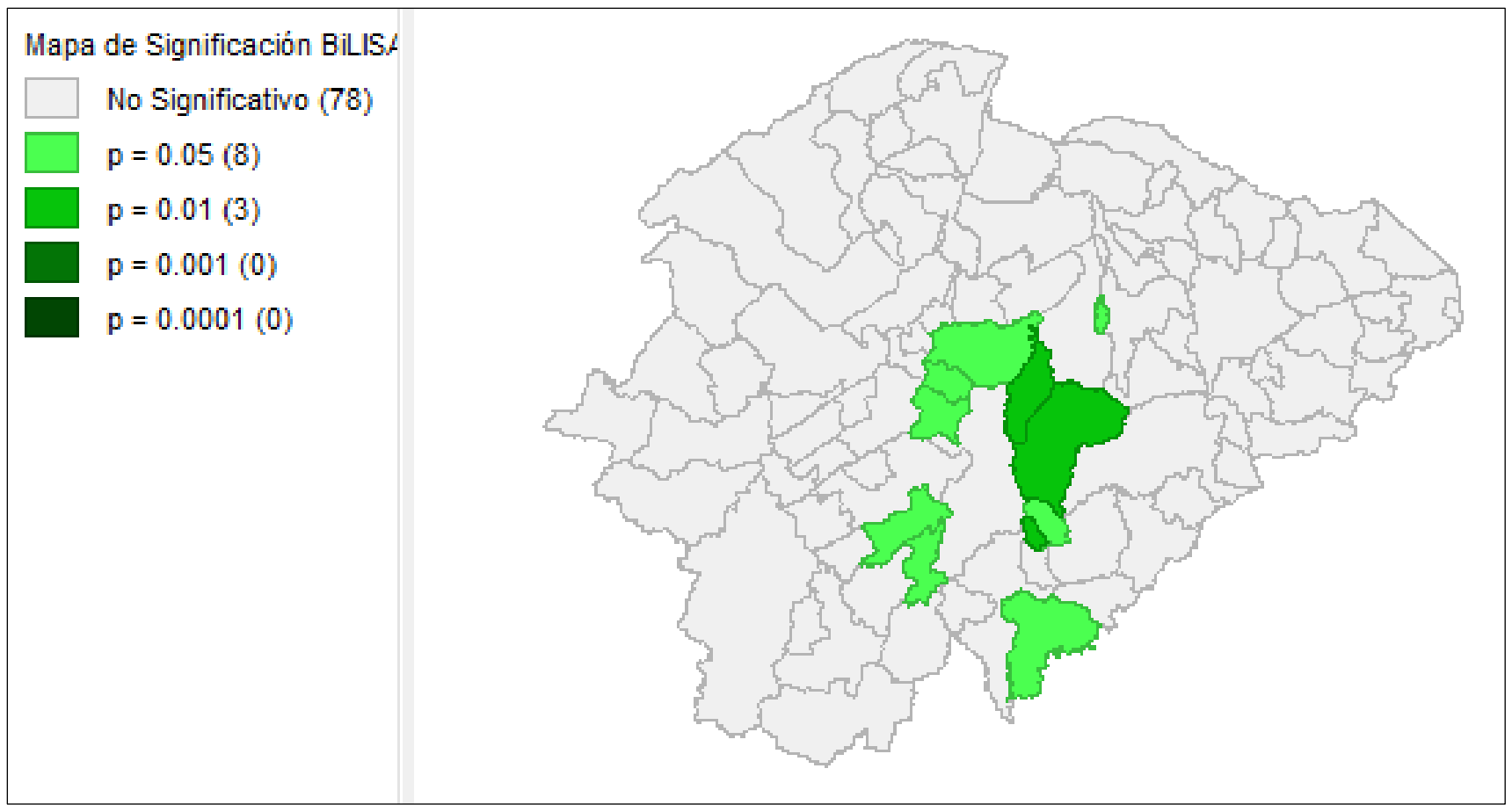

Para el 2017, las áreas de impacto significativo con un valor $p < .01$ que ejercieron las empresas con el rubro de informática, tecnología y comunicaciones en relación con el producto interno bruto municipal se focalizó en los Municipios de Francisco Sa (MC) con un valor $p = .006$, Capitán Eneas (MC) con un valor $p = .001$, y Glaucilandia (MC) con un valor $p = .009$. Confirmando de esta manera la evidencia empírica del estudio, ya que los Municipios marcados con la abreviatura (MC) pertenecen a la región de Montes Claros el cual alcanzó un valor predictivo alto entre los indicadores ET - EPIB de $R^2 = .999$, valor $p = .010$. Lo cual confirma y valida la hipótesis H3 estableciendo que la actividad económica digital impacta significativamente el crecimiento económico regional.

De la misma forma se logró identificar las áreas de impacto significativo con un valor $p < .05$ que ejercieron las empresas de tecnología en relación con el producto interno bruto municipal se focalizó en los Municipios de Sao Joao de Lagoa (MC) con un valor $p = .048$, Claro dos Pecoes (MC) con un valor $p = .012$, Mirabela (MC) con un valor $p = .01$, Patis (MC) con un valor $p = .014$, Sao Joao da Ponte (MC) con un valor $p = .023$, Olhos de Agua (MC) con un valor $p = .033$, Juramento (BC) con un valor $p = .037$, Nova Porterinha (JN[6]) con un valor $p = .048$. Las mismas están rodeados de regiones con alta influencia contigua como es el Municipio de Montes Claros.

Confirmando de esta manera la evidencia empírica del estudio, ya que los Municipios marcados con la abreviatura (MC) pertenecen a la región de Montes Claros el cual alcanzo un valor predictivo alto entre los indicadores ET - EPIB de $R^2 = .999$, valor $p = .010$. Lo cual confirma y valida la hipótesis H3 estableciendo que la actividad económica digital impacta significativamente el crecimiento económico regional. En los casos de los Municipios de Bocaiúva con un valor predictivo alto entre los indicadores ET - EPIB de $R^2 = .989$ y el Municipio de Janauba ET - EPIB de $R^2 = .913$; no se pudo validar la hipótesis ya que, aunque tiene un valor alto predictivo entre los indicadores, su valor de significancia sobrepasa el lumbral de $p < 0.1$.

[6] NP representa a la región de Janauba.

#### 4.6.5.1.2 Impacto de las ocupaciones relacionadas el rubro de informática, tecnología y comunicación

Mapa 26

*Análisis de correlación espacial local entre ECBO y EPIB 2009 (elaboración propia)*

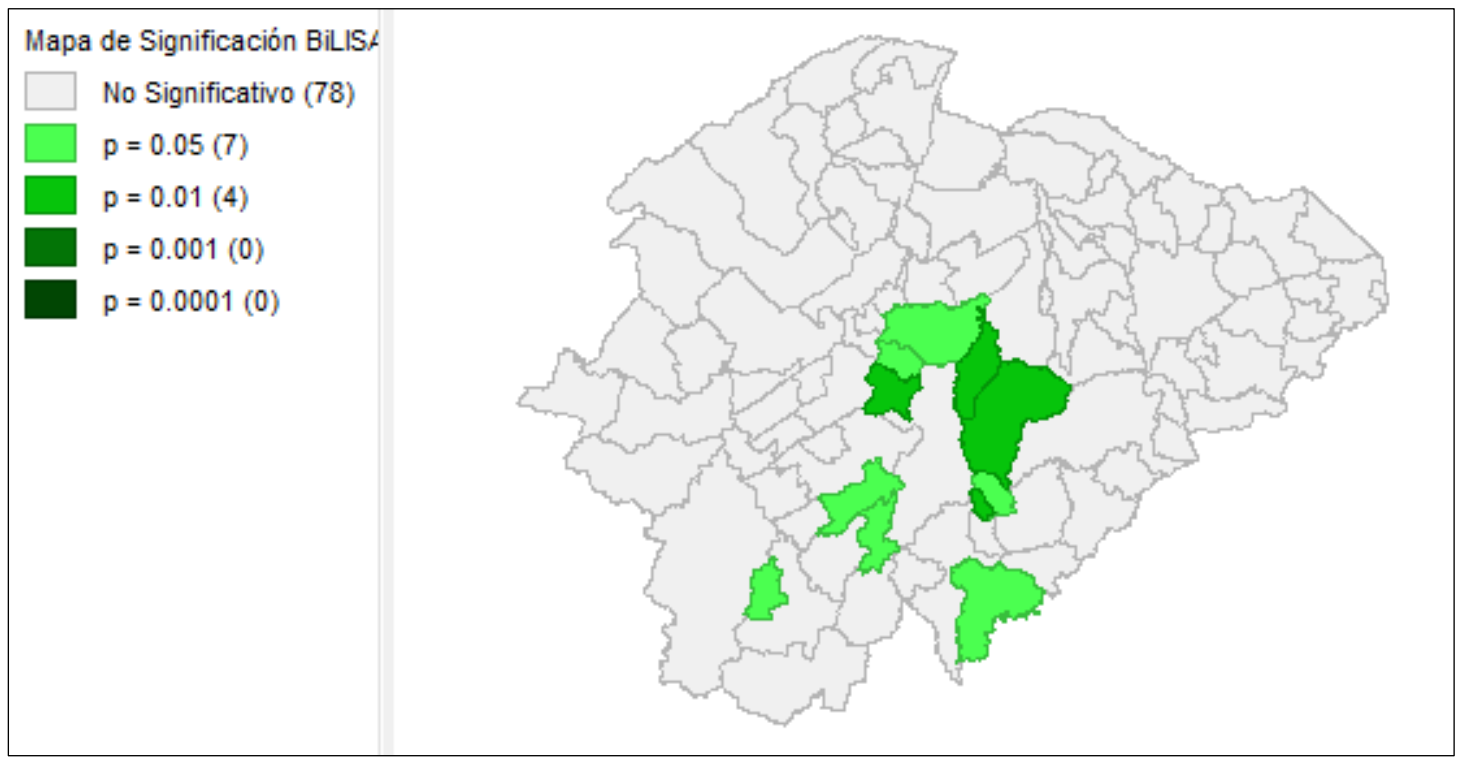

Para el 2009, las áreas de impacto significativo con un valor $p < .01$ que ejercieron las ocupaciones relacionadas al rubro de informática, tecnología y comunicaciones, en relación con el producto interno bruto municipal se focalizó en el Municipio de Mirabela (MC) con un valor $p = .009$, Capitán Eneas (MC) con un valor $p = .001$, Francisco Sa (MC) con un valor $p = .007$, y Glaucilandia (MC) con un valor $p = .009$. Confirmando de esta manera la evidencia empírica del estudio, ya que los Municipios marcados con la abreviatura (MC) pertenecen a la región de Montes Claros el cual alcanzo un valor predictivo alto entre los indicadores ECBO - EPIB de $R^2 = .997$, valor $p = .012$. Lo cual confirma y valida la hipótesis H3 estableciendo que la actividad económica digital impacta significativamente el crecimiento económico regional.

De la misma forma, se logró identificar las áreas de impacto significativo con un valor $p < .05$ que ejercieron las ocupaciones relacionadas al rubro de informática, tecnología y comunicaciones en relación con el producto interno bruto municipal en los

Municipios de Sao Joao de Lagoa (MC) con un valor *p* = .039, Claro dos Pecoes (MC) con un valor *p* = .014, Patis (MC) con un valor *p* = .012, Sao Joao da Ponte (MC) con un valor *p* = .026, Olhos de Agua (MC) con un valor *p* = .036, y Juramento (BC) con un valor *p* = .029. Las mismas están rodeados de regiones con alta influencia contigua como es el Municipio de Montes Claros.

Confirmando de esta manera la evidencia empírica del estudio, ya que los Municipios marcados con la abreviatura (MC) pertenecen a la región de Montes Claros el cual alcanzó un valor predictivo alto entre los indicadores ECBO - EPIB de $R^2$ = .997, valor *p* = .012; de la misma manera el Municipio marcado con la abreviatura (BC) perteneciente a la región de Bocaiúva el cual alcanzó un valor predictivo alto entre los indicadores ECBO - EPIB de $R^2$ = .899, valor *p* = .038. Lo cual confirma y valida la hipótesis H3 estableciendo que la actividad económica digital impacta significativamente el crecimiento económico regional.

Mapa 27

*Análisis de correlación espacial local entre ECBO y EPIB 2017 (elaboración propia)*

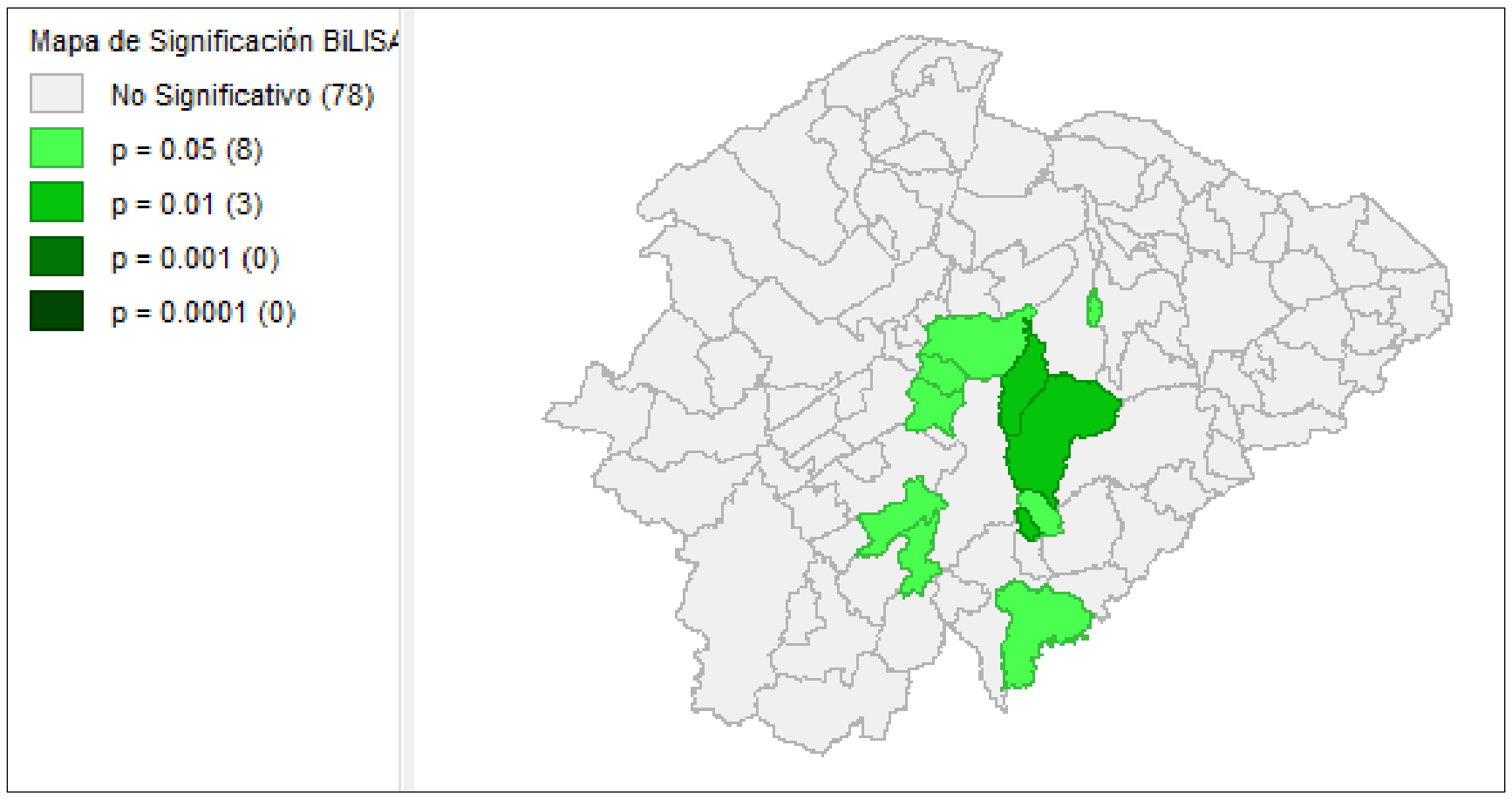

Para el 2017, las áreas de impacto significativo con un valor $p < .01$ que ejercieron las ocupaciones relacionadas al rubro de informática, tecnología y comunicaciones, en relación con el producto interno bruto municipal se focalizó en los Municipios de Capitán Eneas (MC) con un valor $p = .001$, Francisco Sa (MC) con un valor $p = .006$, y Glaucilandia (MC) con un valor $p = .009$. Confirmando de esta manera la evidencia empírica del estudio, ya que los Municipios marcados con la abreviatura (MC) pertenecen a la región de Montes Claros el cual alcanzó un valor predictivo alto entre los indicadores ECBO - EPIB de $R^2 = .998$, valor $p = .035$. lo cual confirma y valida la hipótesis H3 estableciendo que la actividad económica digital impacta significativamente el crecimiento económico regional.

De la misma forma se logró identificar las áreas de impacto significativo con un valor $p < .05$ que ejercieron las ocupaciones relacionadas al rubro de informática, tecnología y comunicaciones en relación con el producto interno bruto municipal, las mismas estuvieron focalizadas en los Municipios de Sao Joao de Lagoa (MC) con un valor $p = .04$, Claro dos Pecoes (MC) con un valor $p = .01$, Mirabela (MC) con un valor $p = .01$, Patis (MC) con un valor $p = .01$, Sao Joao da Ponte (MC) con un valor $p = .02$, Olhos de Agua (MC) con un valor $p = .03$, Juramento (BC) con un valor $p = .037$ y Nova Porterinha (JN) con un valor $p = .048$, las mismas están rodeados de regiones con alta influencia contigua como es el Municipio de Montes Claros.

Confirmando de esta manera la evidencia empírica del estudio, ya que los Municipios marcados con la abreviatura (MC) pertenecen a la región de Montes Claros el cual alcanzo un valor predictivo alto entre los indicadores ECBO - EPIB de $R^2 = .998$, valor $p = .035$; de la misma manera el Municipio de Bocaiúva con un valor predictivo alto entre los indicadores ECBO - EPIB de $R^2$ = .997, valor $p$ = .065. Lo cual confirma y valida la hipótesis H3 estableciendo que la actividad económica digital impacta significativamente el crecimiento económico regional. En el caso del Municipio de Nova Porterinha que pertenece a la región de Janauba no se pudo validar la hipótesis ya que, aunque tiene un valor alto predictivo entre los indicadores, su valor de significancia sobrepasa el lumbral de $p < 0.1$.

### 4.6.6 Prueba de autocorrelación espacial de los municipios de la mesoregión norte de Minas Gerais

La autocorrelación espacial sirve para analizar la variabilidad y el efecto contagio de un fenómeno a través del espacio geográfico, buscando identificar patrones significativos en las unidades espaciales (Siabato & Guzmán, 2019). Estos patrones nos permitieron identificar el grado de concentración o dispersión mediante ubicaciones espaciales significativas. Esto significa que una variable estará espacialmente autocorrelacionada cuando los valores observados en un municipio dependan de los valores observados en municipios vecinos (Mexiquense et al., 2005) permitiéndonos observar un proceso de multidireccionalidad de las relaciones basado en dirección-sentido y complejidad entre el conjunto de unidades geográficas de análisis.

De acuerdo a Siabato & Guzmán (2019) la autocorrelación espacial mide el grado en el que una variable está correlacionada con ella misma en zonas diferentes dentro del área de estudio, por lo tanto, cuando los resultados son positivos significa que los valores similares ocurren en unidades geográficas cercanas favoreciendo con su comportamiento a regiones próximas, lo cual permite evidenciar la existencia de clústeres (Sabogal, 2013); por otro lado, el índice de Moran negativo establece que los valores altos de un municipio se relacionan con valores bajos de los vecinos, donde el comportamiento de una variable está produciendo efectos contrarios en sus vecinos (Moreno y Vayá, 2000).

#### 4.6.6.1 Relación de autocorrelación espacial entre actividad económica digital y crecimiento económico regional (9999 permutaciones)

##### 4.6.6.1.1 Impacto de las empresas registradas en el rubro de informática, tecnología y comunicación

Mapa 28

*Análisis de autocorrelación espacial local de TIC para 2009 (elaboración propia)*

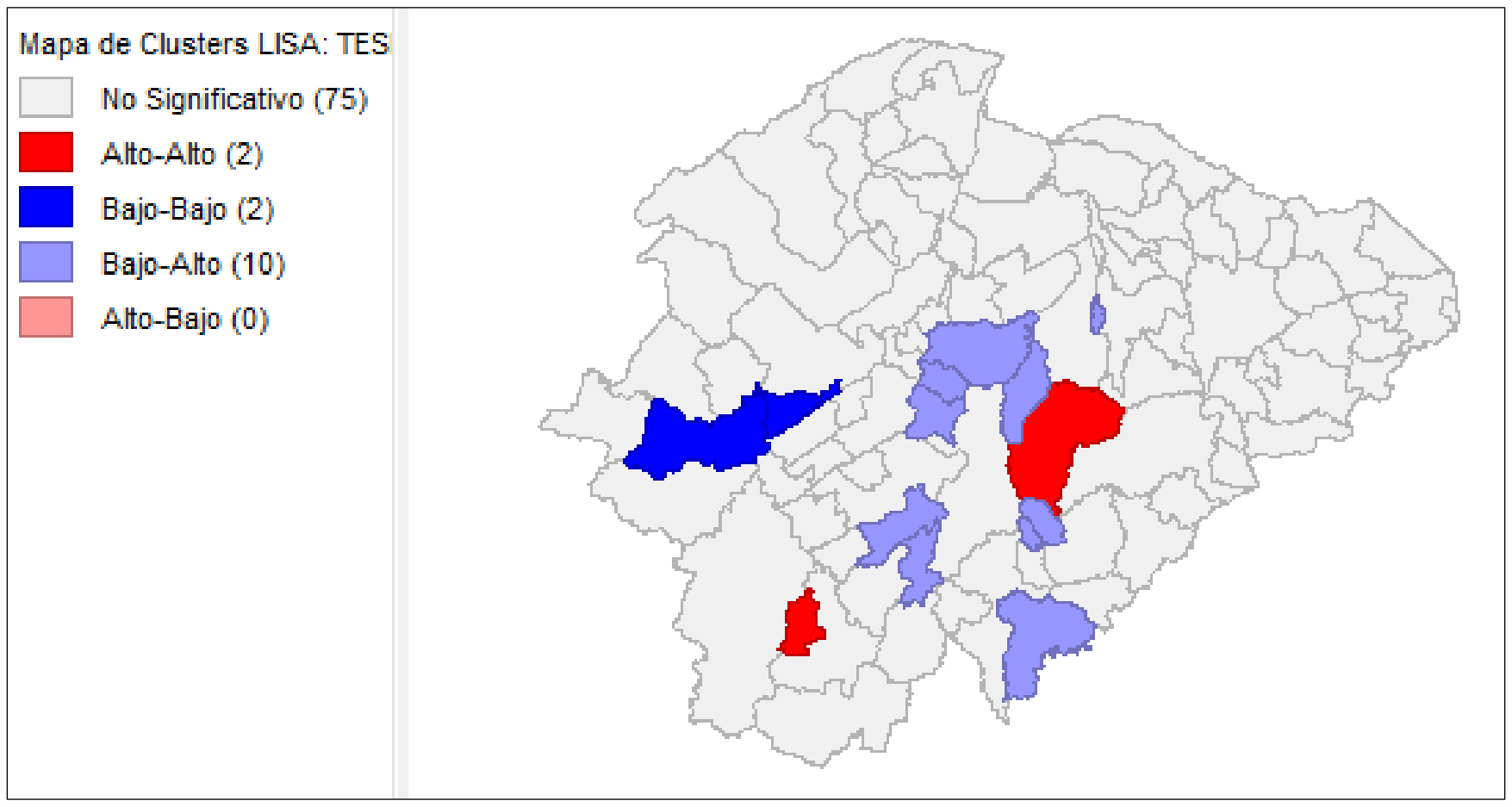

Ilustración 4

*Gráfico de I-Moran LISA de TIC para 2009 (elaboración propia)*

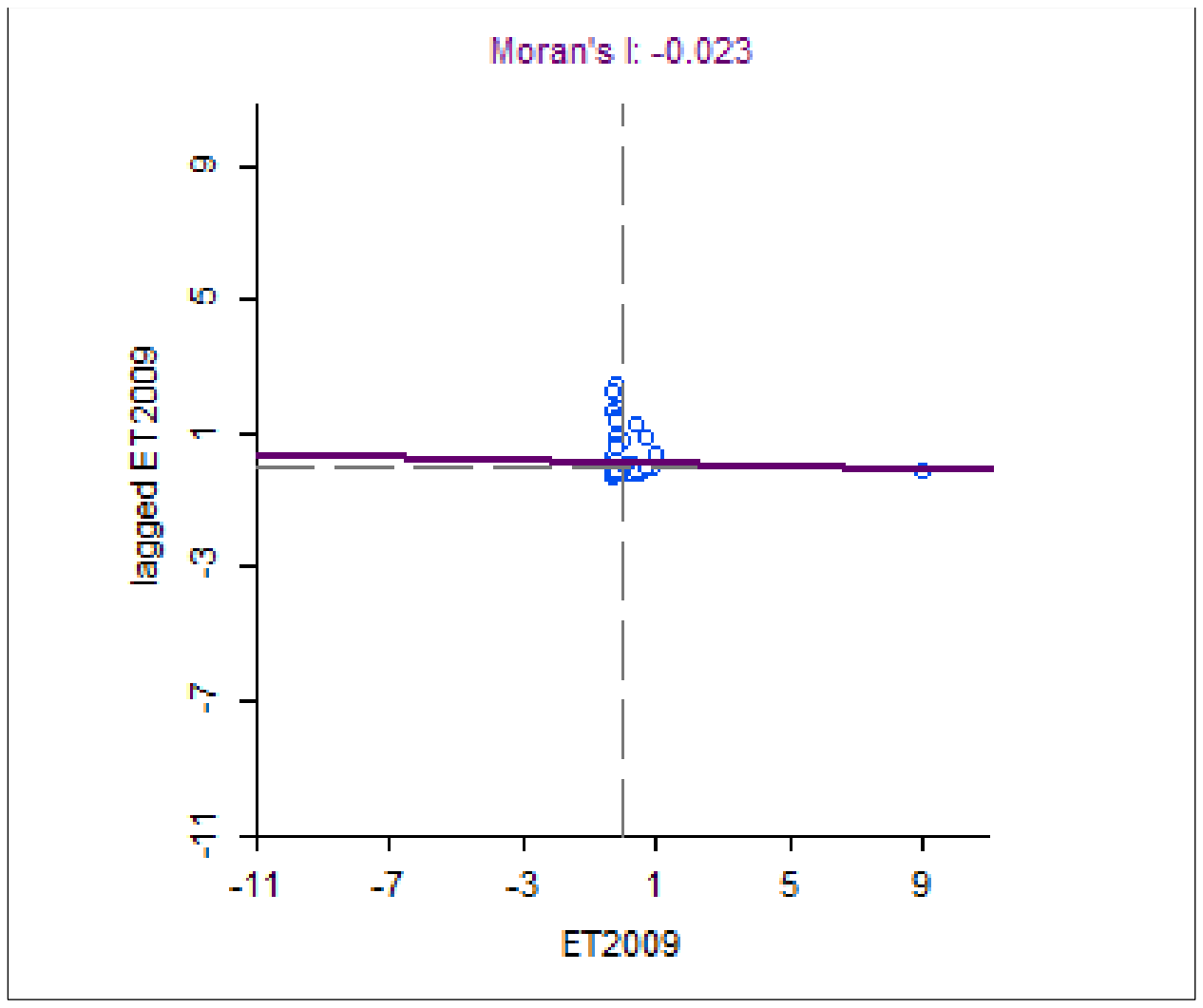

Para el 2009, las áreas de impacto altamente positivo que ejercieron las empresas con el rubro de informática, tecnología y comunicaciones en relación con el marco geográfico de estudio identificaron a los Municipios de Pirapora con *I Moran* = .97 y Francisco Sa con *I Moran* = .35 los mismos con impacto alto-alto y autocorrelación espacial positiva, lo que significa que la influencia (Fenómeno Contagio) de Montes Claros con un $R^2$ = .998 como muestra la evidencia en este estudio, se expande a contigüidad de segundo orden como es el caso del Municipio de Pirapora quien alcanzo un $R^2$ = .972.

De la misma forma se identificó a los Municipios de Sao Romao con *I Moran* = -.29 e Icai de Minas con *I Moran* = -.25 con impacto bajo-bajo y autocorrelación espacial negativa. Lo que sugiere que estos Municipios son influenciados significativamente por Municipios contiguos con bajo impacto de las empresas de tecnología en el producto interno bruto municipal, como es el caso de Januária que alcanzo un $R^2$ = .525.

De igual modo, se identificó diez Municipios compuestos por Sao Joao de Lagoa con *I Moran* = -.25, Claro dos Pecoes con *I Moran* = -.19, Mirabela con *I Moran* = -.19, Patis con *I Moran* = -.28, Sao Joao da Ponte con *I Moran* = -.15, Olhos de Agua con *I Moran* = -.217, Capitán Eneas con *I Moran* = -.23, Glaucilandia con *I Moran* = -.27, Juramento con *I Moran* = -.24 y Nova Porterinha con *I Moran* = -.23 con impacto bajo-alto y autocorrelación negativa; lo que significa que estos Municipios están rodeados de regiones con alta influencia contigua de las empresas de tecnología de la información, como es el Municipio de Montes Claros con un $R^2$ = .998, Pirapora con un $R^2$ = .972 y Bocaiúva con un $R^2$ = .997; para lo cual, el efecto contagio se expande a contigüidad de segundo orden también.

Mapa 29

*Análisis de autocorrelación espacial local de TIC para 2017 (elaboración propia)*

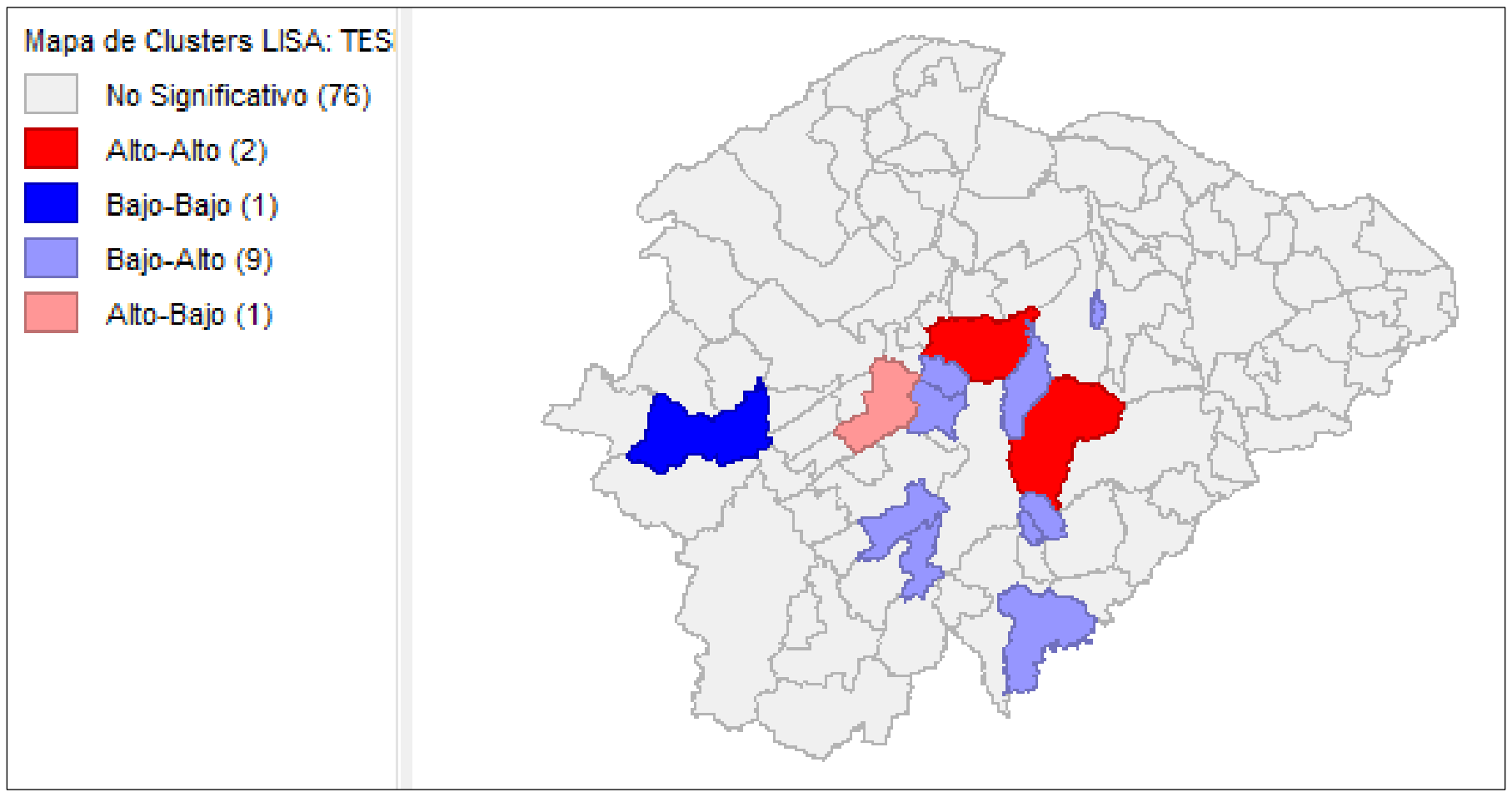

Ilustración 5

*Gráfico de I-Moran LISA de TIC para 2017 (elaboración propia)*

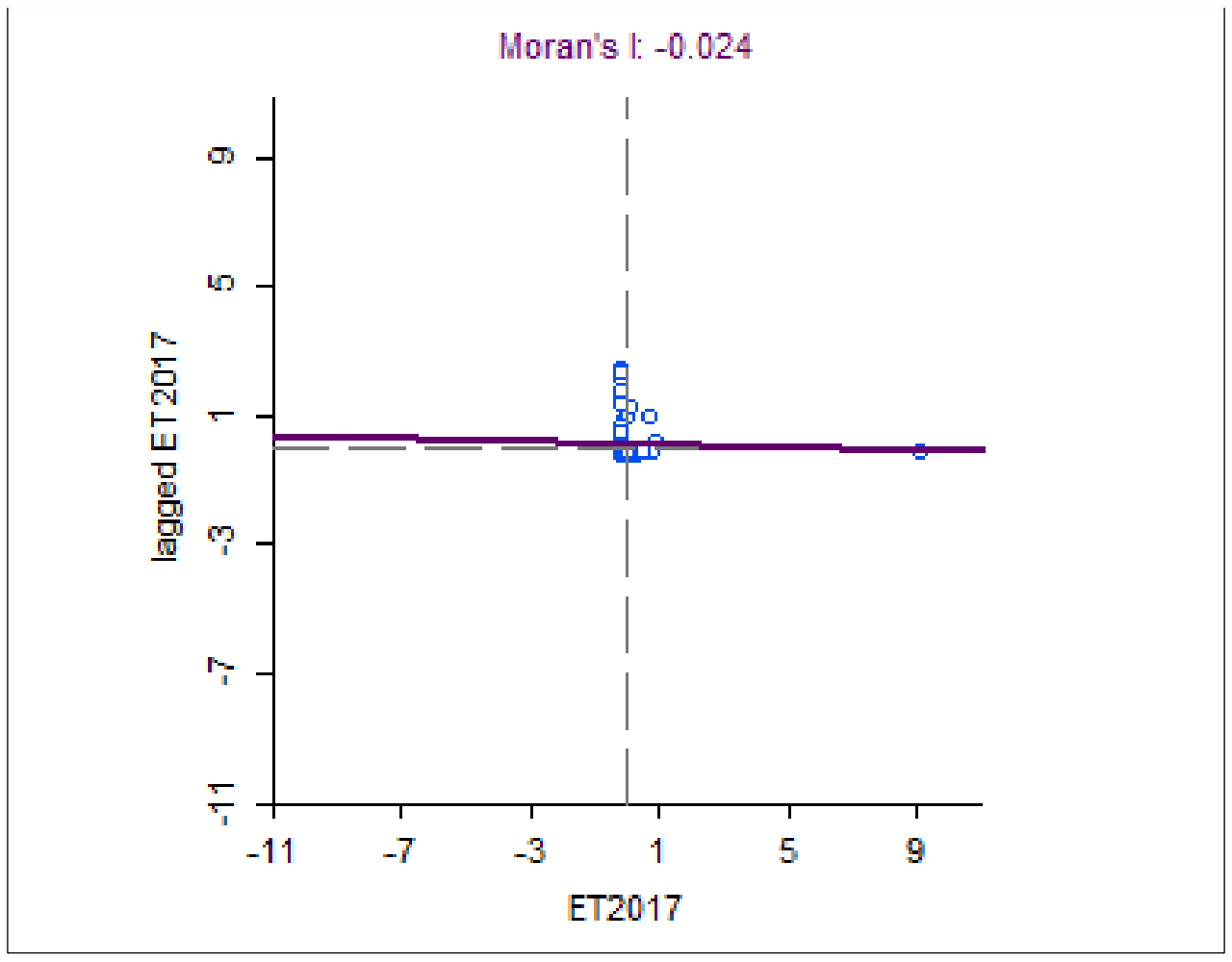

Para el 2017, las áreas de impacto positivo que ejercieron las empresas con el rubro de informática, tecnología y comunicaciones en relación con el marco geográfico de estudio identificaron a los Municipios de Francisco Sa con *I Moran* = .10 y Sao Joao da Ponte con *I Moran* = .01 las cuales tuvieron un impacto alto-alto y autocorrelación espacial positiva. Lo que significa que la influencia (Fenómeno Contagio) de Montes Claros con un $R^2$ = .999 se expande a contigüidad de primer orden afectando positivamente a sus Municipios vecinos.

También tenemos al Municipio de Brasilia de Minas con *I Moran* = .087 con impacto alto-bajo y autocorrelación espacial positiva. Lo que sugiere que este Municipio fue influenciado significativamente por Municipios contiguos con bajo impacto de las empresas de tecnología como es el caso de Januária donde el impacto de las empresas de tecnología en el producto interno bruto municipal que apenas alcanzo un $R^2$ = .498.

De la misma forma se identificó al Municipio de Sao Romao con *I Moran* = -.16 con impacto bajo y autocorrelación espacial negativa. No obstante, en este estudio se encontró que en el Municipio de Icai de Minas para el año 2009 las empresas de tecnología ejercieron un impacto bajo-bajo sobre esa región en comparación con el 2017 donde no existió impacto.

De igual modo se identificó diez Municipios compuestos por Sao Joao de Lagoa con *I Moran* = -.228, Claro dos Pecoes con *I Moran* = -.19, Mirabela con *I Moran* = -.19, Patis con *I Moran* = -.24, Olhos de Agua con *I Moran* = -.16, Capitán Eneas con *I Moran* = -.16, Glaucilandia con *I Moran* = -.21, Juramento con *I Moran* = -.22 y Nova Porterinha con *I Moran* = -.15 con impacto bajo-alto y autocorrelación negativa; lo que significa que estos Municipios están rodeados de regiones con alta influencia contigua de las empresas de tecnología de la información, como es el Municipio de Montes Claros con un $R^2$ = .999, Pirapora con un $R^2$ = .979 y Bocaiúva con un $R^2$ = .989; para lo cual, el efecto contagio se expande a contigüidad de segundo orden también.

#### 4.6.6.1.2 Impacto de las ocupaciones relacionadas el rubro de informática, tecnología y comunicación

Mapa 30

*Análisis de autocorrelación espacial local de CBO para 2009 (elaboración propia)*

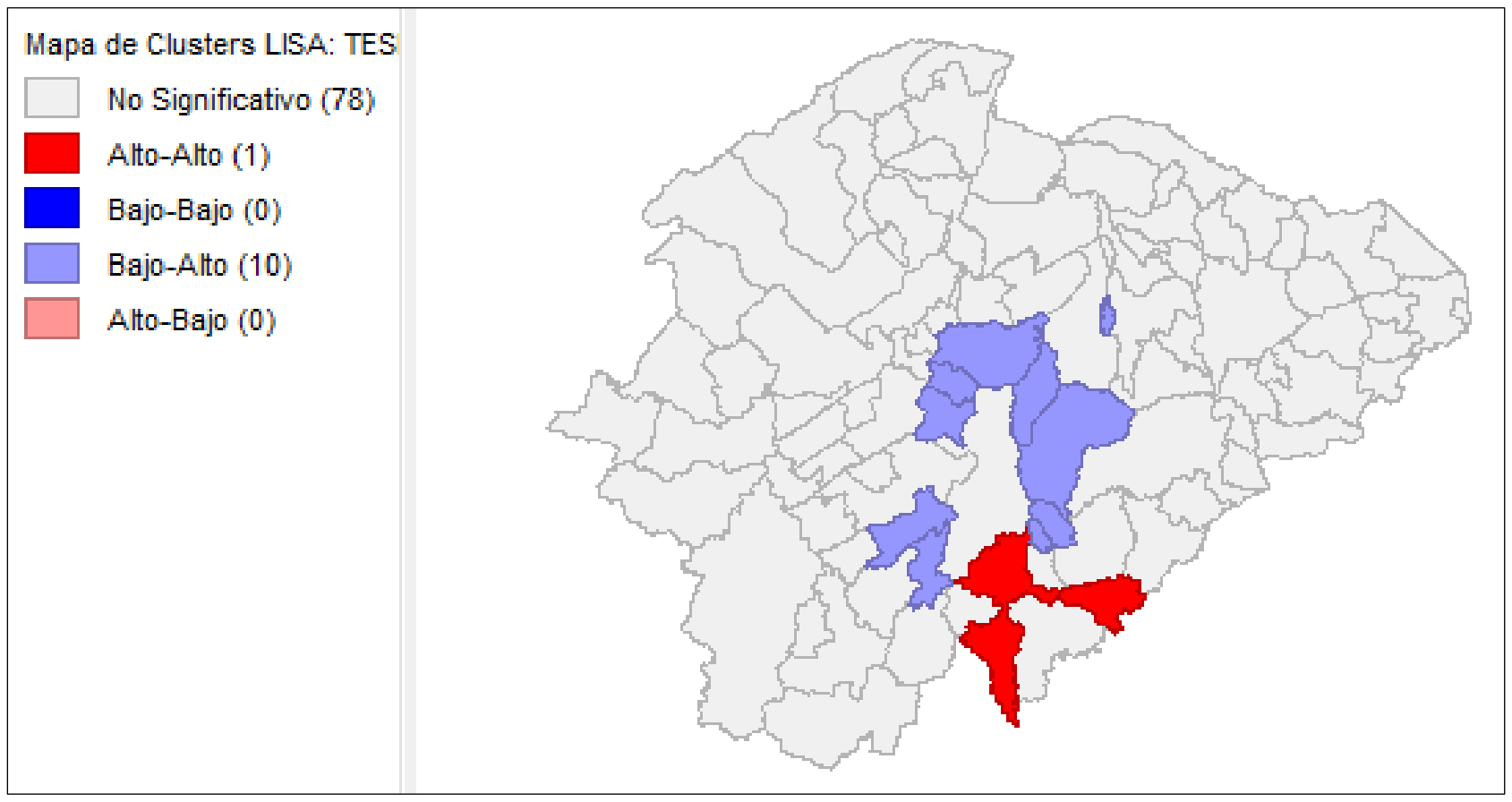

Ilustración 6

*Gráfico de I-Moran LISA de CBO para 2009 (elaboración propia)*

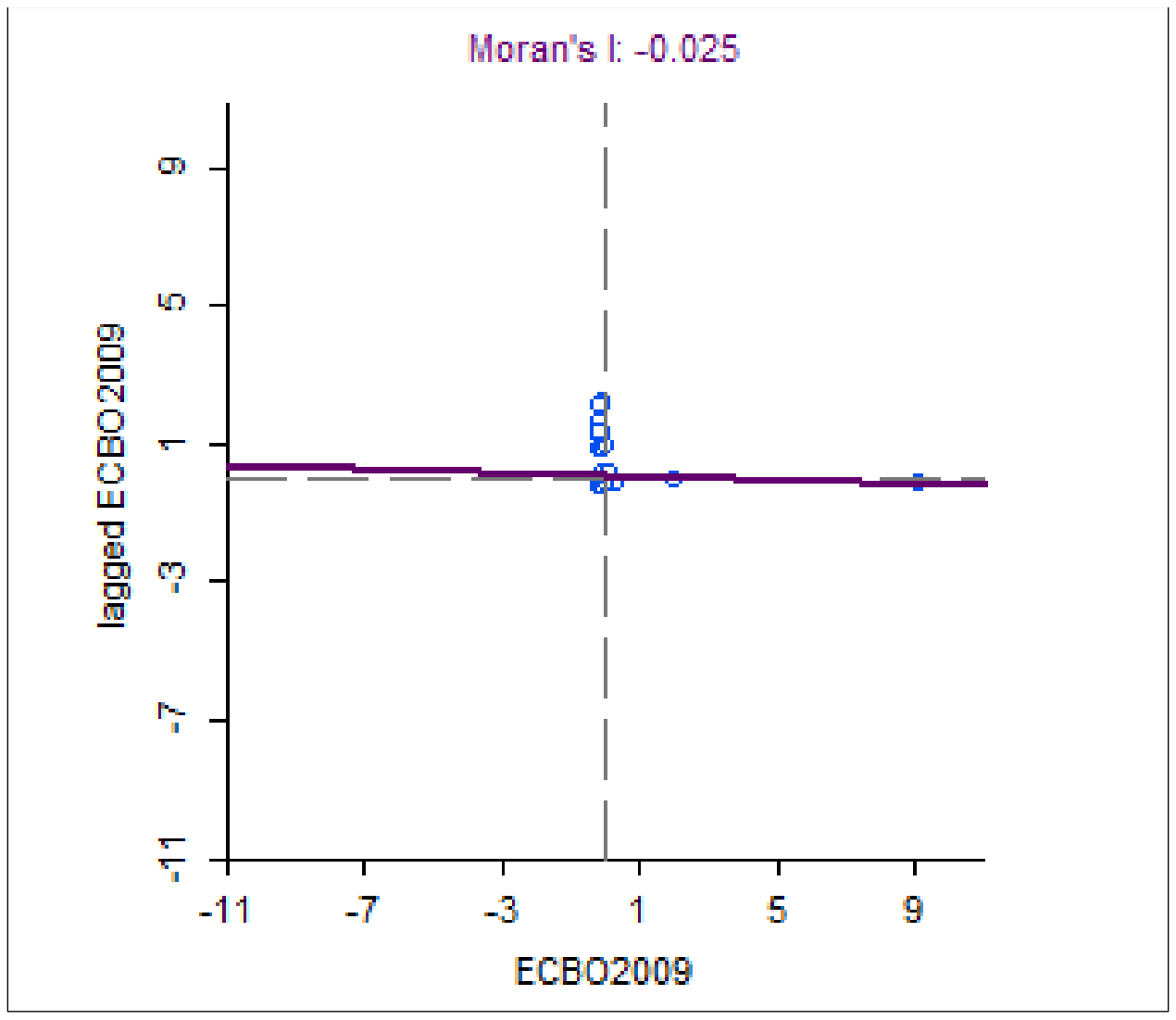

Para el 2009, las áreas de impacto positivo que ejercen las ocupaciones relacionadas al rubro de informática, tecnología y comunicaciones en relación con el marco geográfico de estudio identifico al Municipio de Bocaiúva con *I Moran* = .01 con impacto alto y autocorrelación espacial positiva. Lo que significa que la influencia (Fenómeno Contagio) de Montes Claros con un $R^2$ = .998 se expande a contigüidad de primer orden afectando positivamente a sus Municipios vecinos.

De igual modo se identificó diez Municipios compuestos por Sao Joao de Lagoa con *I Moran* = -.15, Claro dos Pecoes con *I Moran* = -.99, Mirabela con *I Moran* = -.099, Patis con *I Moran* = -.15, Sao Joao da Ponte con *I Moran* = -.07, Capitán Eneas con *I Moran* = -.12, Francisco Sa con *I Moran* = -.157, Glaucilandia con *I Moran* = -.157, Juramento con *I Moran* = -.157 y Nova Porterinha con *I Moran* = -.157 con impacto bajo-bajo y autocorrelación negativa.

Esto significa que estos Municipios están rodeados de un Municipio con alta influencia contigua de las ocupaciones relacionadas al rubro de tecnología de la información, como es el Municipio de Montes Claros con un $R^2$ = .998; para lo cual, el efecto contagio se expande a contigüidad de primer orden.

Mapa 31

*Análisis de autocorrelación espacial local de CBO para 2017 (elaboración propia)*

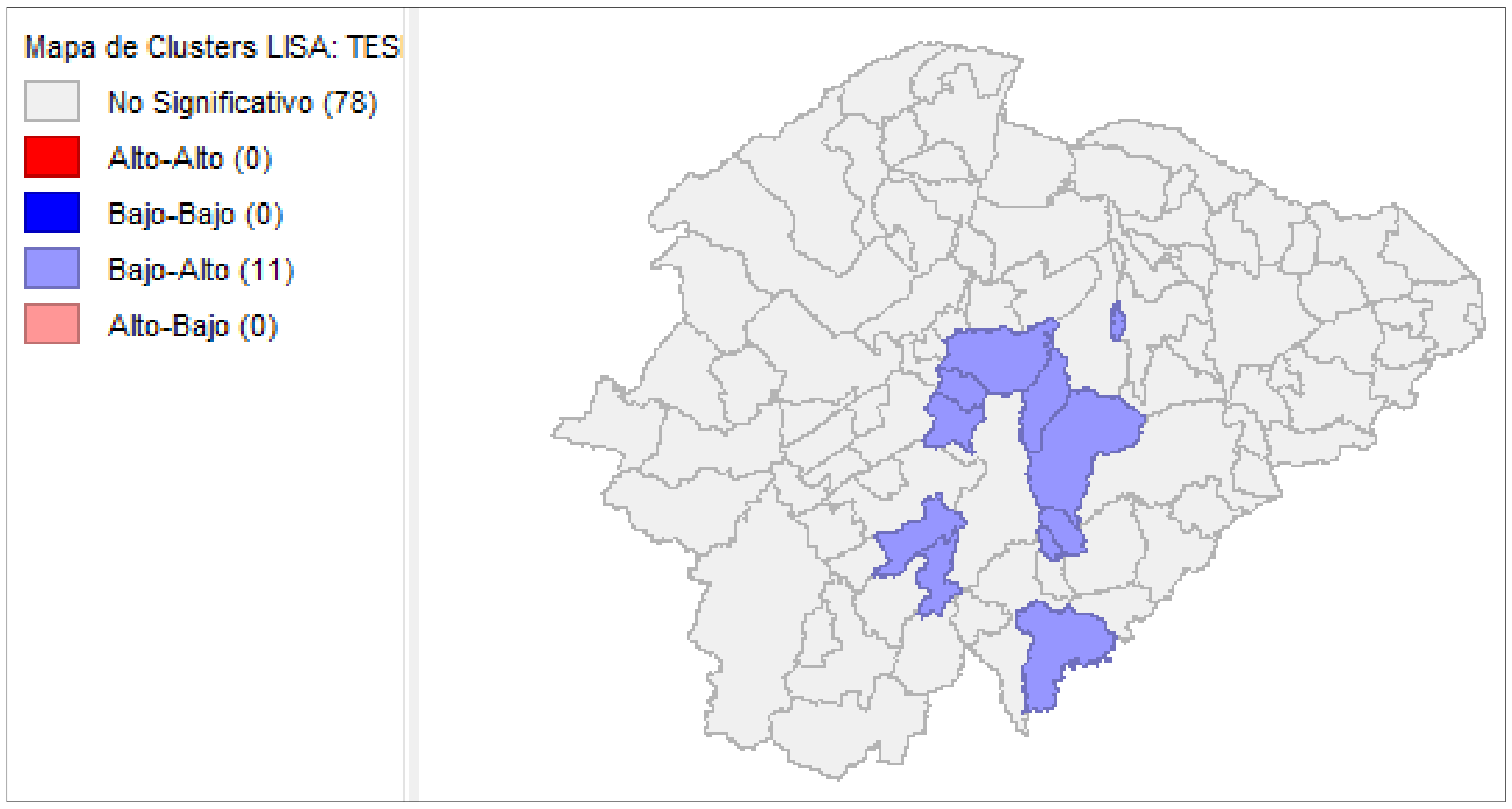

Ilustración 7

*Gráfico de I-Moran LISA de CBO para 2017 (elaboración propia)*

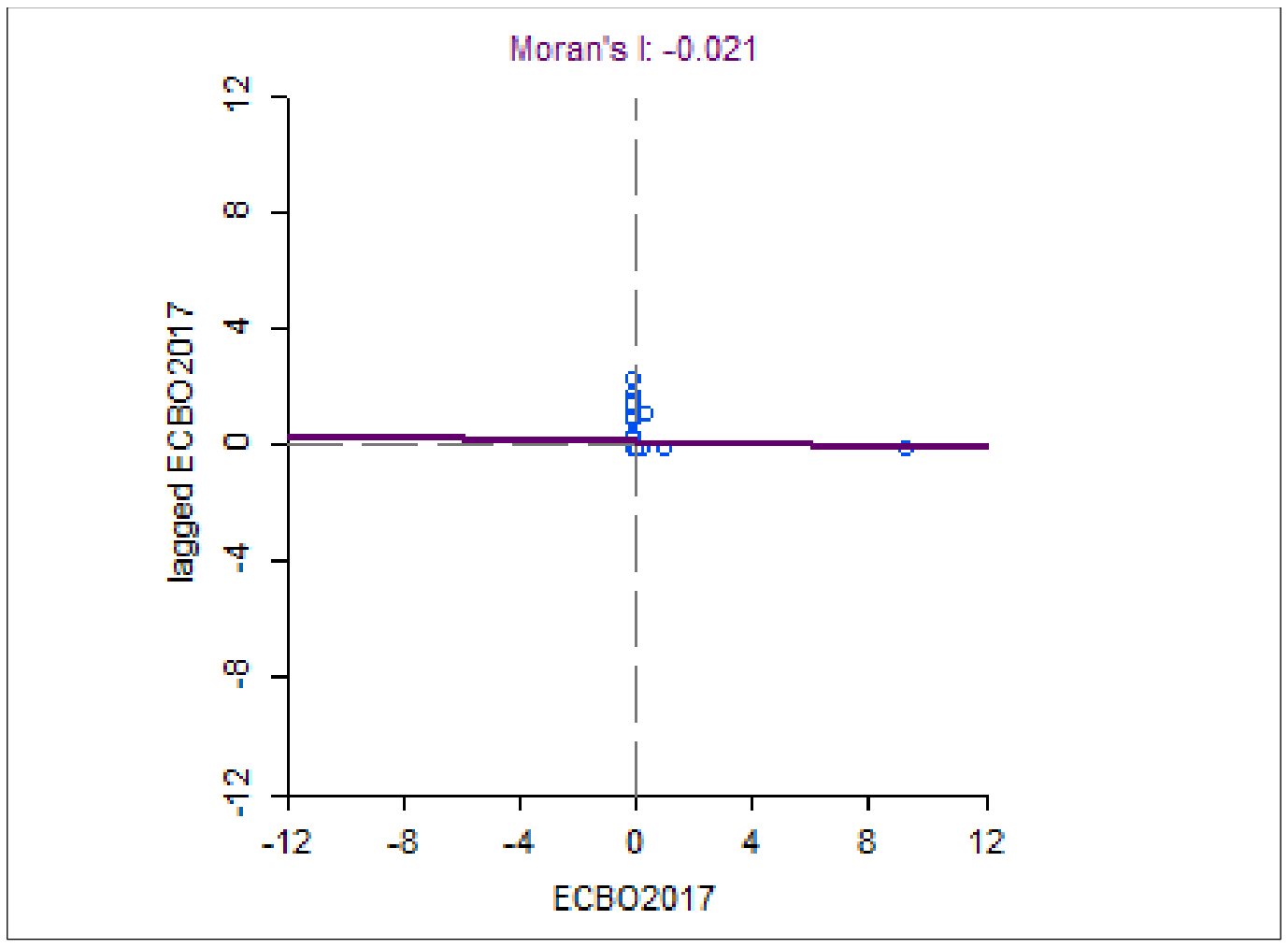

Para el 2017, no se identificó municipios con impacto alto-alto y autocorrelación positiva en relación con las ocupaciones relacionadas al rubro de informática, tecnología y comunicaciones, a diferencia del año 2009 donde se identificó al Municipio de Bocaiúva. No obstante, dentro de los hallazgos que once Municipios compuestos por Sao Joao de Lagoa con *I Moran* = -.14, Claro dos Pecoes con *I Moran* = -.14, Mirabela con *I Moran* = -.12, Patis con *I Moran* = -.14, Sao Joao da Ponte con *I Moran* = -.14, Olhos de Agua con *I Moran* = -.12, Capitán Eneas con *I Moran* = -.106, Francisco Sa con *I Moran* = -.106, Glaucilandia con *I Moran* = -.14, Juramento con *I Moran* = -.14 y Nova Porterinha con *I Moran* = -.12 tuvieron un impacto bajo-alto y autocorrelación negativa.

Lo que significa que estos Municipios están rodeados de regiones con alta influencia contigua de las ocupaciones relacionadas al rubro de tecnología de la información, como es el Municipio de Montes Claros con un $R^2$ = .999 y Bocaiúva con un $R^2$ = .989; para lo cual, el efecto contagio se expande a contigüidad de primer orden.

# CAPITULO V

# DISCUSIÓN Y CONCLUCIONES

## 5.1 Conclusiones

Este capítulo presenta las conclusiones, contribuciones, implicaciones académicas y prácticas, limitaciones, recomendaciones y resumen de conclusiones. Identificándose nuevas y desafiantes líneas de investigación vinculadas a las variables de estudio, así como recomendaciones para el desarrollo de nuevos proyectos de gobernanza regional.

Esta investigación identificó y estudió al conocimiento especializado, asentamiento humano y actividad económica digital como los factores que están relacionados al crecimiento económico regional en 89 municipios de la mesorregión norte del Estado de Minas Gerais. Este estudio midió al conocimiento especializado por medio de los indicadores: instituciones universitarias privadas y públicas, docentes con grados de doctorado, maestría y técnica, así como la aportación a municipios con fondos federales para la educación, y encontró que hubo un impacto significativo alto en las siguientes cuatro [4] regiones: Salinas, Pirapora, Bocaiúva y Montes Claros.

No obstante, en dos [2] regiones Janauba y Januária los hallazgos encontraron un impacto bajo; de igual forma se identificó que la región de Grao Mogol no fue impactada; fenómeno que se discutirá a profundidad posteriormente. Por lo tanto, estos hallazgos confirman y validan el argumento de la primera hipótesis de este estudio [$H_1$], la misma establece la gran importancia que tienen los individuos especializados con conocimiento científico tecnológico en el desarrollo de las empresas de tecnología de información y comunicaciones como lo argumenta Mcadam & Mcadam (2016); Vega, Miranda, Mera, & Mayo (2018); Giones & Brem (2018).

Estos hallazgos también son cónsonos con Vinod Kumar & Dahiya (2016), quienes establecen que los trabajadores urbanos en los países en vías de desarrollo están involucrados en empleos vulnerables, a menudo llamados informales o basados en rubros tradicionales como la agropecuaria. Esto debido a que están en un entorno socioeconómico con limitaciones de educación y capacitación vocacional formal y continua, lo cual inhibe la expansión de las economías urbanas.

Este factor contribuyó a problemas de aglomeración de recursos y capital reflejados en el asentamiento humano entre 2009 y 2018 por los Municipios de Montes Claros y Pirapora, El problema de esta aseveración es que en tu marco conceptual no se establece la relación de asentamiento humano con crecimiento económico y desarrollo social, de sus municipios contiguos (Rahman, Mohiuddin, Kafy, Sheel, & Di, 2018). Este argumento es apoyado por Ma & Zhao (2019), quienes sugieren que en los procesos de industrialización y urbanización las ciudades vinculadas al desarrollo de innovación tecnológica como es el caso de Montes Claros y Pirapora, se convierten en polos regionales de desarrollo económico (Nagy, Oláh, Erdei, Máté, & Popp 2018).

Esto debido a que las características de infraestructura con innovación tecnológica e inversión de capital pernoctan como motores de crecimiento y desarrollo económico, promovidos por la importancia que tienen los individuos con conocimiento especializado y la tendencia a concentrarse en ubicaciones geográficas (Rahman et al., 2018). Por consiguiente, este estudio midió al asentamiento humano por medio de la cantidad de habitantes en los 89 municipios de la región estudiada por nueve años, y encontró que hubo impacto positivo por parte de la actividad económica digital al asentamiento humano validando la segunda hipótesis del estudio [$H_2$], el cual permite observar al crecimiento de zonas pobladas como lugares de innovación social, mediante el impulso de los servicios de las tecnologías de la información que contribuyen a la obtención de ventaja competitiva, transformando las ciudades en unidades regionales con desarrollo económico, cónsono con lo que establecen Ma & Zhao (2019); Chowdhury, Bhaduri, & McKee (2018); Rahman et al. (2018); Meijer & Bolívar, 2016; Li, & Nuccialleri (2016).

Por último, esta investigación midió el impacto de la actividad económica digital por medio de los indicadores: cantidad de empresas y ocupaciones pertenecientes al rubro de la informática, tecnología y comunicaciones, entre el 2009 y 2017 en el producto interno bruto municipal, donde se encontró impacto significativo alto en el crecimiento económico regional en sus siete [7] regiones: Salinas, Montes Claros, Janauba, Januária, Grao Mongol, Pirapora y Bocaiúva que componen esta mesoregión.

No obstante, los hallazgos también permitieron descubrir que en seis [6] de las siete [7] regiones las empresas con el rubro de informática, tecnología y comunicaciones tuvieron un mayor impacto significativo en el crecimiento económico regional que las ocupaciones relacionadas al rubro; este hallazgo quedó en contraste en la región de Januária, la cual experimento su impacto más alto con el área de las ocupaciones que en las empresas con el rubro de informática, tecnología y comunicaciones.

Esta evidencia valida y confirma la tercera hipótesis del estudio [$H_3$] la cual establece la capacidad que tiene la actividad económica digital para generar, productividad, crecimiento económico y bienestar, mediante la construcción de polos regionales de desarrollo económico (PRDE); así como el desenvolvimiento de nuevos modelos de negocios basados en innovación tecnológica y conocimiento especializado. Esta capacidad de la actividad económica digital impulsa el crecimiento y desarrollo económico de ciudades que han integrado progresivamente servicios de tecnologías de la información y comunicaciones, así como otras infraestructuras y servicios urbanos, enunciado respaldado por Brynjolfsson et al. (2018); Rahman et al. (2018); dos Santos (2017); Gomez (2017); Mendez (2016); Manuel et al. (2016); Century et al. (2007) & Cortright (2001).

## 5.2 Contribuciones e implicaciones del estudio

La revolución de las tecnologías de la información han producido efectos de difusión del conocimiento al incorporar la innovación tecnológica en el capital humano mediante el conocimiento especializado (Agrawal et al., 2019). Debido a lo cual la actividad económica digital ha generado productividad, crecimiento económico y bienestar, mediante la construcción de economías regionales (Brynjolfsson, Collis, Diewert, Eggers, & Fox, 2018; Mendez, 2016), las cuales han adoptado nuevos modelos de negocios radicalmente diferentes, estableciendo nuevas formas de organización (Li, 2006, 2014).

Esto contribuye al aprovechamiento de conocimiento especializado (Manesh, Pellegrini & Marzi, 2019), manteniendo una ventaja competitiva a largo plazo generado por los efectos en la difusión del conocimiento por parte de los centros de investigación universitarios (Du et al., 2018), cómo se pudo observar en el estudio; lo cual no es nuevo, ya que Agrawal et al. (2019), sostuvo que el conocimiento especializado en relación con la actividad económica digital minimiza la dependencia a externalidades tecnológicas.

Esto ha llevado a organizaciones de diferentes sectores, a reevaluar sus estrategias y operaciones existentes para enfatizar sus objetivos en la necesidad de construir economías regionales basadas en la innovación y conocimiento (Méndez, 2016), como es el caso de los Municipios de Montes Claros, Pirapora, con resultados altos en procesos productivos no tradicionales.

Lo que no sucedió con el Municipio de Grao Mogol, el cual tuvo un impacto nulo entre el conocimiento especializado, la actividad economía digital y el crecimiento económico municipal dentro del marco regional de estudio. Por consiguiente, para encontrar las causas que contribuyan a los hallazgos de este estudio, se descompuso el valor añadido bruto (VAB) de dicho municipio y se descubrió poca generación en los sectores de servicios como es la educación e industria llevándolo a resultados desfavorables económicamente en relación con otros Municipios como Montes Claros y Pirapora.

Tabla 43

*Matriz de valor añadido bruto (VAB) de tres ciudades del norte de Minas Gerais (Unidad: R$ x1000).*

| **Municipio** | **2017 Agropec.** | **2009 Agropec.** | **2017 Educ/otros** | **2009 Educ/otros** | **2017 Industria** | **2009 Industria** |
|---|---|---|---|---|---|---|
| Grao Mogol | 82,836.31 | 8,855 | 43,569.44 | 18,827 | 84,205.77 | 167,114 |
| Pirapora | 30,134.29 | 17,800 | 618,643,59 | 251,077 | 586,648.81 | 311,239 |
| Montes Claros | 112,319.70 | 78,625 | 4,769.729.31 | 2,041,404 | 1,556.646.34 | 879,817 |

Este estudio descubrió mediante el análisis exploratorio espacial microrregiónal que el desarrollo económico municipal tiende hacer afectado por modelos tradicionales como el sector agropecuario identificándose al grado de escolaridad de los empresarios y trabajadores, como un factor relacionado a nuestra hipótesis [$H_1$]. En esta investigación tenemos el caso del Municipio de Grao Mogol en relación con el municipio de Pirapora, el cual tuvo una marcada diferencia para el 2017 en la generación de la economía basado en el sector agropecuario; de forma similar lo encontramos en su relación con Montes Claros, donde la existencia de valores altos se refleja en todos los años y en todos los sectores.

Esto llevó al análisis de los resultados del impacto significativo que ejercieron las empresas con el rubro de informática, tecnología y comunicaciones entre 2009 al 2017 en relación con el producto interno bruto en tres (3) Municipios compuestos por: Capitán Eneas, Francisco Sa, y Glaucilandia. Así como el análisis de los resultados del impacto significativo de las ocupaciones relacionadas al rubro de informática, tecnología y comunicaciones entre 2009 al 2017 en relación con el producto interno bruto en tres [3] Municipios compuestos por: Capitán Eneas, Francisco Sa, y Glaucilandia. Concluyendo que las causas relacionadas a los hallazgos se encuentran en la poca generación económica de los sectores de servicios como es la educación e industria tal como lo establecen Brynjolfsson et al. (2018); dos Santos (2017); Gómez (2017); Méndez (2016); Manuel et al. (2016); Century et al. (2007) & Cortright (2001).

Ilustración 8

*Valor adicionado bruto a precios corrientes de actividad económica, Educación con otros servicios (izq.) y agricultura (der.) respectivamente (Unidad: R$ x1000*

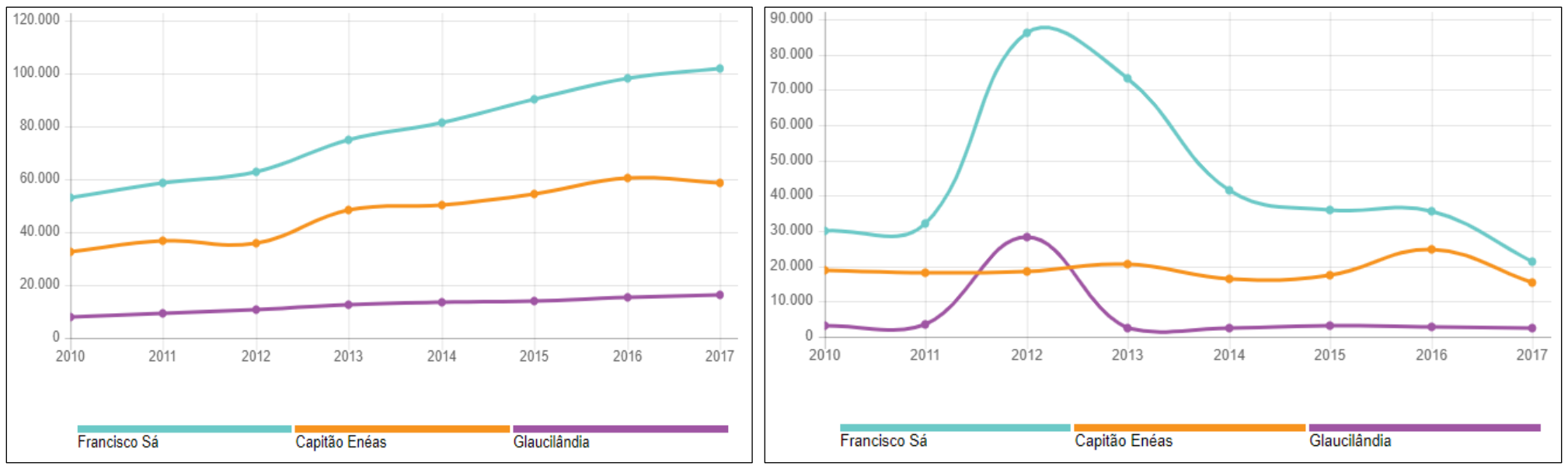

Por lo tanto, dentro de las contribuciones está el sugerir a los gobiernos municipales y empresa privada la necesidad de un cambio de paradigma en el desarrollo de nuevos modelos de negocios vinculados con la innovación y conocimiento especializado, apoyado por los resultados de este estudio empírico el cual encontró que la actividad económica digital a diferencia de las actividades tradicionales como la agropecuaria impacta significativamente el crecimiento económico regional.

Otra contribución de este estudio fue haber identificado a los polos regionales de desarrollo económico (PRDE), compuestas por Municipios que tienden a concentrar empresas en ubicaciones geográficas como lo establecen Krugman (1991) & Marshall (1920); ya que de esta manera aprovechan una serie de beneficios como: la mano de obra especializada, suministro, insumos, conocimiento técnico y más. Los trabajadores se concentran en los (PRDE) por la demanda de recursos humanos calificados, y a su vez otras empresas que requieran mano de obra con alguna calificación específica se establecerán en esa región por dicha razón, como lo menciona Krugman (1993) &

Soriano (2008), evitando los costos económicos y de tiempo en el reentrenamiento e inversión en conocimiento especializado.

Esto demostró lo que sucede con los municipios de Montes Claros y Pirapora en la región norte de Minas Gerais los cuales frenan el desarrollo de los municipios pequeños, como es el caso de Grao Mogol, siendo este hallazgo una aportación para el desarrollo de una nueva discusión al planteamiento expuesto por Lu, Zhang, Sun, & Li (2018), quienes establecen que la descentralización urbana aliviaría las presiones urbanas en las ciudades más grandes promoviendo el desarrollo de ciudades pequeñas.

A la luz de este estudio se puede observar que existen otros factores adicionales a la descentralización urbana que pueden promover el desarrollo, y estos factores están asociados al conocimiento especializado y a la actividad económica digital, debido a que el impacto que ocurren en unidades geográficas cercanas favorece el comportamiento de regiones próximas, evidenciándose la existencia de clústeres como lo argumentan Sabogal (2013); Siabato & Guzmán (2019). Este es el caso del efecto contagio de los Municipios de Pirapora y Francisco Sa; a diferencia de Grao Mogol donde la inversión en educación, industria y servicios es mínima.

Ilustración 9

*Valor adicionado bruto a precios corrientes de actividad económica, actividad educación con otros servicios (Unidad: R$ x1000)*

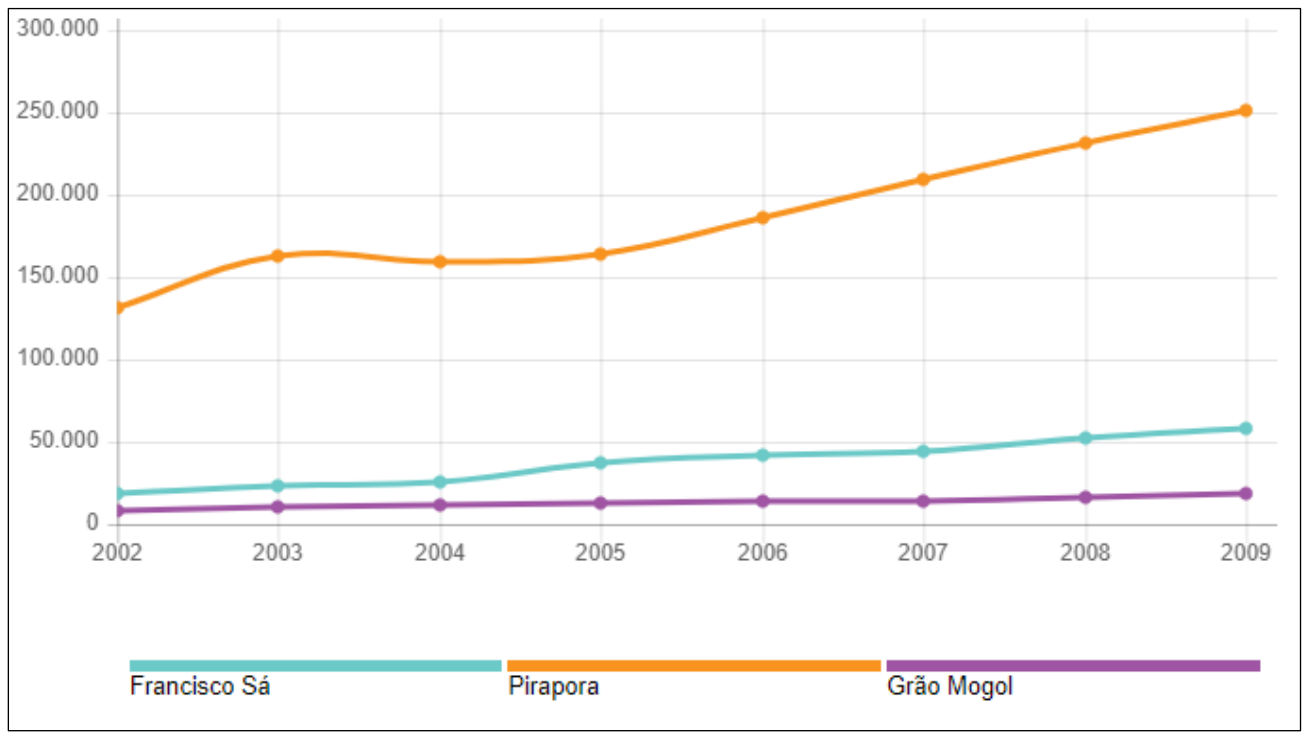

Para concluir, la contribución de este estudio en el desarrollo económico regional demostró que un municipio depende de los valores observados en municipios vecinos (Mexiquense et al., 2005), como es el caso de los Municipios de Montes Claros y Pirapora, los cuales son los principales polos regional de desarrollo económico favoreciendo a regiones próximas de primer orden como es el caso de los Municipios de Francisco Sa, Sao Joao da Ponte, Mirabela, Capitán Eneas, Glaucilandia, Claro dos Pecoes y Patis como lo argumentan Sabogal (2013); Siabato & Guzmán (2019).

Ilustración 10

*Valor adicionado bruto a precios corrientes de actividad económica, actividad educación con otros servicios (Unidad: R$ x1000)*

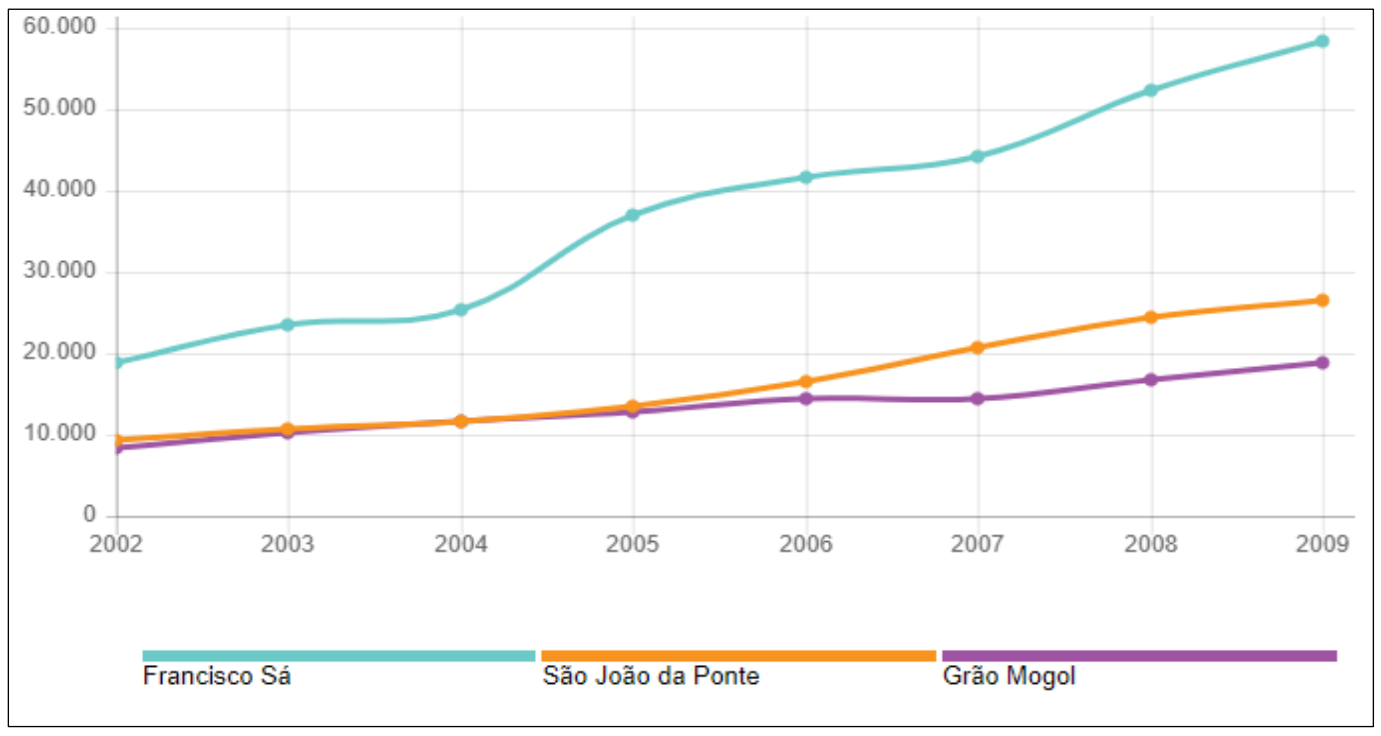

En relación con las ocupaciones vinculadas con el rubro mencionado anteriormente los hallazgos identificaron efecto contagio significativo en el municipio de Bocaiúva, esto significa que el impacto que ocurren en unidades geográficas cercanas favorece al comportamiento de regiones próximas como lo argumentan Sabogal (2013); Siabato & Guzmán (2019). Aquí se puede ver que el efecto contagio de segundo orden llega al municipio de Bocaiúva, el cual es capital de región y se encuentra ubicado entre Montes Claros y Pirapora polos regionales de crecimiento económico.

Ilustración 11

*Valor adicionado bruto a precios corrientes de actividad económica, actividad educación con otros servicios (Unidad: R$ x1000)*

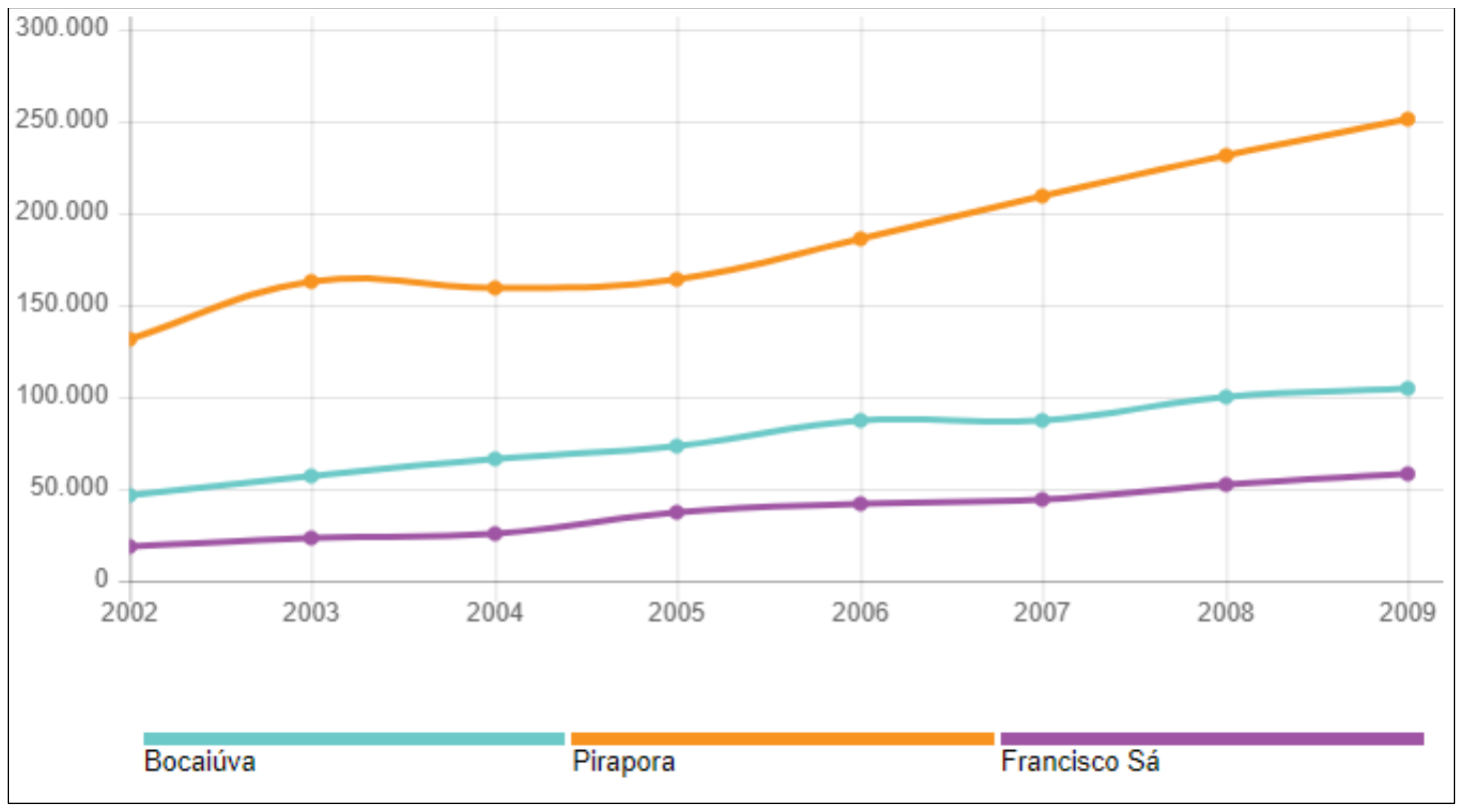

Por lo tanto, el comportamiento de una variable está produciendo efectos contrarios en sus vecinos como lo argumenta Moreno y Vayá (2000) debido al efecto contagio de polos regionales de desarrollo económico (PRDE), compuesto por las empresas que tienden a concentrarse en ubicaciones geográficas.

### 5.2.1 Implicaciones académicas

Aunque la aceleración urbanística puede ser visto como un factor en el crecimiento económico (Cvetanović, Filipović, Nikolić, & Belović, 2015). La teoría de crecimiento económico usada en este estudio establece que el nivel de producción aumenta por la inversión en el capital humano, la transferencia del conocimiento y las tecnologías de la información, parte fundamental del algoritmo creado para este estudio, el cual está basado en la función Cobb-Douglas (ver apéndice 1).

Dentro de las implicaciones académicas también es importante recalcar que en este estudio se usó un modelo de ecuaciones estructurales (PLS-SEM) en combinación con técnicas de análisis espacial y algoritmos de clasificación de tercera generación, lo que nos permitió encontrar hallazgos sobre patrones o “spots” de concentración económica compuestas por las empresas que tienden a concentrarse en ubicaciones geográficas.

### 5.2.2 Implicaciones prácticas

Las implicaciones prácticas de este estudio buscan fomentar la aplicación de los resultados a nivel de gobiernos municipales, ya que sus decisiones involucran a las comunidades productivas como son los micro emprendedores individuales o MEI, micro y pequeñas empresas o MPE; en la búsqueda de alternativas a la solución de problemas mediante estímulos contributivos a nuevos modelos de negocios basados en innovación y conocimiento especializado.

También se pudo identificar ciertos problemas futuros que pueden presentarse dentro del asentamiento humano ya que el envejecimiento gradual de la población en Minas Gerais para el 2050 es un factor que exige grande atención en regiones con población de edad avanzada; considerando la merma en la población laboral y el alza en los gastos dirigidos hacia la atención de salud geriátrica.

Por último, a nivel región en el norte de Minas Gerais se identificaron polos regionales de desarrollo económico cuyo efecto han producido efectos contrarios en sus vecinos dificultando el crecimiento y desarrollo en algunos municipios de la región y fomentando la desigualdad en la región como lo establecen Ma & Zhao (2019); Forga & Valiente (2014), debido a lo cual los gobiernos municipales afectados por este fenómeno deben tomar decisiones que deberían estar basadas en las recomendaciones de este estudio.

No obstante, el reto de las políticas públicas y privadas a nivel territorial deben basarse en ser "sujetos competitivos" (Boisier, 2016), mediante el uso de la tecnología con la aportación formativa del conocimiento especializado; lo cual es reconocido como motor principal en el desarrollo económico territorial por la literatura existente. Esto, debido a que con una mejor estructura en innovación y tecnología los municipios podrían atraer a nuevas industrias (dos Santos, 2017); ya que regiones con estructuras débiles en innovación tecnológica y conocimiento especializado tienden a tener pocas posibilidades de alcanzar los beneficios y avances generados por regiones más desarrolladas.

## 5.3 Limitaciones del estudio

Esta investigación reconoce a la Tecnología como su mayor limitación por el dinamismo con que cambian con el tiempo. En consecuencia, para investigaciones futuras se debe actualizar los indicadores de estudio considerando el desarrollo de la ontología cibernética. En adición, este estudio se enfrentó a la limitación de la ley No. 13.709 de sigilo de datos económicos y personales en Brasil, que impide obtener datos financieros de las empresas, lo cual limitó el acceso a los mismos para la construcción de más indicadores.

## 5.4 Recomendaciones y Futuros Estudios

Las políticas públicas son decisiones, que involucran a las comunidades y a los ciudadanos en la búsqueda de alternativas a la solución de problemas mediante el accionar gubernamental, dentro de nuestras recomendaciones en política pública se encuentran; primero: el incentivo contributivo a nuevos modelos de negocios basados en innovación y conocimiento especializado; segundo: el desarrollo en las secretarias de educación municipal de nuevos programas de formación especializada no tradicional; y tercero: la creación de clusters de negocios con el fin de compartir experiencias y conocimiento especializado buscando combatir las externalidades tecnológicas.

Debido a que la reinvención del gobierno provoca un cambio de paradigmas en la formulación, implantación, evaluación y control de políticas públicas delimitadas por la innovación, eficiencia y la efectividad que busca como resultado una sana y transparente gestión (Osborne, 1995). Se plantea la siguiente pregunta: ¿De qué manera los gobiernos municipales pueden realizar esta transformación? A la luz de los teóricos de la reinvención del gobierno como Osborne, se cree firmemente en la posibilidad de aplicar los modelos de éxito del sector privado en la administración pública (Manso, 2002).

No obstante, esta recomendación académica mediante el uso de un nuevo modelo catalizador (promotor, coordinador, activador y armonizador) de iniciativas privadas y comunitarias, sugiere un cambio del modelo tradicional de proveedor directo (Osborne, 1995b). Por lo tanto, a nivel de política pública se deben abordar ciertas estrategias que revitalicen a las economías municipales y macrorregionales observando los patrones que envuelven al crecimiento económico regional identificados en este estudio.

Por lo tanto, los Gerentes Públicos, deben establecer estrategias de administración de obtención de resultados (Bryson et al., 2014) para de esa manera brindar servicios gubernamentales de una manera más eficiente especialmente en los procesos que envuelven al gobierno electrónico y la gobernabilidad digital. El gran reto en la economía digital debe estar enfocado en un nuevo modelo de gobernanza, el cual asido recomendado en este estudio, donde gobernar, no es solo gestionar, sino que envuelven los más diversos y complejos procesos de generación y análisis de datos basados en la arquitectura de información (AI) dentro de las soluciones ofrecidas por el gobierno electrónico municipal.

Los municipios pueden realizar estas funciones múltiples a medida que ejercen influencia mediante mejores infraestructuras y servicios (en comparación con sus contrapartes regionales), lo que ayudará a sus economías de aglomeración y los procesos de producción (Kumar & Dahiya, 2016). Esto debiéndose considerar a las tecnologías inteligentes, la colaboración inteligente, una población altamente educada y un gobierno eficiente, lo cual es imprescindible para enfrentar los desafíos de las ciudades modernas, con entornos de transformación humana a través del aprendizaje continuo (Kumar & Dahiya, 2016), lo cual permitirá a las ciudades justificar su propia existencia.

Dentro de futuras líneas de investigación que se sugiere en este estudio se encuentran; primero: La identificación y definición de las áreas de conocimiento especializado que tuvieron mayor impacto en el desarrollo económico regional; para mediante Política Pública promover nuevos modelos académicos que impulsen la actividad económica digital; segundo: Definir las características que agrupan a los polos regionales de desarrollo económico (PRDE) en una región, y como estos contribuyen al incremento de la migración interna y al desarrollo de la concentración urbana; y tercero: El factor de proximidad geográfica en las modalidades "Online2Offline (O2O)" y las implicaciones en los polos regionales de desarrollo económico.

### 5.5 Resumen de conclusiones

Concluyendo con el resumen de este capítulo y cónsonos con los hallazgos encontrados en esta investigación, se puede señalar que esta investigación utilizó una nueva metodología de medición la cual permitió medir el impacto de la actividad económica digital en el crecimiento económico regional.

Además, se identificaron ciertos fenómenos como los polos de crecimiento económico regional que envuelven a las regiones económicamente florecientes con sus vecinos tal como lo establece la teoría del efecto vecino de Tobler quien menciona que en el espacio geográfico todo se encuentra relacionado (Darmofal, 2006).

La notoriedad de los municipios con altos grados de crecimiento económico debido al impacto de la actividad económica digital y conocimiento especializado, hallazgo que a lo largo de estos últimos setenta años fue apoyado por economistas como Solow (1956), Kuznets (1973), Lucas (1988) y Barro (1991), apuntan a la evolución tecnológica como el principal factor del crecimiento económico moderno (Ramfla, 2015).

## APÉNDICES

**Apéndices 1** *Algoritmo de análisis de crecimiento basado en Cobb Douglas*

```
# -*- coding: utf-8 -*-
"""
Edited on Thu Feb 20 16:37:12 2020

@author: Cesar R. Salas
"""
import numpy as np
import matplotlib.pyplot as plt
# En esta Funcion de Crecimiento Regional basado en Cobb-Douglass:
# La expresión A representa a la actividad económica digital; y
# La expresión K representa al conocimiento especializado; sugerido por
# por Pack (1994) y la expresión CER representa al
# crecimiento económico regional, por lo tanto:
# CER[mt] = ( K*[ce]**a) ) * ( A[aed]**(1-a) )
def cobb_douglas(K,A):
    NPOINTS = 100
    A = np.linspace(0,A,NPOINTS)
    y = A**K
    plt.plot(A,y)
    return A
"""
  Para usar este algoritmo los datos deben estar previamente
  estandarizados
  Este algoritmo cuenta con un gráfico bidimensional plot
  que representa a las expresiones A,K
"""
"""
  Introduzca los valores usando la definicion cobb_douglas
  recuerde que los valores ingresadas en las expresiones A,L
  deben estar estandarizadas
"""
def cobb_douglas_y(K,A):
    y = A**K
    return y
```

## Referencias